\documentclass[a4paper,fleqn,usenatbib,useAMS,usedcolumns]{mnras}

\usepackage[T1]{fontenc}
\usepackage{ae,aecompl}
\usepackage{comment}
\usepackage{threeparttable}
\usepackage{booktabs, caption, makecell}

\usepackage{graphicx}	
\usepackage{amsmath}	
\usepackage{gensymb}
\usepackage{siunitx}
\usepackage{comment}

\usepackage{xr}
\newcommand{\mgii}{\mbox{Mg\,{\sc ii}}}

\newcommand{\civ}{\mbox{C\,{\sc iv}}}

\newcommand{\nv}{\mbox{N\,{\sc v}}}

\newcommand{\siiv}{\mbox{Si\,{\sc iv}}}

\newcommand{\Anand}[1]{{\color{red}[Anand: #1]}}

\newcommand{\Aromal}[1]{{\color{magenta}[Aromal: #1]}}
\newcommand{\RN}[1]{%
  \textup{\uppercase\expandafter{\romannumeral#1}}%
}

\def\h2{$\rm H_2$}
\def\Nh2{$N$(H${_2}$)}

\def\kms{km\,s$^{-1}$}

\def\zabs{$z_{\rm abs}$}
\def\zem{$z_{\rm em}$}

\def\21{21-cm}

\def\t0{T$_{0}$}

\def\c21{$C_{21}$}

\def\nv{N~{\sc v}}
\def\civ{C~{\sc iv}}
\def\siiv{Si~{\sc iv}}

\def\J13{J$1322+0524$}

\def\dw{$\Delta W$}
\def\fdw{$\frac{\Delta W}{W}$}
\def\afdw{\big{|}$\frac{\Delta W}{W}$\big{|}}
\def\mbh{$M_{BH}$}
\def\lbol{$L_{bol}$}
\def\redd{$\lambda_{\rm Edd}$}

\title[SALT UFO BAL sample]{Time variability  of ultra fast BAL outflows using SALT: \civ\ equivalent width analysis  \thanks{Based on observations collected at Southern African Large Telescope (SALT; Programme IDs 2015-1-SCI-005, 2018-1-SCI-009, 2019-1-SCI-019 and 2020-1-SCI-011) and the European Organisation for Astronomical Research in the Southern Hemisphere under ESO programme 093.A-0255.}}

\author[Aromal et al.]{
P. Aromal$^{1}$\thanks{E-mail: aromal@iucaa.in (PA)},
R. Srianand$^{1}$,
and P. Petitjean$^{2}$
\\
$^{1}$IUCAA, Postbag 4, Ganeshkind, Pune 411007, India\\
$^{2}$ Institut d'Astrophysique de Paris, Sorbonne Universit\'e and CNRS, 98bis boulevard Arago, 75014 Paris, France\\
}

\date{Accepted XXX. Received YYY; in original form ZZZ}

\pubyear{2015}

\begin{document}
\label{firstpage}
\pagerange{\pageref{firstpage}--\pageref{lastpage}}
\maketitle

\begin{abstract}
We study the time variability (over $\le$7.3 yrs) of ultra fast outflows (UFOs) detected in a sample of 64 \civ\ broad absorption line (BAL) quasars 
(with 80 distinct BAL components) monitored using the Southern African Large Telescope. 
By comparing the properties of the quasar in our sample with those of a control sample of non-BAL quasars
we show that the distributions of black hole mass are different and the bolometric luminosities and optical photometric variations of UFO BAL quasars are slightly smaller compared to that of non-BAL quasars.
The detection fraction of \civ\ equivalent width (W) variability ($\sim$95\%), the fractional variability amplitude (\fdw) and the fraction of ``highly variable" BAL (i.e., \afdw$>$0.67) components ($\sim$ 33\%) are higher in our sample compared to the general BAL population. The scatter in \fdw\ and the fraction of ``highly variable" BALs  
increase with the time-scale probed. 
The \fdw\ distribution is asymmetric at large time scales. We attribute this to the BAL strengthening time scales being shorter than the weakening time scales. The BAL variability amplitude correlates strongly with the BAL properties compared to the quasar properties.
BALs with low W, high-velocity, shallow profiles, and low-velocity width tend to show more variability. 
When multiple BAL components are present a correlated variability is seen  
between low- and high-velocity components with the latter showing a larger amplitude
variations.
We find an anti-correlation between the fractional variations in the continuum flux and W. While this suggests photoionization-induced variability, the scatter in continuum flux is much smaller than that of W. 
\end{abstract}




\section{Introduction}

In the spectra of 10\%–20\% of optically selected QSOs, 
strong absorption features, so-called Broad Absorption Lines (BAL), 
of \siiv, \civ\ and \nv\ \citep[occasionally of Fe~{\sc ii} and Mg~{\sc ii,}][]{Wampler1995}  
are observed blueshifted up to 0.2c from the corresponding rest-frame emission lines \citep[][]{Weymann1991}. It has been suggested that the intrinsic BAL fraction, after accounting for dust and other observational biases, could be as high as $\sim$40\% \citep[][]{dai2008,allen2011}.
Rare examples of QSO spectra exhibiting broad absorption lines redshifted with respect to the emission lines are also known \citep[e.g.,][]{Hall2013}.  Based on large velocity widths and blue shifts it is believed that the absorbing gas is related to the central regions of quasars.
The BAL fraction among quasars reflects either the covering factor of outflows that prevail in all quasars \citep[e.g.,][]{Elvis2000} or a precise AGN evolutionary phase \citep[e.g.,][]{Wang2013,chen2022}. 
Based on the measured properties like blackhole mass (\mbh),  Eddington ratio (\redd) etc.,  hot dust emission, and overall optical-IR  spectral energy distribution (SED),
it appears that both BAL and non-BAL QSOs are drawn from the same parent
population \citep{Reichard2003,rankine2020,Yi2020,temple2021}.
While this  may favor the first possibility, it does not rule out the BAL being part of an evolutionary sequence. It is also widely believed that these outflows could play an important role in the central black hole growth, the host galaxy evolution \citep{ostriker2010,kormendy2013}, and the chemical enrichment of the intergalactic medium
\citep[][]{Wampler1995,Dunn2010,Capellupo2012,borguet2013}.

Very large velocity outflows ($v_{out}\sim 0.1-0.2c$) are seen  both in the form of X-ray \citep[e.g., Ultra fast outflows (UFOs) studied by][]{Tombesi2010} and UV  \citep[e.g., Extremely high-velocity outflows, EHVOs, studied by] []{Rodriguez2020} absorption. 
BAL outflows with such large velocities are interesting as to achieve large terminal
velocities, the launching power ($\propto v_{\rm out}^3$) must be very large and the wind must be launched very close to the central engine \citep[see for example,][]{murray1995}. 
Hence these outflows that carry large momentum can significantly influence the properties of gas in host galaxies and be part of what is known as AGN feedback \citep{hopkins2010}. Also, the origin and evolution of the UFO BALs can be considerably different from their lower velocity counterparts, hence a study which focuses on absorption as a function of its velocity  will provide additional clues on their launching and acceleration mechanism. A similar study done by \citet{tombesi2013} for X-ray outflows concludes that the X-ray UFOs and the comparatively lower velocity warm absorbers  could actually represent parts of a single large-scale stratified outflow observed at different locations from the black hole where the X-ray UFOs are likely
launched from the inner accretion disc and the WAs at larger distances, such as the outer disc and/or torus.

In particular,
the exact mechanism that can accelerate the absorbing gas to such large velocities is still debated. 
Line-driven radiative acceleration is generally preferred  based on the evidence found through observations of line-locking \citep[e.g.,][]{Srianand2002} and the Ly$\alpha$ ghost signature \citep[e.g.,][]{arav1996}. 
Also, note that \citet{bowler2014} found line-locking in both BAL and non-BAL populations with similar frequency suggesting line-driving is important in both populations.
Given the high luminosity of the quasar, an extended uniform-density absorber will be too highly ionized
to produce \civ\ absorption out to $\sim$1 kpc \citep{Baskin2014}.  Therefore it will 
be very difficult to accelerate the gas through line-driven 
acceleration as it is expected to be highly ionized. In the framework of the traditional disk-wind model, \citet{murray1995} suggested that this over-ionization problem can be avoided by
introducing a highly ionized gas component (i.e., the so-called shielding gas) located close
to the ionizing source that acts as a screen to prevent the high energy photons from heating the gas producing BAL \citep[see also][]{proga2000,Higginbottom2013,Matthews2016}. Such a scenario can also be useful in understanding the X-ray weakness of BAL quasars \citep{gibson2009,stalin2011}.
Alternatively, magnetic driving can also
explain high velocity and highly ionized gas \citep{dekool1995}.

The validity of a given outflow model and/or constraints on the parameters of 
models can be obtained using high-resolution spectroscopic observations and/or 
spectroscopic monitoring with adequate time sampling.  While the former can be used to constrain density, covering factor, chemical enrichment, and location of the gas 
the latter is useful in constraining variability time scales that can be linked to density and or time scales of either the gas ejection event and/or transverse motions.
The distance measurement obtained for some of the BALs using fine-structure excitation tend to be very large
\citep[see for example,][]{arav2018}. While large distances make BAL outflows an important contributor to the AGN feedback, it poses a problem to simulations as distance scales are much smaller in simulations. Time dependence, velocity dependence, and correlated velocity scale of absorption line variability can also provide interesting constraints for the disk wind models. 
Here we mainly focus on the variability of \civ\ BALs having high ejection velocities.

The existence of BAL variability has been known for over
three decades 
\citep[e.g.,][]{Foltz1987,Barlow1992} and surveys of
increasing number of objects have become common 
\citep{Lundgren2007,Gibson2008, Gibson2010, filiz2012, capellupo2011, Capellupo2012, Vivek2014, Filiz2013,mcgraw2017,rogerson2018,vivek2018,cicco2018,aromal2022}. 
In general, the BAL variability includes extreme optical depth variations like emergence, disappearance, and kinematic shifts. 
Possible origins of BAL variability are: (i) large variations in the quasar ionizing flux, (ii) changes 
in the covering factor ($f_c$) of the outflow with respect to the background source, (iii) the transverse motion of 
the outflow perpendicular to our line of sight.  Detailed investigation of the time variability 
of BAL profiles can yield tight constraints on the lifetime and the location of the outflow, and provides 
significant insights on the origin and physical mechanisms driving the flow. 
If the BALs are formed in
the vicinity of the launching region, then the timescale for wind
material to cross the region of interest is about 1–10 yr, and this is
a reasonable characteristic timescale over which flow structures are expected to change \citep{capellupo2013}. This is also the characteristic
timescale for significant angular rotation of the accretion disk
at the wind-launching radius.
%
%

For the past several years, we have been carrying out a low-resolution spectroscopic monitoring of a sample of 64 quasars that show \civ\ BALs with outflow velocities ($v_{out}$) greater than 15000 \kms (refer to as UFO BAL in this work) using the Southern African Large Telescope (SALT). Detailed analysis of two of these quasars was presented in our previous papers \citep{Aromal2021,aromal2022}.
In this paper, we focus on the statistical analysis of the full sample. In particular, we study the \civ\ equivalent width variability of  UFO BALs over  time scales up to $\sim$7.3 yrs. 
Our main aim is to quantify, the \civ\ equivalent width variability, its dependence on time-scale, physical parameters of quasar and the BAL absorption profile, emergence/disappearance and acceleration signatures and their correlation to broad emission lines and continuum variability.

This article is arranged as follows.
In Section~\ref{sec:sample} we present our sample of UFO BALs, methods for identifying BAL regions and their variations, and also construct a control sample of non-BAL quasars matched in r-band magnitude and redshift. Section~\ref{sec:observations} provides  details of spectroscopic and photometric data used in this study and  compares certain photometric and spectral properties of UFO BALs to that of the quasars in the control sample. Section~\ref{sec:results} presents our results on BAL variability and its dependence on time-scale, quasar and BAL properties, properties of \civ\ emission line and photometric variability nature of the quasars.
In section~\ref{sec:discussions}, we discuss our main results and their implications.  Throughout this paper we use the flat $\Lambda$CDM cosmology with  $H_0$ = 70 \kms\ Mpc$^{-1}$ and $\Omega_{m,0}$ = 0.3. 



\section{UFO BAL sample and control samples}
\label{sec:sample}

\begin{table*}
\begin{threeparttable}
\caption{Details of quasars with UFOs in our sample}
\begin{tabular}{cccccccccccc}
\hline
QSO & \zem  & \zabs  & $v_{min}$  & $v_{max}$  & BI  & Class  & log($M_{BH}$)  & log($L_{Bol}$) & log(\redd) & Number  & $\Delta t$ (yr) \\
 & & &(\kms) &(\kms) & (\kms) & &(M$_\odot$) & (erg s$^{-1}$) & & of & (min,max)\\
  &  & &  &  &   & &  &  & & epochs & \\
  (1) & (2) & (3) & (4) & (5) & (6) & (7) & (8) & (9) & (10) & (11) & (12)  \\

\hline
     &  \\
     J0028-0539 	&	 2.5584 	&	 2.36 	&	 10378 	&	 25096 	&	 4138 	&	 1 	&	 9.50 	&	 46.98 	&	 -0.62 	&	 3 	&	 0.53, 2.44 \\ 
    J0034+0309 	&	 2.3329 	&	 2.14 	&	 10799 	&	 26307 	&	 4442 	&	 1 	&	 9.46 	&	 46.88 	&	 -0.68 	&	 4 	&	 0.29, 3.18 \\ 
    J0046+0104 	&	 2.1492 	&	 2.01 	&	 10202 	&	 19450 	&	 4649 	&	 2 	&	 9.21 	&	 47.30 	&	 -0.01 	&	 13 	&	 0.00, 6.64 \\ 
    J0052+0909 	&	 2.6625 	&	 2.43 	&	 10672 	&	 30000 	&	 8866 	&	 1 	&	 9.41 	&	 47.05 	&	 -0.46 	&	 3 	&	 0.29, 2.67 \\ 
    J0054+0027 	&	 2.5142 	&	 2.42 	&	 4426 	&	 11048 	&	 1249 	&	 2 	&	 9.48 	&	 47.16 	&	 -0.42 	&	 8 	&	 0.03, 5.96 \\ 
     	&	 	&	 2.31 	&	 12220 	&	 25899 	&	 1246 	&	 2 	&	 9.48 	&	 47.16 	&	 -0.42 	&	 8 	&	 0.03, 5.96 \\ 
    J0138+0124 	&	 2.5441 	&	 2.42 	&	 3000 	&	 28896 	&	 15374 	&	 3 	&	 9.21 	&	 47.19 	&	 -0.12 	&	 4 	&	 0.24, 3.04 \\ 
    J0152+0929 	&	 2.1794 	&	 2.04 	&	 6840 	&	 26925 	&	 5516 	&	 3 	&	 9.31 	&	 46.93 	&	 -0.48 	&	 3 	&	 0.32, 3.33 \\ 
    J0200-0037 	&	 2.1422 	&	 1.99 	&	 6846 	&	 29777 	&	 12261 	&	 3 	&	 9.29 	&	 47.11 	&	 -0.27 	&	 7 	&	 0.70, 6.64 \\ 
    J0216+0115 	&	 2.2310 	&	 2.19 	&	 3191 	&	 6227 	&	 1337 	&	 999 	&	 8.98 	&	 47.02 	&	 -0.06 	&	 6 	&	 0.31, 6.79 \\ 
     	&	 	&	 2.02 	&	 7538 	&	 29943 	&	 6501 	&	 999 	&	 8.98 	&	 47.02 	&	 -0.06 	&	 6 	&	 0.31, 6.79 \\ 
    J0220-0812 	&	 2.0095 	&	 1.86 	&	 9509 	&	 29709 	&	 6129 	&	 1 	&	 9.70 	&	 47.11 	&	 -0.69 	&	 5 	&	 0.35, 7.00 \\ 
    J0224-0528 	&	 2.0845 	&	 1.91 	&	 14234 	&	 20623 	&	 1473 	&	 1 	&	 9.85 	&	 47.47 	&	 -0.48 	&	 3 	&	 0.34, 3.19 \\ 
     	&	 	&	 1.85 	&	 22059 	&	 27995 	&	 455 	&	 1 	&	 9.85 	&	 47.47 	&	 -0.48 	&	 3 	&	 0.34, 3.19 \\ 
    J0229-0034 	&	 2.1382 	&	 2.03 	&	 3083 	&	 28332 	&	 13129 	&	 3 	&	 9.39 	&	 47.02 	&	 -0.48 	&	 14 	&	 0.00, 2.55 \\ 
    J0242+0049 	&	 2.0573 	&	 1.88 	&	 10573 	&	 26859 	&	 3059 	&	 2 	&	 9.91 	&	 47.09 	&	 -0.93 	&	 9 	&	 0.02, 6.85 \\ 
    J0244-0108 	&	 3.9856 	&	 3.68 	&	 10283 	&	 27536 	&	 5335 	&	 1 	&	 9.96 	&	 47.47 	&	 -0.59 	&	 6 	&	 0.11, 4.37 \\ 
    J0814-0004 	&	 2.5936 	&	 2.37 	&	 10636 	&	 29162 	&	 2830 	&	 1 	&	 9.62 	&	 47.32 	&	 -0.40 	&	 4 	&	 0.28, 4.39 \\ 
    J0817+0717 	&	 2.4428 	&	 2.33 	&	 4641 	&	 18133 	&	 6032 	&	 3 	&	 9.30 	&	 47.33 	&	 -0.06 	&	 2 	&	 2.63, 2.63 \\ 
    J0831+0354 	&	 2.0761 	&	 1.95 	&	 8823 	&	 27670 	&	 6475 	&	 1 	&	 9.41 	&	 46.94 	&	 -0.57 	&	 3 	&	 2.67, 6.18 \\ 
    J0837+0521 	&	 2.3624 	&	 2.31 	&	 3000 	&	 7547 	&	 1647 	&	 2 	&	 9.60 	&	 47.12 	&	 -0.58 	&	 3 	&	 2.62, 5.34 \\ 
     	&	 	&	 2.10 	&	 22073 	&	 28994 	&	 1502 	&	 2 	&	 9.60 	&	 47.12 	&	 -0.58 	&	 3 	&	 2.62, 5.34 \\ 
    J0845+0812 	&	 2.3545 	&	 2.13 	&	 15183 	&	 27269 	&	 3778 	&	 1 	&	 9.38 	&	 47.08 	&	 -0.40 	&	 3 	&	 0.04, 2.44 \\ 
    J0911+0550 	&	 2.7933 	&	 2.58 	&	 12300 	&	 21914 	&	 1641 	&	 1 	&	 9.66 	&	 47.52 	&	 -0.24 	&	 3 	&	 1.14, 3.48 \\ 
    J0924-0128 	&	 2.4461 	&	 2.38 	&	 4427 	&	 7118 	&	 1764 	&	 2 	&	 8.91 	&	 47.16 	&	 0.14 	&	 2 	&	 3.15, 3.15 \\ 
     	&	 	&	 2.22 	&	 9527 	&	 28598 	&	 4602 	&	 2 	&	 8.91 	&	 47.16 	&	 0.14 	&	 2 	&	 3.15, 3.15 \\ 
    J0932+0237 	&	 2.1679 	&	 1.96 	&	 13042 	&	 30000 	&	 7982 	&	 1 	&	 9.50 	&	 47.09 	&	 -0.51 	&	 4 	&	 0.27, 6.59 \\ 
    J0951-0157 	&	 3.2553 	&	 2.97 	&	 14909 	&	 27271 	&	 3953 	&	 1 	&	 9.73 	&	 47.24 	&	 -0.59 	&	 3 	&	 0.26, 2.62 \\ 
    J1006+0119 	&	 2.3030 	&	 2.16 	&	 6674 	&	 24194 	&	 8743 	&	 3 	&	 9.39 	&	 47.22 	&	 -0.27 	&	 2 	&	 2.97, 2.97 \\ 
    J1007+0304 	&	 2.1245 	&	 1.92 	&	 15770 	&	 24885 	&	 1986 	&	 1 	&	 9.44 	&	 47.05 	&	 -0.49 	&	 6 	&	 0.30, 6.13 \\ 
    J1054+0150 	&	 2.2366 	&	 2.08 	&	 13434 	&	 17152 	&	 1978 	&	 1 	&	 9.43 	&	 46.85 	&	 -0.68 	&	 4 	&	 0.30, 6.14 \\ 
     	&	 	&	 2.01 	&	 19284 	&	 23542 	&	 784 	&	 1 	&	 9.43 	&	 46.85 	&	 -0.68 	&	 4 	&	 0.30, 6.14 \\ 
    J1110-0140 	&	 2.8192 	&	 2.63 	&	 8252 	&	 24514 	&	 6785 	&	 1 	&	 9.80 	&	 47.29 	&	 -0.61 	&	 3 	&	 0.00, 2.88 \\ 
    J1116+0808 	&	 3.2429 	&	 2.97 	&	 16170 	&	 24210 	&	 663 	&	 1 	&	 9.29 	&	 47.40 	&	 0.01 	&	 2 	&	 2.04, 2.04 \\ 
    J1135-0240 	&	 2.4611 	&	 2.23 	&	 15141 	&	 25096 	&	 1281 	&	 1 	&	 9.51 	&	 47.05 	&	 -0.55 	&	 3 	&	 0.28, 3.47 \\ 
    J1143+0912 	&	 2.3253 	&	 2.20 	&	 5665 	&	 16556 	&	 3775 	&	 3 	&	 9.32 	&	 46.92 	&	 -0.49 	&	 2 	&	 2.69, 2.69 \\ 
    J1156+0856 	&	 2.1077 	&	 1.94 	&	 6919 	&	 30000 	&	 12496 	&	 3 	&	 8.92 	&	 46.75 	&	 -0.27 	&	 3 	&	 2.84, 6.13 \\ 
    J1205+0134 	&	 2.1523 	&	 1.96 	&	 12909 	&	 29458 	&	 6180 	&	 1 	&	 9.57 	&	 47.09 	&	 -0.59 	&	 5 	&	 0.32, 6.28 \\ 
    J1208+0355 	&	 2.0210 	&	 1.95 	&	 4995 	&	 9738 	&	 1286 	&	 2 	&	 9.40 	&	 46.88 	&	 -0.62 	&	 3 	&	 2.95, 6.23 \\ 
     	&	 	&	 1.82 	&	 12972 	&	 26326 	&	 2808 	&	 2 	&	 9.40 	&	 46.88 	&	 -0.62 	&	 3 	&	 2.95, 6.23 \\ 
    J1215-0034 	&	 2.6987 	&	 2.54 	&	 4151 	&	 22790 	&	 6568 	&	 3 	&	 9.94 	&	 47.64 	&	 -0.40 	&	 3 	&	 2.46, 5.63 \\ 
    J1301+0314 	&	 2.1115 	&	 1.91 	&	 16175 	&	 25380 	&	 1656 	&	 1 	&	 9.58 	&	 47.05 	&	 -0.63 	&	 3 	&	 3.22, 6.75 \\ 
    J1317+0100 	&	 2.6961 	&	 2.59 	&	 3816 	&	 18015 	&	 5987 	&	 3 	&	 9.17 	&	 47.32 	&	 0.05 	&	 6 	&	 0.01, 6.01 \\ 
     	&	 	&	 2.45 	&	 19596 	&	 25568 	&	 427 	&	 3 	&	 9.17 	&	 47.32 	&	 0.05 	&	 6 	&	 0.01, 6.01 \\ 
    J1331+0042 	&	 2.4341 	&	 2.35 	&	 6555 	&	 9453 	&	 498 	&	 2 	&	 9.37 	&	 47.09 	&	 -0.38 	&	 4 	&	 0.01, 3.16 \\ 
     	&	 	&	 2.27 	&	 11383 	&	 16893 	&	 468 	&	 2 	&	 9.37 	&	 47.09 	&	 -0.38 	&	 4 	&	 0.01, 3.16 \\ 
    J1341-0115 	&	 2.7682 	&	 2.55 	&	 8502 	&	 25835 	&	 7132 	&	 1 	&	 9.72 	&	 47.24 	&	 -0.59 	&	 3 	&	 2.33, 5.02 \\ 
    J1343+0351 	&	 2.8686 	&	 2.68 	&	 12032 	&	 19532 	&	 2192 	&	 1 	&	 9.92 	&	 47.23 	&	 -0.79 	&	 3 	&	 0.25, 2.81 \\ 
    J1350+0843 	&	 2.6157 	&	 2.44 	&	 13629 	&	 17632 	&	 953 	&	 1 	&	 9.72 	&	 47.30 	&	 -0.52 	&	 3 	&	 0.25, 2.77 \\ 
    J1357+0055 	&	 2.0081 	&	 1.80 	&	 11851 	&	 30000 	&	 2757 	&	 1 	&	 9.27 	&	 47.32 	&	 -0.05 	&	 4 	&	 0.27, 7.26 \\ 
    J1359+0310 	&	 2.6778 	&	 2.52 	&	 8662 	&	 19833 	&	 4439 	&	 1 	&	 9.49 	&	 47.28 	&	 -0.31 	&	 2 	&	 2.70, 2.70 \\ 
    J1400+0507 	&	 2.3015 	&	 2.16 	&	 7038 	&	 25561 	&	 7335 	&	 3 	&	 9.34 	&	 47.01 	&	 -0.43 	&	 3 	&	 0.29, 3.04 \\ 
    J1405+0229 	&	 2.8266 	&	 2.67 	&	 6487 	&	 24861 	&	 7057 	&	 3 	&	 9.43 	&	 47.25 	&	 -0.28 	&	 4 	&	 0.25, 5.52 \\ 
    J1424+0441 	&	 2.2762 	&	 2.05 	&	 11006 	&	 28320 	&	 2923 	&	 1 	&	 9.44 	&	 47.12 	&	 -0.42 	&	 4 	&	 0.29, 3.39 \\ 
    J1445-0023 	&	 2.2296 	&	 2.05 	&	 7822 	&	 27489 	&	 8866 	&	 3 	&	 9.68 	&	 47.40 	&	 -0.39 	&	 4 	&	 0.02, 4.04 \\ 
    J1452+0932 	&	 2.4607 	&	 2.29 	&	 5694 	&	 28047 	&	 8540 	&	 3 	&	 9.40 	&	 47.24 	&	 -0.26 	&	 4 	&	 0.31, 4.65 \\ 
    J1500+0033 	&	 2.4394 	&	 2.23 	&	 9914 	&	 27649 	&	 4574 	&	 1 	&	 9.88 	&	 47.27 	&	 -0.70 	&	 6 	&	 0.30, 6.22 \\ 
    J1547+0606 	&	 2.0188 	&	 1.95 	&	 5432 	&	 9467 	&	 2587 	&	 2 	&	 9.64 	&	 47.27 	&	 -0.48 	&	 3 	&	 2.31, 5.61 \\ 
     	&	 	&	 1.87 	&	 9984 	&	 24909 	&	 7467 	&	 2 	&	 9.64 	&	 47.27 	&	 -0.48 	&	 3 	&	 2.31, 5.61 \\ 
    J1606+0718 	&	 2.0766 	&	 1.99 	&	 3835 	&	 22135 	&	 6819 	&	 3 	&	 9.46 	&	 46.90 	&	 -0.66 	&	 4 	&	 0.34, 5.24 \\ 
    J1606+0746 	&	 2.3687 	&	 2.20 	&	 8414 	&	 23723 	&	 2005 	&	 1 	&	 9.72 	&	 47.08 	&	 -0.74 	&	 3 	&	 0.29, 2.74 \\ 
    J1609+0526 	&	 2.3802 	&	 2.27 	&	 4115 	&	 26115 	&	 9608 	&	 3 	&	 9.29 	&	 47.10 	&	 -0.29 	&	 4 	&	 0.28, 4.07 \\

 \hline
  \end{tabular}
  \end{threeparttable}

  \end{table*}
    
    \begin{table*}
    \begin{threeparttable}

    \ContinuedFloat
    \centering
    \caption{Continued}
   \begin{tabular}{ccccccccccccc}
\hline
QSO & \zem  & \zabs  & $v_{min}$  & $v_{max}$  & BI  & Class  & log($M_{BH}$)  & log($L_{Bol}$) & log(\redd) & Number  & $\Delta t$ (yr) \\
 & & &(\kms) &(\kms) & (\kms) & &(M$_\odot$) & (erg s$^{-1}$) & & of & (min,max)\\
  &  & &  &  &   & &  &  & & epochs & \\
  (1) & (2) & (3) & (4) & (5) & (6) & (7) & (8) & (9) & (10) & (11) & (12)  \\
  \hline
     &  \\ 
    J1621+0758 	&	 2.1394 	&	 1.98 	&	 12355 	&	 19811 	&	 1016 	&	 1 	&	 9.58 	&	 47.10 	&	 -0.58 	&	 9 	&	 0.02, 5.46 \\ 
    J2126-0211 	&	 2.4629 	&	 2.23 	&	 12298 	&	 26593 	&	 2717 	&	 1 	&	 9.33 	&	 47.18 	&	 -0.25 	&	 4 	&	 0.28, 3.03 \\ 
    J2130+0115 	&	 2.5275 	&	 2.26 	&	 9697 	&	 30000 	&	 5839 	&	 1 	&	 9.62 	&	 47.54 	&	 -0.18 	&	 9 	&	 0.00, 4.99 \\ 
    J2137+0844 	&	 2.1893 	&	 2.12 	&	 4314 	&	 9046 	&	 3094 	&	 2 	&	 9.32 	&	 47.13 	&	 -0.28 	&	 5 	&	 0.01, 3.68 \\ 
     	&	 	&	 2.05 	&	 10082 	&	 23576 	&	 4260 	&	 2 	&	 9.32 	&	 47.13 	&	 -0.28 	&	 5 	&	 0.01, 3.68 \\ 
    J2209-0126 	&	 2.4227 	&	 2.27 	&	 5076 	&	 30000 	&	 12293 	&	 3 	&	 9.48 	&	 47.15 	&	 -0.43 	&	 5 	&	 0.00, 2.84 \\ 
    J2221-0103 	&	 2.6754 	&	 2.44 	&	 9723 	&	 29032 	&	 5237 	&	 1 	&	 9.94 	&	 47.49 	&	 -0.55 	&	 6 	&	 0.02, 4.77 \\ 
    J2239-0047 	&	 2.2200 	&	 2.01 	&	 14375 	&	 25370 	&	 1456 	&	 1 	&	 9.68 	&	 47.13 	&	 -0.64 	&	 6 	&	 0.24, 6.18 \\ 
    J2256+0105 	&	 2.2680 	&	 2.21 	&	 3000 	&	 8535 	&	 2893 	&	 2 	&	 9.37 	&	 47.06 	&	 -0.41 	&	 5 	&	 0.00, 6.07 \\ 
     	&	 	&	 2.05 	&	 15840 	&	 26759 	&	 2161 	&	 2 	&	 9.37 	&	 47.06 	&	 -0.41 	&	 5 	&	 0.00, 6.07 \\ 
    J2310+0746 	&	 2.3109 	&	 2.23 	&	 6302 	&	 9269 	&	 1584 	&	 2 	&	 9.35 	&	 46.95 	&	 -0.50 	&	 4 	&	 0.28, 2.95 \\ 
     	&	 	&	 2.19 	&	 10000 	&	 13108 	&	 540 	&	 2 	&	 9.35 	&	 46.95 	&	 -0.50 	&	 4 	&	 0.28, 2.95 \\ 
     	&	 	&	 2.08 	&	 14232 	&	 29382 	&	 2197 	&	 2 	&	 9.35 	&	 46.95 	&	 -0.50 	&	 4 	&	 0.28, 2.95 \\ 
    J2352+0105 	&	 2.1513 	&	 1.92 	&	 14600 	&	 29706 	&	 2182 	&	 2 	&	 9.70 	&	 47.31 	&	 -0.49 	&	 8 	&	 0.03, 6.93 \\ 
    J2355-0357 	&	 2.4448 	&	 2.26 	&	 11721 	&	 28163 	&	 1513 	&	 1 	&	 9.76 	&	 47.03 	&	 -0.83 	&	 3 	&	 0.27, 2.26 \\ 
    J1322+0524 	&	 2.0498 	&	 1.97 	&	 5680 	&	 9830 	&	 2116 	&	 2 	&	 9.75 	&	 47.01 	&	 -0.84 	&	 12 	&	 0.00, 6.61 \\ 
     	&	 	&	 1.93 	&	 10731 	&	 13059 	&	 288 	&	 2 	&	 9.75 	&	 47.01 	&	 -0.84 	&	 12 	&	 0.00, 6.61 \\ 
     	&	 	&	 1.85 	&	 15600 	&	 26632 	&	 2003 	&	 2 	&	 9.75 	&	 47.01 	&	 -0.84 	&	 9 	&	 0.00, 2.28 \\ 
     	\\
\hline     
\end{tabular}

 \begin{tablenotes}\footnotesize
    \item  Column 3 : The absorption redshift of the BAL calculated using the optical depth weighted velocity centroid; 
    Columns 4, 5 and 6 : The minimum velocity ($v_{min}$), maximum velocity ($v_{max}$) and balnicity index (BI) of the identified BAL region respectively; Column 7 : The class in which the source belongs according to the shape of the BAL profile as mentioned in Section~\ref{subsec:class}. The BAL that could not be fitted into the classification scheme is indicated with `999'; Columns 8, 9 and 10 : The estimated quasar properties black hole mass (\mbh), bolometric luminosity (\lbol) and Eddington luminosity (\redd) respectively;
    Column 11 : The total number of spectroscopic epochs for the source;
    Column 12 : The minimum and maximum rest-frame time separation between two spectroscopic epochs for the source.
 \end{tablenotes}
 
\label{tab_sampledetails}
\end{threeparttable}
\end{table*}


We have constructed a sample of \civ\ UFO BALs from the SDSS  data release 12 quasar population \citep{paris2017} by applying the following criteria. The BAL parameter in the  catalog of \citet{paris2017} should be set to 1 and the Balnicity index (BI) should be greater than 0 \kms. The observed maximum outflow velocity at the time of observations should be $v_{\text{outflow, max}}$ > 15000 \kms. We restrict our sample to quasars having $z_{\rm em} > 2.0$ to ensure that \civ\ (also \siiv\ in most cases) absorption falls in the optical region and in the most sensitive wavelength range of the spectra in the case of both SDSS and SALT observations. We also restrict our sample to objects with declination $<+10$ deg and magnitude brighter than $m_r$= 18.5 mag. 
The former is to ensure that the source is accessible to SALT and the latter is 
to ensure that we get a sufficiently high spectroscopic signal-to-noise ratio (SNR) even if the 
observing conditions are sub-optimal with SALT.

We end up with a sample of 66 sources.
From these sources, we remove 2 sources (namely, J021119.65-042158.2 and J110100.38+092314.30),  where wrong identifications of BELs have led to inaccurate redshifts and the correct redshifts are found to be  $z_{em}<2$.
After visual inspection of the spectra, we remove another source, namely J002248.46-044510.3, where the identification of BAL is suspected to be wrong.
Hence, we find a total of 63 UFO BAL sources from the SDSS DR12 catalog.
We add to this list another UFO BAL source namely J132216.25+052446.3, an interesting BAL quasar we have 
been monitoring for the past 7 years using SALT. This source is identified as a BAL QSO in the SDSS DR12 BAL quasar sample, but the UFO BAL is absent in the SDSS spectrum and has only emerged during our other observing programme \citep{aromal2022}.
Our final sample consists of 64 UFO BAL quasars which are studied in 
detail in this paper.  
The median values of the r-band magnitude and \zem\ are 18.25 and 2.3437 respectively for quasars in our sample.
{ The list of sources, log of observations, and details of spectra obtained at different epochs are given in Table B1 in the online material.}

In Table~\ref{tab_sampledetails} we provide some physical characteristics of the quasars in our sample. 
Columns 8,  9 and 10 of Table~\ref{tab_sampledetails} give, respectively, the mass of the central
black hole (\mbh), the bolometric luminosity ($L_{Bol}$) and the Eddington ratio (\redd).
The emission redshift (\zem, given in column 2) 
is taken from \citet{Hewett2010} (which derives the systemic redshift from the fit of the C~{\sc iii}] emission line) whenever 
available and if not from \citet{paris2017}.

\subsection{BAL identification }
\label{sec:BALID}

\begin{figure}
    \centering
    \includegraphics[viewport=85 50 2350 1225, width=\textwidth,clip=true]{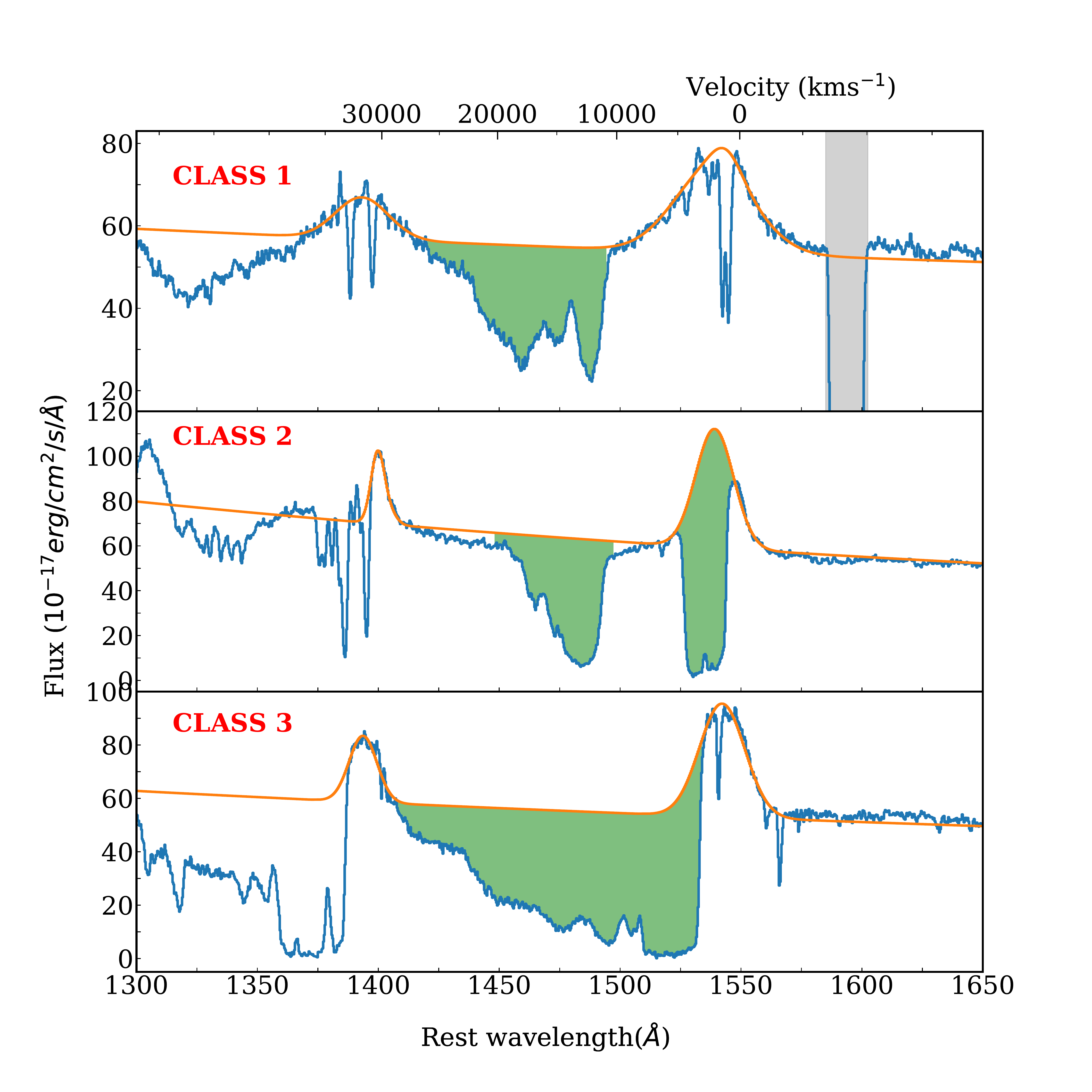}
    \caption{
     Examples of observed spectra and the best fits to the continuum and broad emission lines are shown in blue and orange respectively. Each panel shows absorption profiles of UFO BALs (green shaded regions) belonging to different classes as defined later in Section~\ref{subsec:class}. The rest wavelength
    (bottom) and velocity (top) scales are defined with respect to \zem\ given in Table~\ref{tab_sampledetails}. The gray-shaded region in the top panel corresponds to the CCD gap in SALT spectra. } 
    \label{fig:classification}
\end{figure}

We used the publicly available
multi-component spectral fitting code {\sc PyQSOFit}\footnote{https://github.com/legolason/PyQSOFit} \citep{guo2018a} 
to fit the continuum and broad emission lines (BEL) of all 
quasar spectra in our sample.
We visually inspected each spectrum and identified wavelength ranges devoid of any absorption lines.
We then fitted these line-free regions with a power-law + multiple Gaussian (for BELs) model which provided fairly good fits using $\chi^2$ minimization as shown in Fig~\ref{fig:classification} for a few sources in our sample. 
For sources where \civ\ BEL is not severely contaminated by 
absorption lines, we typically needed two Gaussians to fit the line whereas for all other cases, a single Gaussian 
was used. Similarly, for other prominent BELs in the spectra like \siiv, \nv\ etc., we used single Gaussian 
to fit the lines.
We then identified BAL troughs
using the conventional definition given by \citet{Weymann1991} where a BAL is defined as a continuous 
absorption wider than 2000 \kms\ below 90 $\%$ of the continuum level. Note that a spectrum can contain several distinct BAL troughs.
Here, we searched for \civ\ BALs in the region
between 3,000 to 30,000 km~s$^{-1}$ from the emission redshift. 
We do not consider BALs beyond 30,000~km~s$^{-1}$ 
as they could be contaminated by the \siiv\ absorption from lower velocities.
We consider only BALs beyond 3,000~km~s$^{-1}$ to avoid contamination by narrow associated absorption systems (that appear broad due to blending and moderate resolution spectra used here) that are not part of the BAL flow. 

For the identification of BAL complexes, we adopt the same method as in \citet{Filiz2013} (refer to their Section 3.2) 
as this method helps in efficiently quantifying the BAL variability across different spectroscopic epochs. 
In short, we define the BAL regions for each source after assigning a minimum and maximum velocity associated with these BALs considering all epochs. These velocities (denoted as $v_{\rm min}$ and $v_{\rm max}$) for each source are listed in columns 5 and 6 of Table~\ref{tab_sampledetails}.
After visually inspecting the results, we clearly identify 80 distinct BAL complexes in our UFO BAL sample. 
The redshifts of individual BAL complexes and the minimum and maximum velocity spanned by them are provided in columns 3, 4 and 5 respectively in  
Table~\ref{tab_sampledetails}. The measured 
maximum Balnicity Index (BI) for individual BAL components 
are provided in column 6 of the same table.

\subsection{BAL variations}

Once the BAL troughs were identified, we estimated different \civ\ BAL properties such as equivalent width (W) and the maximum depth of the 
BAL (d$_{BAL}$). 
 The BAL properties at different epochs including W, d$_{BAL}$ etc. are given in Table B2 in the online material.
Using the equivalent width as a measure of the strength of the BAL, we quantified the BAL variability by 
calculating the variations and the fractional variations 
in W as defined below \citep[see also][]{Filiz2013} :

\begin{equation}
    \Delta W = W_{2} - W_{1}, \quad \sigma_{\Delta W} = \sqrt{\sigma_{W_1}^2 + \sigma_{W_2}^2}
\end{equation}

\begin{equation}
    \frac{\Delta W} {W} = \frac{W_{2} - W_{1}}{ (W_{1} + W_{2}) \times 0.5 }
\end{equation}

\begin{equation}
    \sigma_{\frac{\Delta W}{W}} = \frac{4 \times (W_1 \sigma_{W_2} + W_2 \sigma_{W_1})}{(W_1 + W_2)^2}
\end{equation}
where $W_1$ and $W_2$ are equivalent widths measured at $t_1$ and $t_2$ respectively with $t_1$ < $t_2$. 
Thus an increase (or decrease) of W with time results in a  positive (or negative) $\Delta W$.
Based on the above equation, variations in equivalent width by a factor of 2 and 3 will correspond 
to $\Delta W/W$ of 0.67 and 1.0 respectively. An emerging (disappearing) absorption will also correspond to a $\Delta W/W$ of $+2$ 
(respectively $-2$). 
$\sigma_{\Delta W }$ and $\sigma_{\Delta W / W}$
are errors in $\Delta W$ and \fdw\ respectively.

\subsection{Control sample}

For comparing different physical properties of UFO BAL quasars in our sample with that of non-BAL quasars,
we made a control sample of non-BAL quasars having similar emission redshift ($\Delta z_{\rm em}\le0.2$) and r-band magnitude ($\Delta m_{\rm r}\le 0.3$ mag) distributions.
For Non-BAL quasars, we selected sources that are included in both SDSS DR7 \citep{shen2011} and SDSS DR12 catalogs since 
\citet{shen2011} provides several quasar parameters such as black-hole mass (\mbh), bolometric luminosity of 
the quasar (\lbol) and Eddington ratio (\redd).  
Hence, we build a sample of 320 non-BAL sources 
(i.e., 5 non-BAL quasars for each quasar in our UFO sample) following the above criteria. 
For these objects, we used the parameters provided by \citet{shen2011} and for the UFO BAL sample, we derived the
corresponding quantities from the results of the fits 
performed with {\sc PyQSOFit} and listed them  in columns 8-10 of Table~\ref{tab_sampledetails}. Details of how we measure these quantities are provided in the Appendix-A.

\begin{figure}
    \centering
    \includegraphics[viewport=35 0 1175 645, width=0.47\textwidth,clip=true]{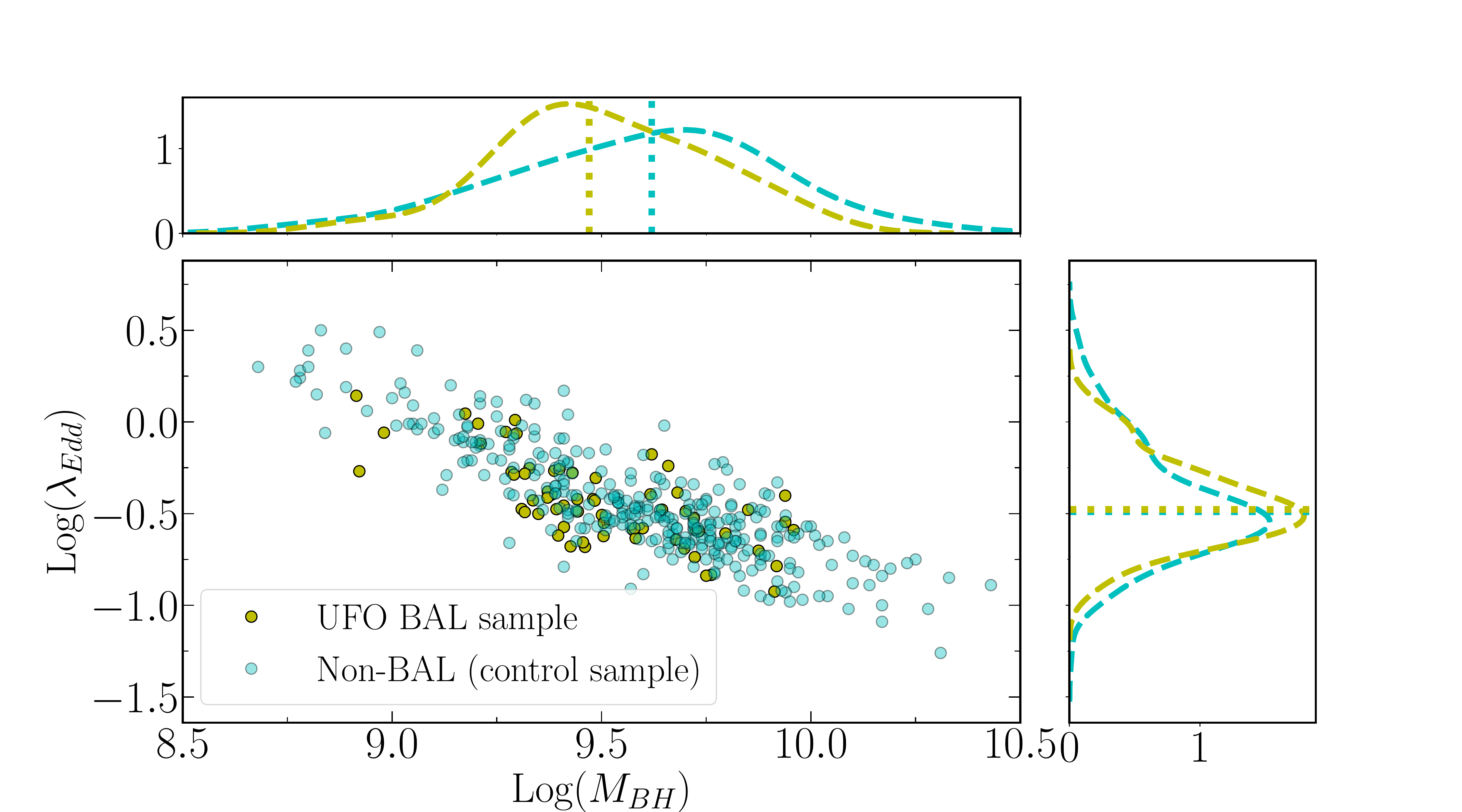}
    \caption{\mbh\ vs. \redd\ for the UFO BAL (yellow points and curves) and non-BAL control sample (green points and curves). A strong anti-correlation is evident for both samples. 
    {\sl Top and right-hand side panels}: PDFs of  \mbh\ and
    \redd\ respectively for the two samples. Dashed lines indicate
    the median values.
    }
    \label{fig:mbh_vs_redd}
\end{figure}

\citet{Yi2020} reported a correlation between \redd\ and \mbh\ for their high-$z$ (i.e., $3\le z\le 5$) BAL as well as non-BAL comparison samples. 
However, the correlation is found to be only tentative in the case of the BAL sample. They attribute this possible difference to the presence of substantial outflows in BAL QSOs as testified by large blueshift in the \civ\ emission lines. In Fig.~\ref{fig:mbh_vs_redd}, we plot \mbh\ $vs.$ 
\redd\ for our UFO sample as well as for the non-BAL control sample. A strong anti-correlation is apparent in this figure. 
Visually, the UFO BAL sample traces the same region spanned by the data from the control sample. 
The Spearman's coefficients are found to be r = $-$0.68
and r = $-$0.82 for UFO BAL and non-BAL samples respectively. In both cases, the p-values are $<10^{-3}$ indicating the correlation to be significant.


In the top panel of Fig.~\ref{fig:mbh_vs_redd}, we compare the probability distribution function (PDF) for \mbh\ between the UFO BAL and non-BAL control samples. These PDFs were estimated using a non-parametric kernel density estimation  \citep[KDE,][]{silverman1986} for a fixed bandwidth Gaussian kernel.  The median \mbh\ for the UFO BAL and control samples are $10^{9.47}$M$_\odot$ and $10^{9.62}$M$_\odot$ respectively. 
The KS test confirms the \mbh\ distribution in the two samples to be significantly different with a p-value of $6\times10^{-3}$. 
This difference comes from the fact that there is a clear lack of objects with log~\mbh$\ge$10.0 (alternatively quasars with FWHM of \civ\ BEL more than 10,000 \kms) and an excess of objects with
log~\mbh$\sim$9.3 in our UFO BAL sample. The lack of objects with large \mbh\ (or FWHM of the \civ\ BEL) may be attributed to the presence of absorption features biasing our Gaussian fits towards lower values.


In the right panel of Fig.~\ref{fig:mbh_vs_redd} we compare the \redd\ 
PDFs of the UFO BAL and non-BAL control samples. We find the distribution of \redd\ to be similar for both samples. The KS-test also confirms the same with a p-value of 0.60. This lack of difference between the UFO BAL and non-BAL samples is consistent with the finding of \citet{Yi2020}, for their high-$z$ ($3\le z\le 5$)
BAL QSO sample where the \mbh\ measurements are based on rest-frame optical emission lines.
Interestingly, we notice that for a given \mbh\ the \redd\ values observed in our UFO BAL and control samples are lower than that found for the sample of \citet{Yi2020}.
Two main differences are (i) QSOs in the sample of \citet{Yi2020} are from higher redshifts and (ii) their \mbh\ measurements are based on rest-frame optical lines that are more reliable compared to \civ\ emission line used in our study.

As a consequence of the above results we find the distributions of Bolometric luminosity are different for non-BAL and the UFO BAL sample 
(i.e., a p-value of $2.8\times10^{-4}$ for the KS test). This is surprising as we have matched the r-band magnitudes while constructing the control sample. As the bolometric luminosity is obtained using the rest frame 1350\AA\ continuum luminosity, the above result suggests a possible difference in the colour distribution of the non-BAL and UFO BAL sub-samples. 
This may  indicate a redder distribution of color for the UFO BAL sample which is consistent with the general understanding that BAL quasars are redder on average than non-BAL quasars \citep{brotherton2001,Reichard2003}.
However, we do find the median Bolometric luminosity to be similar
i.e., $10^{47.22}~erg~s^{-1}$ for the non-BAL control sample and 
$10^{47.13}~erg~s^{-1}$ for the UFO BAL sample.

\section{Details of observations and data used in this study}
\label{sec:observations}
For the detailed analysis presented in this work,
we have used available spectra from the Sloan Digital Sky Survey-I/II (SDSS), Baryon Oscillation Spectroscopic Survey (BOSS) and supplemented them with spectra from our own ongoing spectroscopic monitoring  using the Southern African Large Telescope (SALT) \citep[][]{buckley2005}.
%

For the SALT observations, we used the Robert Stobie Spectrograph \citep[RSS,][]{Burgh2003,Kobulnicky2003} in the long-slit mode with a 1.5" wide slit and the PG0900 grating. This combination gives a typical spectral resolution of $\sim$300 \kms. For each SALT/RSS observation, the GR angle was chosen such that the CCD gaps do not fall in the expected wavelength range of the broad absorption lines. In cases where this was not possible, we tried to observe the target with different GR angles to cover the full wavelength range without any gap.
A detailed log of observations for all the objects in our sample is provided in Table B1. 
In the case of J1621+0758, we also have NTT/ESO observations taken in the year 2014 \citep[details of which can be seen in,][]{Aromal2021}
We have, in total 375 spectra out of which 211 spectra are taken from SDSS and 164 spectra from our SALT observations.

For our SALT observations, the preliminary processing of raw CCD frames were carried out using the SALT data reduction pipeline \citep{Crawford2010}. We used the standard {\sc iraf}\footnote{{\sc iraf} is distributed by the National Optical Astronomy Observatories, which are operated by the Association of Universities for Research in Astronomy, Inc., under co-operative agreement with the National Science Foundation.} procedures to reduce the resulting 2D spectra. Flat-field corrections and cosmic ray zapping were applied to all science frames. We extracted the one-dimensional quasar spectrum from the background subtracted 2D science frames from each epoch using the {\sc iraf} task ``apall". Wavelength calibration was performed using different standard lamp spectra like Ar, ThAr, HgAr and Xe. In addition, skylines from the wavelength calibrated spectrum were matched with the sky line atlas provided by SALT and, if needed, corrections were applied to increase the wavelength accuracy. Similarly, flux calibration was performed using standard reference stars observed close to our observing nights. 
We performed continuum fitting to our SALT spectra using the same method described above in Section~\ref{sec:BALID}. 

\begin{figure}
    \centering
    \includegraphics[viewport=70 100 2360 1335, width=\textwidth,clip=true]{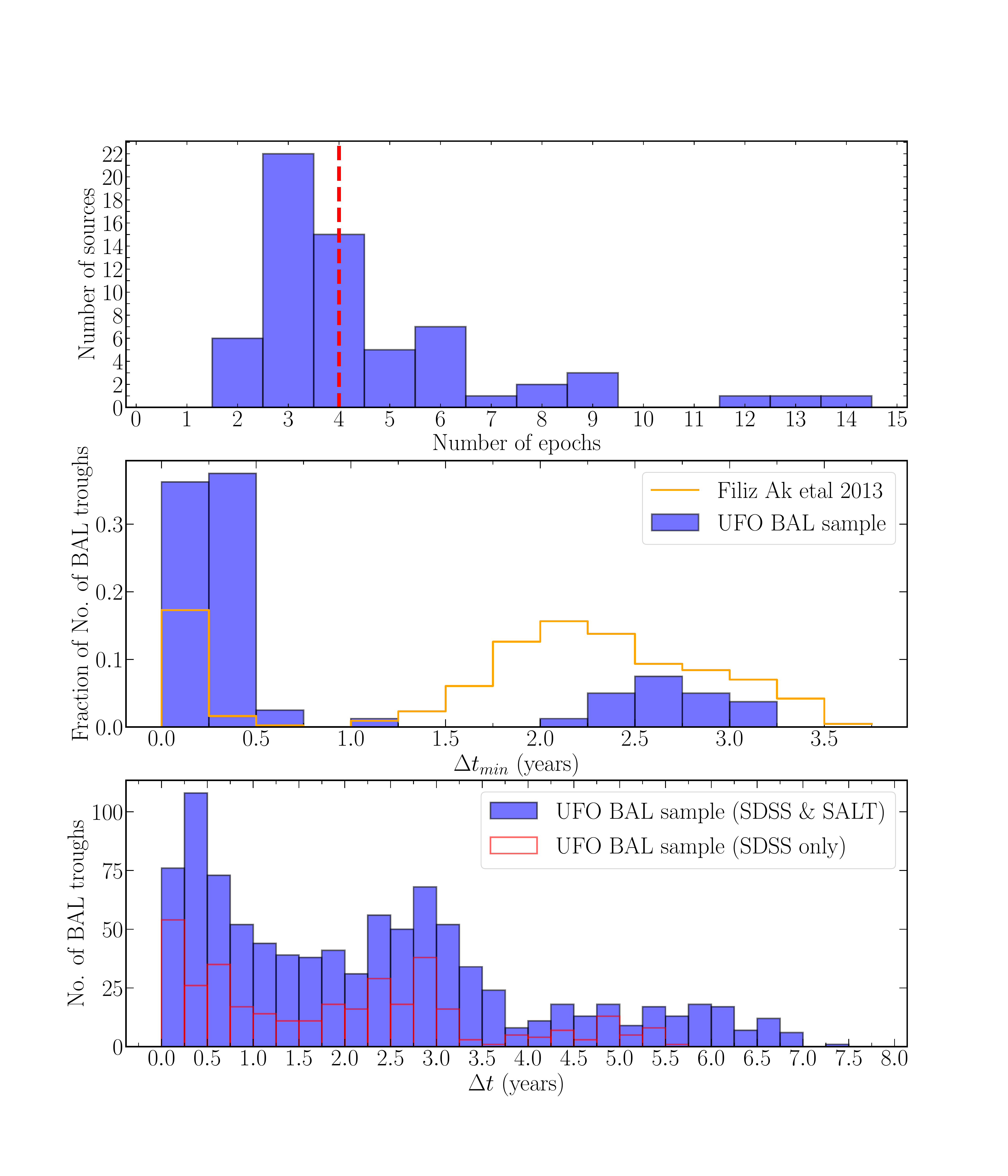}
    \caption{ {\it Top panel}: Distribution of total number of epochs (combining both SDSS and SALT observations) probed for each UFO BAL (blue histogram) with the median of the distribution marked with a red dashed line. {\it Middle panel}: Distribution of minimum sampled rest-frame time scales, $\Delta t_{min}$, for the UFO sample (blue histogram) along with the same distribution for the SDSS sample studied by \citet{Filiz2013} (orange histogram).  {\it Bottom panel}: Time-scales probed by all possible combinations of SDSS and SALT epochs (blue) and only SDSS epochs (red) for each individual source in the sample.
    }
    \label{fig:del_t_min_distr}
\end{figure}

One of the most important aspects of our study is to probe both the short- and long- time scale variability of the UFO BAL sample. Our SALT observations, in addition to the SDSS epochs, have brought great improvement to the time-sampling of these UFO BAL sources as demonstrated in Fig~\ref{fig:del_t_min_distr}. The top panel shows the distribution of the total number of spectroscopic epochs for each source with a median value of 4. For 62 out of 64 sources, at least one epoch was added by SALT observations.
The middle panel shows the fraction of objects in different minimum sampled rest-frame time scale ($\Delta t_{min}$). It can be seen that 
roughly 70 $\%$ of the sources have a minimum separation between the spectroscopic epochs less than 0.5 yr. This is mostly because of the multiple SALT observations performed within a rest frame year which is crucial for characterizing the short-time scale variability. 
This is a great improvement compared to the study by \citet{Filiz2013} which had only  $\sim$20\% of such sources (orange histogram in the figure).
The bottom panel of Fig~\ref{fig:del_t_min_distr} shows the histogram of all possible time-scales as probed by both SALT and SDSS (blue) and SDSS alone (red). Again this clearly shows the improvement brought by SALT epochs at 
different time-scales including $\Delta t$ > 5.5 yr.


\subsection{Photometric variability}
\label{sub:VAS}
As in \citet{aromal2022}, we look for correlations between the absorption line variability and the continuum variability using available broad-band photometric light curves. We have obtained publicly available photometric light curves of almost all the UFO BAL sources from the Panoramic Survey Telescope and Rapid Response System \citep[Pan-STARRS;][]{panstarrs2016}, the Palomar Transient Factory \citep[PTF;][]{Law2009} and the Zwicky Transient Facility \citep[ZTF;][]{zptf2019b,zptf2019a} surveys. Pan-STARRS provides photometric data of quasars in five broad band filters, i.e., g, r, i, z and y whereas ZTF gives the same for the g, r and i bands.
While PTF and Pan-STARRS provide sparsely sampled light curves ZTF provides much better sampling since the year 2018 (i.e., MJD$\ge$58200) for all the sources in our UFO BAL QSO sample. 

\begin{figure}
    \centering
    \includegraphics[viewport=40 55 2400 1150, width=\textwidth,clip=true]{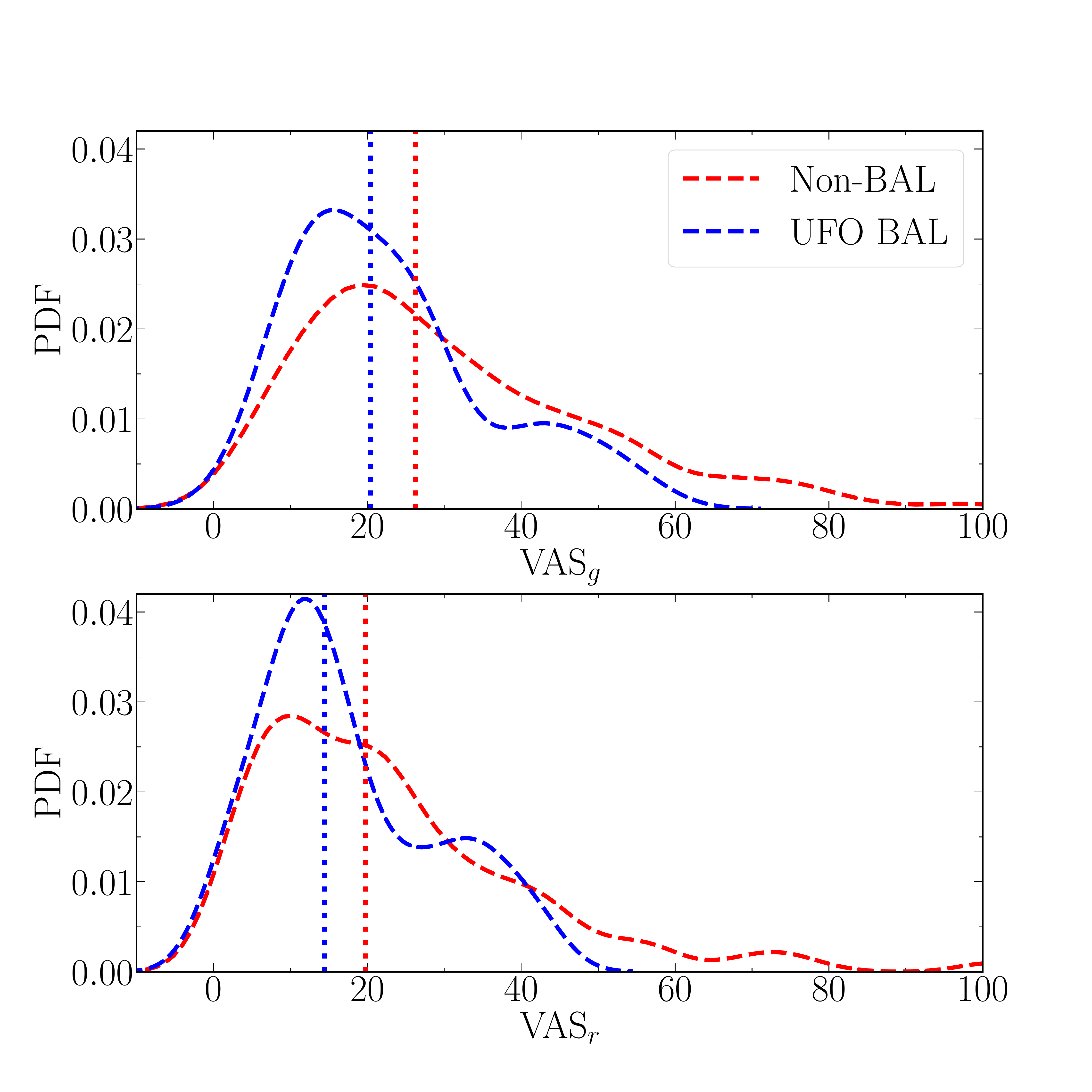}
    \caption{Comparison of variability amplitude strength (VAS) between non-BAL (red) and UFO BAL (blue) samples. Results for the g- and r-bands (VAS$_g$ and VAS$_r$ respectively) are shown in the top and bottom panels respectively. Vertical dashed lines indicate the median values.
    }
    \label{fig:lc_vas_prop_comp}
\end{figure}

We also obtained all the available ZTF photometric measurements for all quasars in the ``non-BAL'' control sample. From the ZTF light curves, we first estimate the variability amplitude ($\sigma_{rms}^2$) and its error ($S_D^2$) using the normalized excess variance method described in \citet{vaughan2003}. The variability amplitude strength (VAS) is then defined as VAS = $\sigma_{rms}^2$ / $S_D^2$. We show the  distributions of VAS  in g- and r-bands for the two samples in Fig~\ref{fig:lc_vas_prop_comp}. 
It is interesting to note (from the indicated median values) that the VAS distribution measured for the g-band is wider compared to that of the r-band. This indicates larger amplitude variations at lower wavelengths for quasars in both samples. 
We also notice that in both the bands the VAS for the UFO BAL sample is smaller than that of the control sample. 
The KS-test confirms this with a p-value of $<0.1$.
Therefore, purely based on optical photometric variability we do not find the quasars in our UFO BAL sample to be more variable compared to the general population of quasars.

\subsection{Properties of the \civ\ emission line}
\label{sec:civprop}
In this section, we mainly focus on the \civ\ BEL properties of the UFO BALs compared to that of the non-BAL control sample. 
We recognize that the BEL of UFO quasars may be affected by strong BALs close to 
the emission redshift.
Therefore, we only consider sources having no BALs at velocities less than 6000 \kms\ with respect to \zem.
{ We also visually inspected the double gaussian fits to the \civ\ BEL profiles in order to check and remove certain epochs that are severely affected by either narrow absorption lines or CCD gaps in the case of SALT epochs.} Thus, we made a sub-sample of 41 sources satisfying the above criteria to compare the emission line properties and study the relationship between UFO BAL and BEL properties.

We estimate two quantities 
namely, the equivalent width ($W_\text{BEL}$) of the emission line and its blueshift (BS$_\text{BEL}$) for the \civ\ BEL. We follow the method by \citet{rankine2020} in order to calculate BS$_\text{BEL}$, i.e.,
\begin{equation}
    BS_\text{BEL} = c \times (\lambda_r - \lambda_{half}) / \lambda_r
\end{equation}
where $c$ is the velocity of light, $\lambda_r$ is the rest-frame 
wavelength of the emission line, 1549.48\AA~ for the CIV doublet
and $\lambda_{half}$ is the wavelength which bisects the cumulative
total line flux.
\begin{figure}
    \centering
    \includegraphics[viewport=20 0 2250 690, width=0.9\textwidth,clip=true]{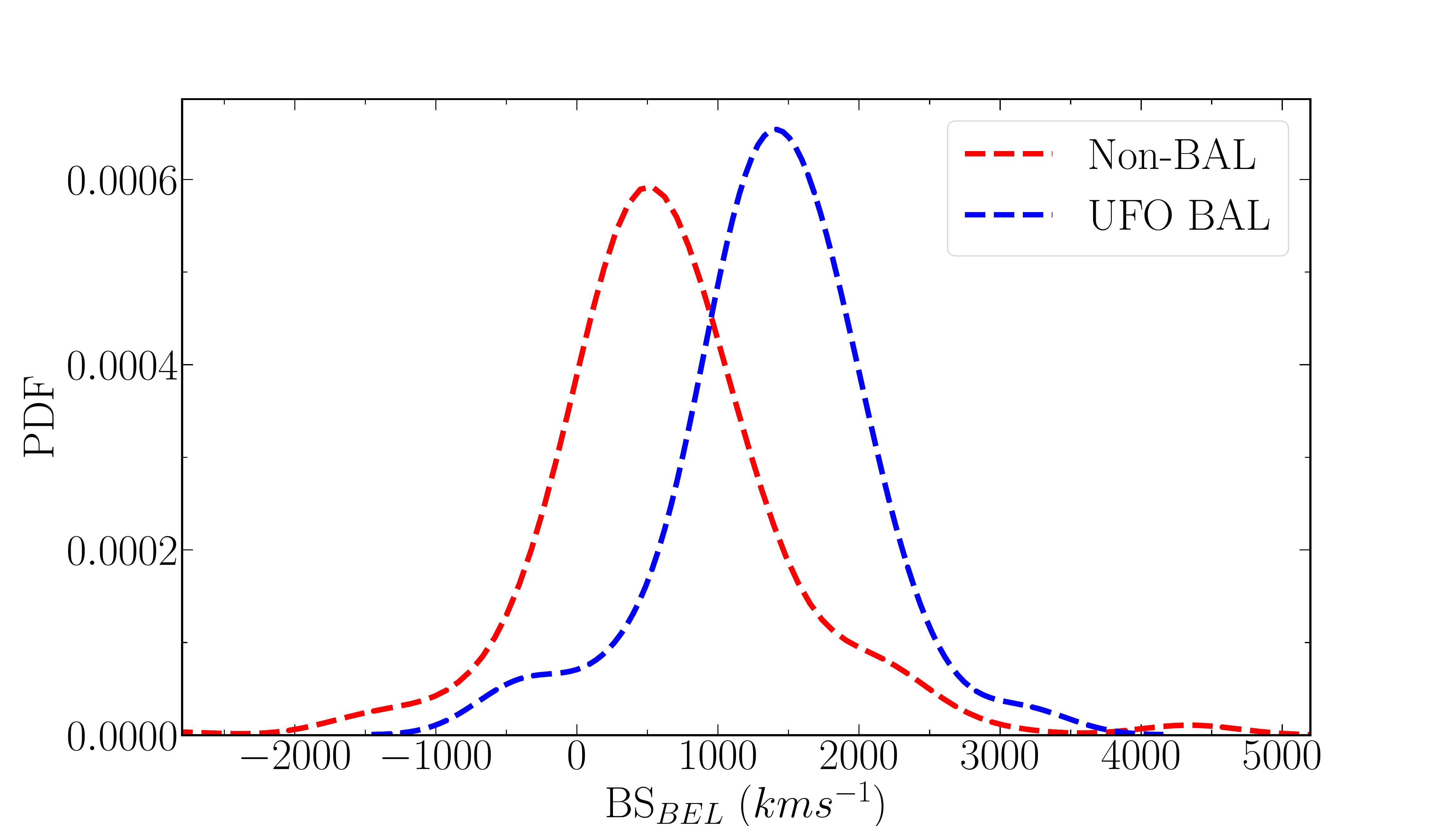}
    \caption{Comparison between \civ\ BEL blue-shifts in 
    spectra from the non-BAL (blue) and UFO QSO BAL (orange) samples.
    }
    \label{fig:blueshift_comp}
\end{figure}
%
 
As shown in Fig~\ref{fig:blueshift_comp}, we confirm the findings by \citet{rankine2020} that \civ\ BELs in sources with BALs reaching high velocities such as our UFO BAL sample tend to show more blue-shifted emission compared to the same of its non-BAL counterparts.
But, it is good to keep in mind that the emission-line outflow and physical properties of BAL and non-BAL quasars overlap, hence demonstrating that (high-ionization) BALs and non-BALs represent different views of the same underlying quasar population.


\section{Results on BAL variability}
\label{sec:results}
\subsection{Frequency of BAL variability}
\label{sec:varfraction}

Following \citet{Filiz2013}, we identify the variable BALs when the  \civ\ BAL equivalent width variation are $>3\sigma$ 
significance level.
Using all possible pairs of epochs for each object, we find that 95.3$^{+14}_{-12}$\%\footnote{The errors are calculated using Poisson statistics as discussed in  \citet{Gehrels1986}.} of UFO BAL QSOs in our sample show 
at least one variable trough. 
Among the 80 UFO BALs identified in Table~\ref{tab_sampledetails}, 95$^{+12}_{-11}$\% of them show $>$3$\sigma$ variability. This confirms that $\sim$95\% of UFO BAL quasars (and UFO BAL components) in our sample show significant variability at least once between possible pairs of epochs probed.
Our results are consistent with that of \citet{Gibson2008} who have found 12/13 of the BALs in their sample to show significant equivalent width variability over 3-6 yrs.

However, smaller variability fractions have also been reported in the literature.
Only 11/32 BAL trough in the sample of \citet{Lundgren2007} show more than $3\sigma$ equivalent width variability over a time-scale of $\le$102 days. 
In their sample of 24 bright BAL quasars
\citet{capellupo2011} have found that 39\% and 65\% were varied in the 
short-term (i.e., 4-9 months) and long-term (i.e., 3.8-7.7 yr), respectively.
%
\citet{Filiz2013} have found that 62.2\% of BAL quasars and 57.9\%  of the BAL trough in their sample show significant \civ\ equivalent width variation when they considered a pair with the shortest time interval for each quasar. 
The shortest time interval varies for each source due to the non-uniform SDSS observations and ranges from as small as 5.9 hr (~10$^{-3}$ yrs) to 3.7 yr with a median of 2.1 yr.
If we apply to our sample the same criteria as \citet{Filiz2013}, i.e., consider only the pair with the shortest time-scale for each object (see last column in Table~\ref{tab_sampledetails}), we get $\sim$70\% of UFO BAL QSOs and UFO BAL components to show variability. 
 We also consider the epoch pairs with the longest time separation (i.e., $\Delta t$ in the range 2.0$-$7.3 yr) for each object and found that  $\sim$89\% and $\sim$87\% of UFO BAL QSOs and UFO BAL components show variability respectively.
Recall the shortest time intervals probed in our sample (see the middle panel in Fig.~\ref{fig:del_t_min_distr}) are shorter than that of \citet{Filiz2013}.  
As the variability amplitudes are larger for longer time scales (see below), 
this may indicate that our UFO BALs are showing more variability compared to the general population of BAL quasars at all time scales (section~\ref{subsec:filiz2013_comp} provides further discussions on this). 

\citet{hemler2019}, studied the very short time-scale ($<$10 quasar rest frame days) 
BAL variability towards 27 BAL QSOs observed as a part of reverberation mapping
studies with a median of 58 spectral epochs per quasar.
Using all possible combinations of epochs and the \civ\ equivalent widths given in Table~4 of \citet{hemler2019}, we find that the variable fraction of sources is 96\% and the variable fraction of BALs is 92\%.
This is consistent with what we find in our sample.

To quantify the fraction of variable objects 
we consider three time bins
corresponding to short (0.0-0.5 yrs), intermediate (0.5-2.0 yrs) and long ($>$ 2.0 yrs) time-scales. In the 
three bins, the percentage of variable sources are 89$^{+16}_{-14}$\%, 100$^{+21}_{-17}$\% and 94$^{+14}_{-12}$\% and the percentage of variable BALs are 84$^{+14}_{-12}$\%, 97$^{+18}_{-15}$\% and 94$^{+12}_{-10}$\%.
%
In the case of the \citet{hemler2019} observations, we find that the variable fraction of sources are 0.96 and 0.94 and the variable fraction of BALs are 0.92 and 0.90 for the short and intermediate time bins respectively. Thus it appears that, with a sufficient number of measurements, \civ\ BALs tend to show significant equivalent width variability at all time scales.
%
%

Next, we consider the fraction of epoch pairs in a given time bin that is showing significant equivalent width variability. For our UFO BAL sample we find 0.78$\pm$0.08, 0.81$\pm$0.05, 0.85$\pm$0.05, 0.89$\pm$0.07 for the time intervals 0.1$-$0.5, 0.5$-$2.0, 2.0$-$3.5, $>$3.5 yrs respectively. 
This once again confirms the highly variable nature of BAL troughs in our UFO BAL sample. When we repeat the same exercise for the full sample of \citet{hemler2019} we find 0.27$\pm$0.01, 0.43$\pm$0.01, 0.47$\pm$0.01 for 0$-$0.1, 0.1$-$0.5, 0.5$-$2.0 yrs bins respectively.
The difference between the two samples is interesting. This could be either due to a difference in the properties of the quasars or to the more frequent sampling at smaller time scales  in the case of \citet{hemler2019}.
The typical sampling time scale between consecutive epochs in this study varies between less than a day to a few days where we expect the absorption line variation to be less.
%
\begin{figure}
    \centering
    \includegraphics[viewport=85 0 2900 960, width=\textwidth,clip=true]{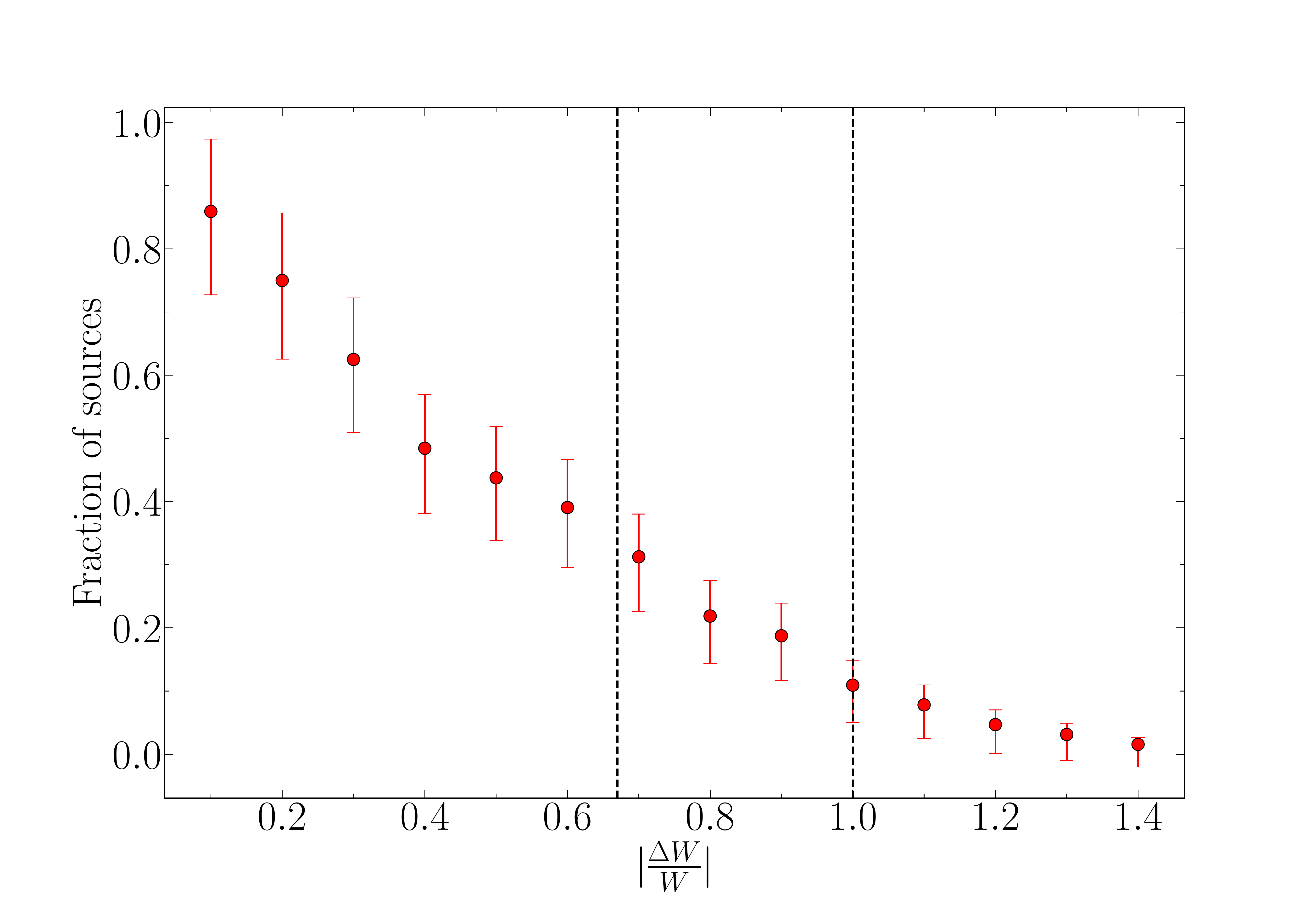}
    \caption{Fraction of sources showing  \civ\ absorption variability larger than a certain threshold value of \afdw\ vs the threshold value \afdw. Vertical dashed lines mark \afdw\ corresponding to equivalent with variations by a factor 2 and 3 respectively. Roughly 33\% of quasars in our sample have UFO BALs showing W changing by a factor of 2 during our monitoring period.
    }
    \label{fig:fdw_threshold_prob_abs}
\end{figure}

In Fig.~\ref{fig:fdw_threshold_prob_abs}, we  plot the fraction of sources showing \civ\ equivalent width variability between two epochs larger than a certain threshold \afdw\ as a function of \afdw. 
This fraction decreases with increasing \afdw\ threshold.
The percentage of sources showing \afdw\ greater than 0.67 and 1.0 (corresponding to equivalent width variations by a factor of 2 and 3 respectively) are 33\% and 11\% respectively as shown in Fig~\ref{fig:fdw_threshold_prob_abs}. 
In the sample of \citet{hemler2019} 26\% of the BAL quasars show \afdw$>$0.67. When we consider only BAL quasars satisfying our definition of UFOs the fraction increases to 40\%.
In the following discussions, we refer to BAL components with \afdw$>$0.67 as ``highly variable" BALs.  In the case of \citet{Lundgren2007} $\sim9$\% of the BALs are ``highly variable" (at time-scales of $\le102$ days) and this becomes $\sim$17\% if we consider the BALs that satisfy our condition to be an UFO BAL. In the sample of \citet{Gibson2008} $\sim8$\% of BALs are ``highly variable" over a time scale of 3-6 yrs. Even in the sample of \citep{Filiz2013} we find only $\sim$8\%\ of the BAL studied are highly variable. Thus it appears that UFO BALs tend to show equivalent width variations more frequently and with large amplitudes compared to the general BAL population.

Since most of the BALs in our sample show significant \civ\ equivalent width variations, it is interesting to study BALs that show roughly stable general profiles with insignificant equivalent width variations. 
%
It is possible that the non-detection of significant equivalent width variability in these types of sources can be attributed to the unavailability of a sufficient number of spectroscopic epochs. Alternatively, the absorption profile variation may be complex (i.e., uncorrelated over the full profile) and not perfectly captured by the equivalent width variations.
%
This may be the case for some of the BALs which are spread over a few ten thousand \kms\  and consist of multiple narrow variable regions varying in an uncorrelated manner.  
To address such cases, we  should consider pixel-based analysis which we plan to present in an upcoming  paper.
In our sample, the most dramatic BAL profile variation which resulted in the largest $W$ change of 35 \AA\ is observed in J1156+0856 (3 epochs).

\subsection{Time dependence of \civ\ BAL variability}
\label{sub:timedependence}

\begin{figure}
    \centering
    \includegraphics[viewport=25 75 2400 1550, width=\textwidth,clip=true]{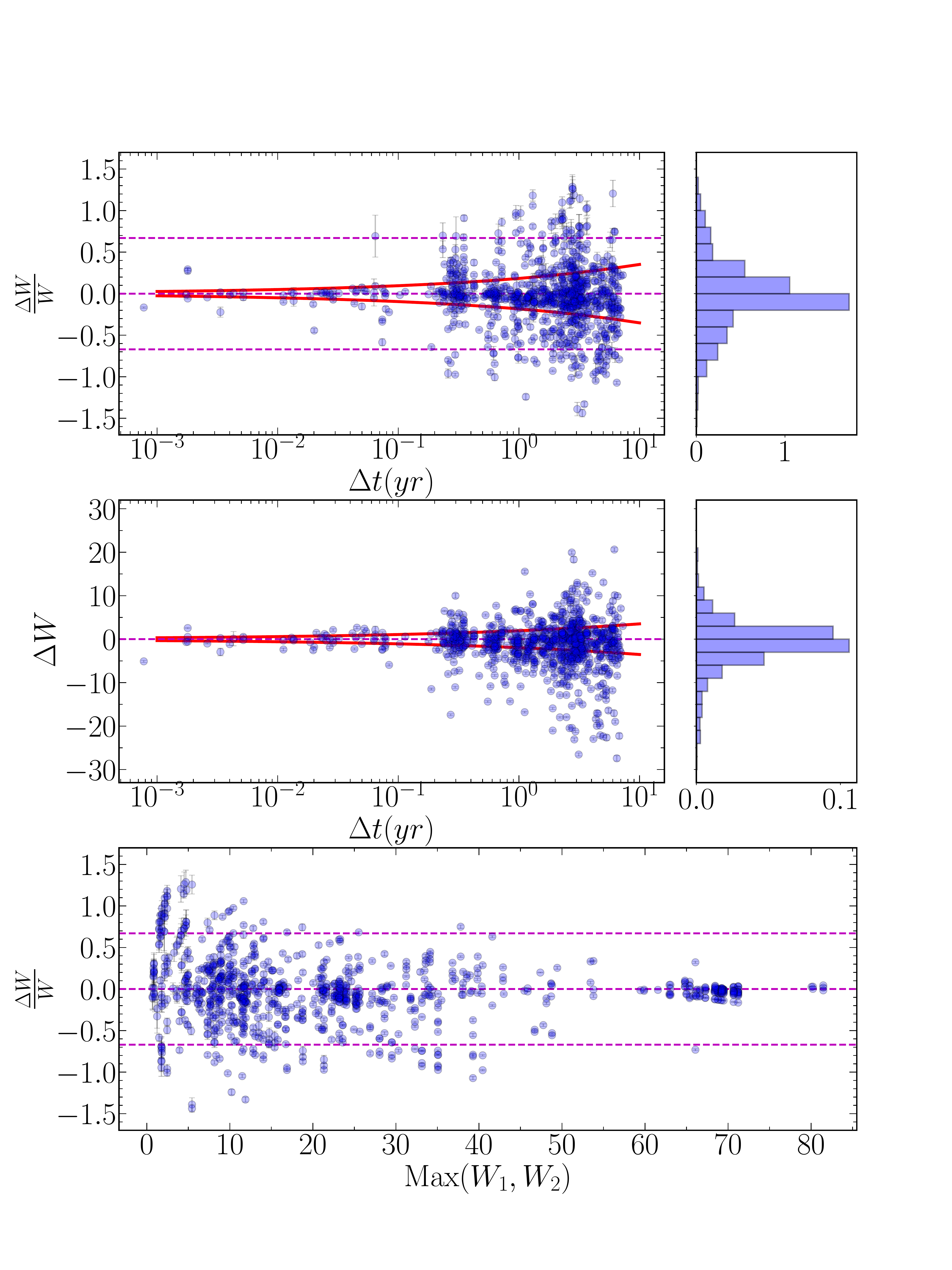}
    \caption{ {\it Top panel : } This figure shows the \fdw\ distribution for the full sample as explained in Section~\ref{sub:timedependence} along with the histogram (right panel).  Horizontal dotted lines mark \fdw = 0, $\pm$0.67 (i.e a factor 2 variation in W$_{CIV}$. 
    {\it Middle panel : } This figure shows the $\Delta W$ distribution for the full sample with histogram (right panel). In both panels the red curves are the best fit relationship obtained by \citet{Filiz2013}.
    {\it Bottom panel : } This figure shows the \fdw\ vs $\Delta W$ distribution for the full sample where horizontal dotted lines mark \fdw = 0, $\pm$0.67. 
    Note that the apparent “track” of points located close to each other mostly arises when the max(W$_1$,W$_2$) remains the same but \fdw\ keeps changing between various epochs for a single object.
    }
    \label{fig:fdw_dw_distr}
\end{figure}

\begin{figure}
    \centering
    \includegraphics[viewport=15 80 1230 1340, width=0.5\textwidth,clip=true]{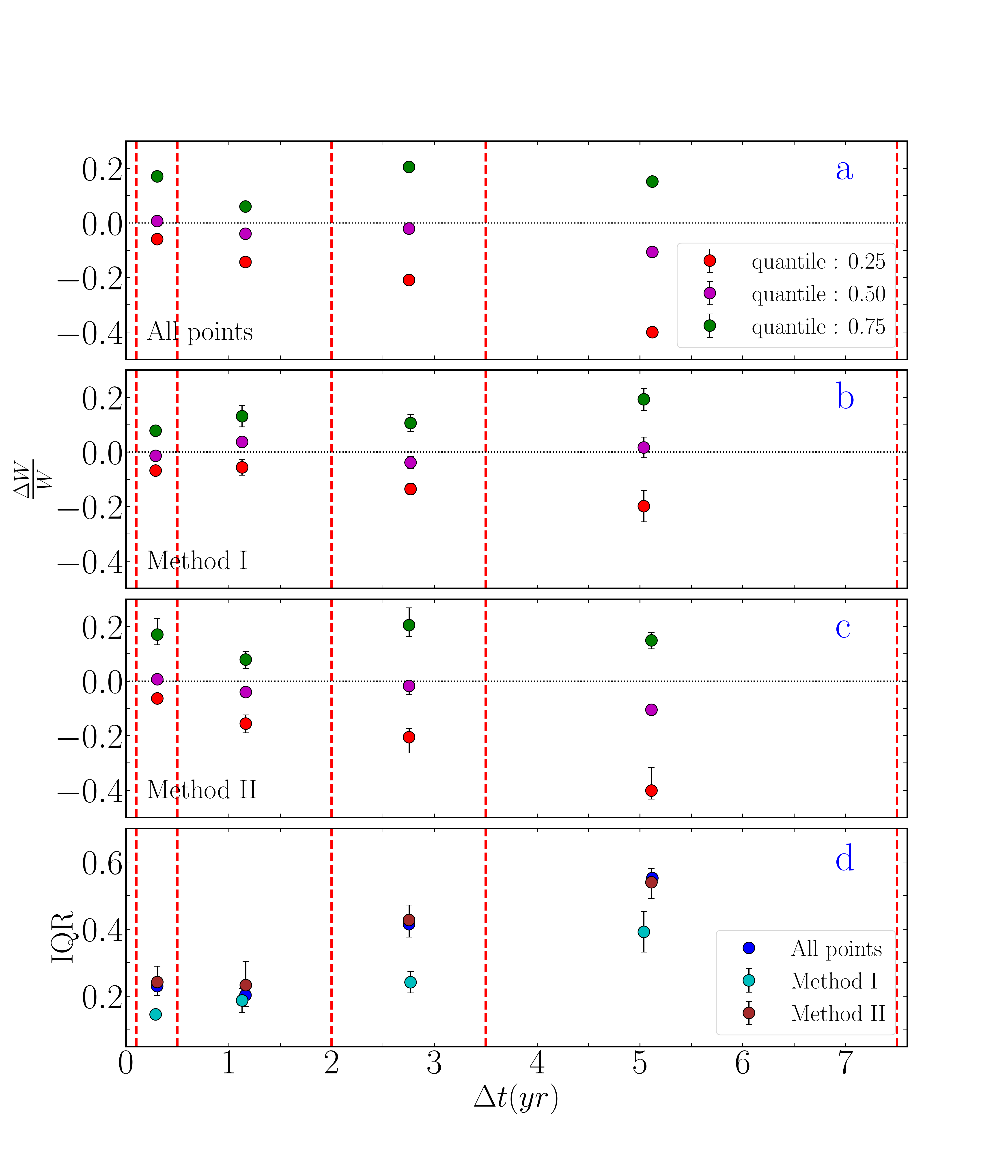}
    \caption{ 
 In the top three panels, we show the mean and sigma of quantiles at 0.25, 0.5 and 0.75 of \fdw\ distribution using all the points in \fdw\ distribution (panel a), using method I (panel b) and method II (panel c) as described in Section~\ref{sub:timedependence}, for four different time bins.
 In panel d, we show the IQR of \fdw\ distribution as calculated by different methods using quantiles shown in the top 3 panels.
 In each panel, the vertical red dashed lines mark the four $\Delta t$ bins used for the analysis.
%
    }
    \label{fig:iqr_final_plot}
\end{figure}

\begin{figure}
    \centering
    \includegraphics[viewport=10 5 2400 770, width=\textwidth,clip=true]{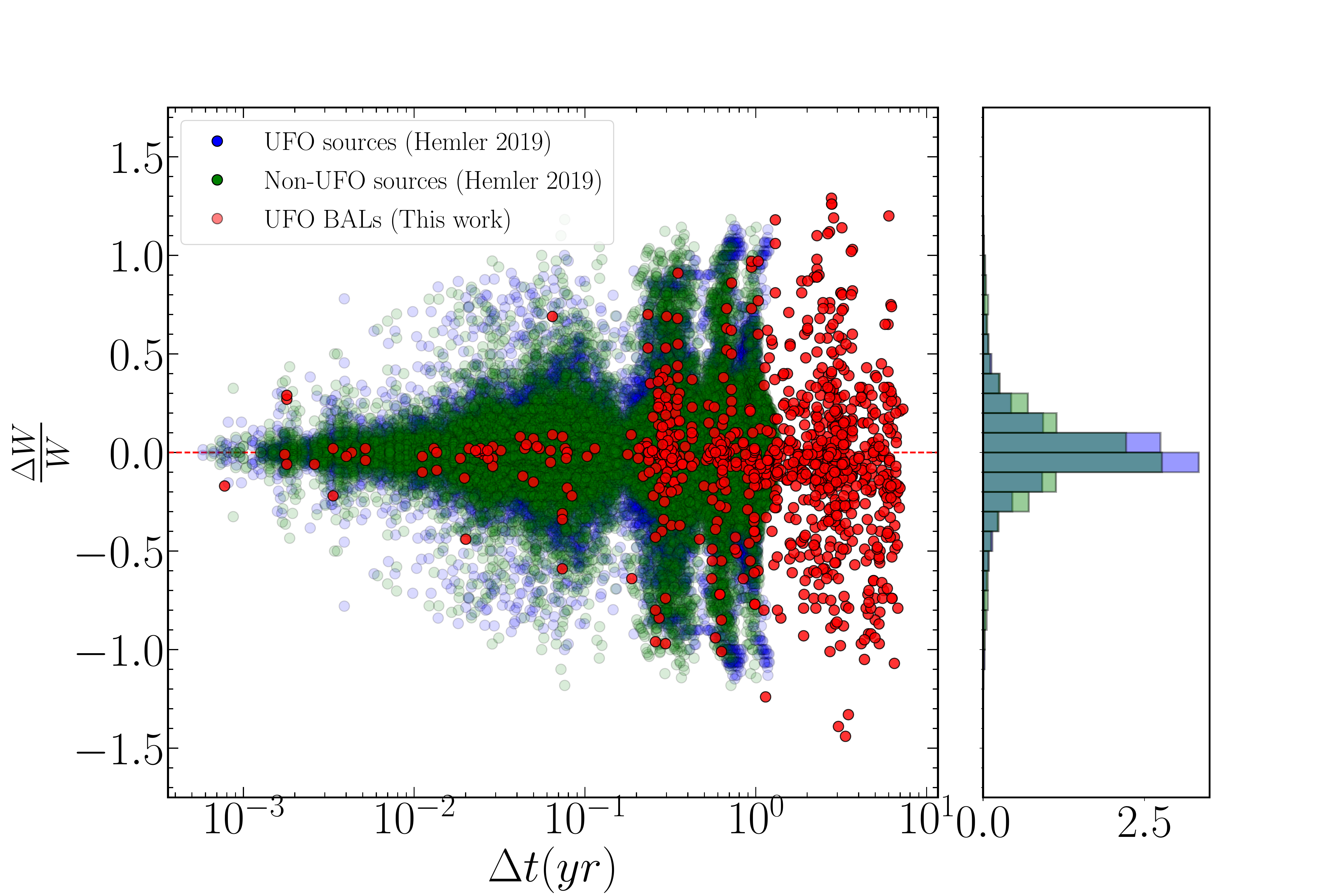}
    \caption{This figure shows the \fdw\ as a function of time for UFO (blue) and non-UFO (green) sources. For comparison, points from our UFO sample (red) is also shown.
    }
    \label{fig:delw_vs_dmjd_hemler2019}
\end{figure}

Next, we study the variability of the \civ\ BAL equivalent width and its dependence on the time interval probed. In the top and middle panels of Fig.~\ref{fig:fdw_dw_distr}, respectively, we plot \fdw\ and \dw\ as a function of the quasar rest frame time interval for all possible pairs of epochs (see Eqs.~1 and 2 for definitions). 
 In the two panels the best-fit relationship obtained by \citet{Filiz2013}
 is over-plotted as red solid curves. We can see that a significant number
 of our data points lie beyond these curves.
  In the bottom panel of Fig.~\ref{fig:fdw_dw_distr}, we show \fdw\ as a function of the maximum \civ\ rest equivalent width between the two epochs considered.
The two horizontal dashed lines in the top and bottom panels indicate a fractional variability of $\pm$0.67 (i.e., ``highly variable" BALs).
 We note that the scatter in both \fdw\ and \dw\ is larger at longer time scales. It is also evident that the \civ\ equivalent width variations by more than a factor of 2  (or \big{|}\fdw\big{|}$\ge$0.67) occur only on large time-scales (i.e., $>$ 0.2 yr) in our sample. 

 From the bottom panel of Fig.~\ref{fig:fdw_dw_distr},
 it is evident that there is very little scatter in \fdw\ when the maximum \civ\ rest equivalent width is more than 55\AA. On the other hand, a large scatter in \fdw\ is seen at small \civ\ equivalent widths (i.e $<$5 \AA). We do find a nearly uniform scatter in \fdw\ in the middle range of maximum \civ\ equivalent widths. Also, we note that the occurrence of epoch pairs with \afdw$>0.67$ (i.e., ``highly variable" BAL) is independent of \civ\ rest equivalent width in this middle $W$ range.

Overall, we have 42 (52), 32 (40), 64 (80), and 36 (46) individual sources (or BALs) contributing 131, 287, 291, and 192 pairs of epochs respectively in 0.1-0.5, 0.5-2.0, 2.0-3.5, and 3.5-7.5 yrs time bins. 
In these 4 time bins the fraction of UFO BAL quasars having at least one BAL component that is ``highly variable" are 0.12, 0.16, 0.23 and 0.28. The same for the BAL components are 0.14, 0.18, 0.20 and 0.24 respectively.
Thus there is a clear indication in our sample that the fraction of ``highly variable" BALs increases with increasing time intervals. For the first three time-bins the fraction of ``highly variable" BALs are 0.01, 0.08 and 0.11 for the BALs studied in \cite{Filiz2013}. 
This once again confirms our finding that UFO BALs are more variable compare to the general BAL population.
%
In the time bin 2.0$-$3.5 yrs (where all quasars in our sample contribute) the number of UFO BAL quasars showing \fdw $>+0.67$ (6 BALs) and \fdw$<-0.67$ (7 BALs) are nearly identical. Three BALs show both positive and negative variations with \afdw $>$0.67 in this time bin. However, in the 3.5$-$7.5 yrs time bin, we find a low fraction of UFO BAL quasars and BAL components showing \fdw $>0.67$ (3 BAL components) compared to that showing  \fdw$<-0.67$ (7 BAL components).  Interestingly, for the 2.0$-$3.5 yrs time-bin,  the number of ``highly variable" BALs showing a negative trend is roughly a factor two higher than those showing positive trend in the sample of \citet{Filiz2013}. This may indicate that the growth and decay of \civ\ equivalent width may have different characteristic time scales.
We will investigate this in more detail below.

We quantify the time dependence of \civ\ equivalent width variability using the inter-quantile range (IQR) as an indicator of the strength of variability for different time scales. 
We consider the same four time bins discussed above
and compute  the quantiles at 0.25, 0.50, and 0.75 of the \fdw\ distribution for each time bin considering all points in the \fdw\ distribution shown in Fig.~\ref{fig:fdw_dw_distr}. The quantiles and 
the IQR estimated from these are shown in panels (a) and (d) of Fig.~\ref{fig:iqr_final_plot}, respectively, for the four time bins. 
We observe an increasing trend in IQR  with increasing time intervals with values of 0.23$^{+0.04}_{-0.04}$, 0.20$^{+0.07}_{-0.06}$, 0.41$^{+0.04}_{-0.05}$, and 0.55$^{+0.04}_{-0.05}$  for the four time bins in the ascending order.
This indicates that the scatter in \fdw\
increases with time 
\citep[as found in past studies e.g.][]{capellupo2011,Filiz2013}.
%

As shown in the top panel of Fig.~\ref{fig:del_t_min_distr}, the number of spectroscopic epochs varies strongly from one source to the other in our sample
(see Table~\ref{tab_sampledetails}).
In order to probe any bias in the \fdw\ distribution due to this non-uniform time sampling, we use two alternate methods to estimate the IQR.
In the first method (hereafter method I), for each time bin, 
we randomly choose one measurement of \fdw\ per BAL and measure the quantiles at 0.25, 0.50, and 0.75 and IQR for the derived \fdw\ distribution.
 We repeat this procedure 1000 times and calculate the mean and $\sigma$ of the resulting IQR distribution. Results using this method are shown in panel (b) of Fig.~\ref{fig:iqr_final_plot}.
The final IQR values are $0.18 \pm 0.03$, $0.22 \pm 0.04$, $0.26 \pm 0.03$, and $0.39 \pm 0.07$ for 0.1-0.5, 0.5-2.0, 2.0-3.5, and 3.5-7.5 yrs time bins respectively.

In the second method (hereafter method II), we randomly choose 50 out of 64 sources and re-sample 64 sources from this sub-sample with replacement. We then calculate IQR for each time bin using all points from the randomly selected sources. We repeat this procedure 100 times and the mean and $\sigma$ of the resulting IQR distribution are estimated.
The final values of quantiles at 0.25, 0.50, and 0.75 and the IQR estimates are shown in panels (c) and (d) of Fig.~\ref{fig:iqr_final_plot}.  
 {The final IQR values are $0.24 \pm 0.04$, $0.23 \pm 0.07$, $0.42 \pm 0.05$ and $0.53 \pm 0.06$ for 0.1-0.5, 0.5-2.0, 2.0-3.5 and 3.5-7.5 yrs time bins respectively. These values are even closer (i.e well within $1\sigma$) to the ones obtained using the full sample.} This exercise confirms that the increase in \fdw\ with time is not biased by non-uniform time sampling in our sample.

 In addition, we looked at the width of the \fdw\ distribution after removing BALs having either very small or large W. 
For this, we considered only BALs with 5 $< W_{max} <$ 55 \AA\ which corresponds to 69 out of the 80 BALs in our sample and carried out the same analysis to derive IQR values. Using method I, we obtained IQR values of 0.25 $\pm$ 0.03, 0.36 $\pm$ 0.08, 0.37 $\pm$ 0.04, 0.61 $\pm$ 0.07 for 0.1-0.5, 0.5-2.0, 2.0-3.5, and 3.5-7.5 yrs time bins respectively. This confirms that the increase in the \civ\ equivalent width variability over time is not dominated by \civ\ BALs with large or small rest equivalent widths.
{ In a similar way, we calculated the IQR for all the sources
in the sample of \citet{hemler2019} which turned out to be 0.11, 0.20, and 0.25 for 0-0.1, 0.1-0.5, and 0.5-2.0 years time bins respectively. This again
suggests an increase with time of the scatter in the \fdw\ distribution as shown in Fig~\ref{fig:delw_vs_dmjd_hemler2019}. 
There are two time bins that are common to our sample for which the IQR measured are consistent within a $1.5\sigma$ range.
%

\subsection{Is the \fdw\ variation symmetric?}
\label{sec:asymmetry}

From Fig~\ref{fig:iqr_final_plot}, we can see that the absolute value 
of the \fdw\ distribution quantiles at 0.25 and 0.75 are not the same 
in the 3.5-7.5 yrs time bin. 
This might suggest a possible asymmetry in our sample. This is what we study in this section.
A symmetric distribution will indicate a statistically  similar 
characteristic time-scale for increase/decrease in the \civ\  BAL 
absorption. 
Interestingly, while studying emerging/disappearing BAL components, several authors have noted \citep[see for example,][]{mcgraw2017, Misra2019} that the average time scales probed are higher for BAL disappearance events compared to the emergence events. Discussions presented above also indicate that highly variable UFO BAL components tend to show a more negative trend in the 3.5$-$7.5 yrs time bin. 

%

First, we carefully looked at the \civ\ rest equivalent width as a function of time for all objects in our sample to identify the  presence of any monotonous trends either in the positive (i.e., increasing) or in the negative (decreasing) direction. Based on visual inspection, we classified the 80 BAL troughs in our sample into four classes: the ones  showing (i) both considerable negative and positive \dw\ variations, (ii) broadly positive \dw\ variations (in these sources, there may be a few epochs with negative EW variations which are not significant compared to the general trend of variations), (iii) broadly negative \dw\ variations and (iv) no systematic trend. The number of BALs in each category is 38, 16, 18, and 8 respectively. Thus, our sample does have 34 objects that show predominantly increasing or decreasing trends in the rest equivalent width as a function of time. 

First, we consider the 3.5$-$7.5 yrs time bin, for which the quantiles plotted in the top 3 panels of Fig.~\ref{fig:iqr_final_plot} suggest a possible asymmetry towards the negative direction (i.e decreasing W). We measure the absolute difference in the values at 0.25 and 0.75 quantiles to be 0.25$\pm$0.09, 0.17$\pm$0.07 and 0.25$\pm$0.09 respectively for the values obtained using the full sample, method I and method II respectively. This suggests an asymmetry towards negative (decreasing W) at 2.4$\sigma$ to 2.8$\sigma$ level.
We note however that 28\% of the UFO BALs contributing to this time bin are ``highly variable" BALs with the tendency to show negative variations. 
We thus probe the symmetry of the distribution after removing the 10 UFO BAL quasars in this bin that show  ``highly variable" BAL components. We find the distribution to be more symmetric with an absolute difference in the values at 0.25 and 0.75 quantiles to be $0.02\pm0.04$ when we use ``method I". The KS-test returns a p-value of 0.71. Thus the large asymmetry seen in this time bin is largely driven by ``highly variable" BAL sources.

Next, we consider the 2.0$-$3.5 yrs time bin where all the 64 UFO BAL quasars in our sample contribute. We measure the absolute difference in the values at 0.25 and 0.75 quantiles to be 0.05$\pm$0.06, 0.01$\pm$0.06 and 0.00$\pm$0.06 respectively for the values obtained using the full sample, method I and method II respectively. This indicates a symmetric distribution in the positive and negative directions. The KS-test results confirm the same with a p-value of 0.36. Next, we ask whether the distribution remains symmetric when we avoid the 10 UFO BAL QSOs that show ``highly variable" BALs in the 3.5$-$7.5 yrs time-bin. In this case we get 
the absolute difference in the values at 0.25 and 0.75 quantiles to be $0.07\pm0.08$ when we use "method I" and the KS-test p-values of 0.41.
The distribution is found to be symmetric in any case.

Since the asymmetry seems to arise from highly variable BALs on longer timescales,
we focus on the time evolution of 21 BAL components from the 20 quasars in our sample  that show high variability (\afdw $>$ 0.67) for $\Delta t > 2$ yr.
To begin with, for each highly variable BAL in this sub-sample, we plot the rest equivalent width, W,  normalized by the maximum W (i.e., $W_{max}$) observed for that BAL against the time difference from the epoch where it attained the maximum W value (see Fig~\ref{fig:dw_doubling_dt}). Hence, the positive (negative) values in the abscissa correspond to epochs after (before) the maximum W is achieved. Here, we clearly see that there are more BALs that show large variability on the decreasing side compared to the increasing side. This has been discussed in detail in Section~\ref{sub:timedependence}. 
Interestingly, we also observe that many BALs on the decreasing side (i.e., positive t-t$_{W_{max}}$) reach values less than $ \frac{W_{max}}{2} $ (as indicated by the red dashed lines in Fig~\ref{fig:dw_doubling_dt}) at larger time scales of more than 4 yr. But, on the increasing side (i.e., negative t-t$_{W_{max}}$), among the comparatively lower number of BALs showing high variability, most of them increase to $W_{max}$ from values less than $ \frac{W_{max}}{2} $  at rather smaller time scales of $\sim$ 3 yr.  We notice that the time scales for the equivalent width to decrease by a factor of 2 are on average higher than that of increasing equivalent width cases (look at the histogram in the top panel of Fig.~\ref{fig:dw_doubling_dt}).
As asymmetry seen in \fdw\ for this time bin is dominated by the ``highly variable" BAL components this could be driven by the fact that
the decreasing/disappearing time scale is 
generally longer than the increasing/appearing time scale.
But it is good to keep in mind that when we study the appearing and disappearing BAL time scales, there can be a certain bias in the results due to the fact that all our BALs are fairly strong in the first epoch itself. This may affect the study of appearing time scales since we have missed the epochs when the BAL was formed.

\begin{figure}
    \centering
    \includegraphics[viewport=50 0 2430 650, width=\textwidth,clip=true]{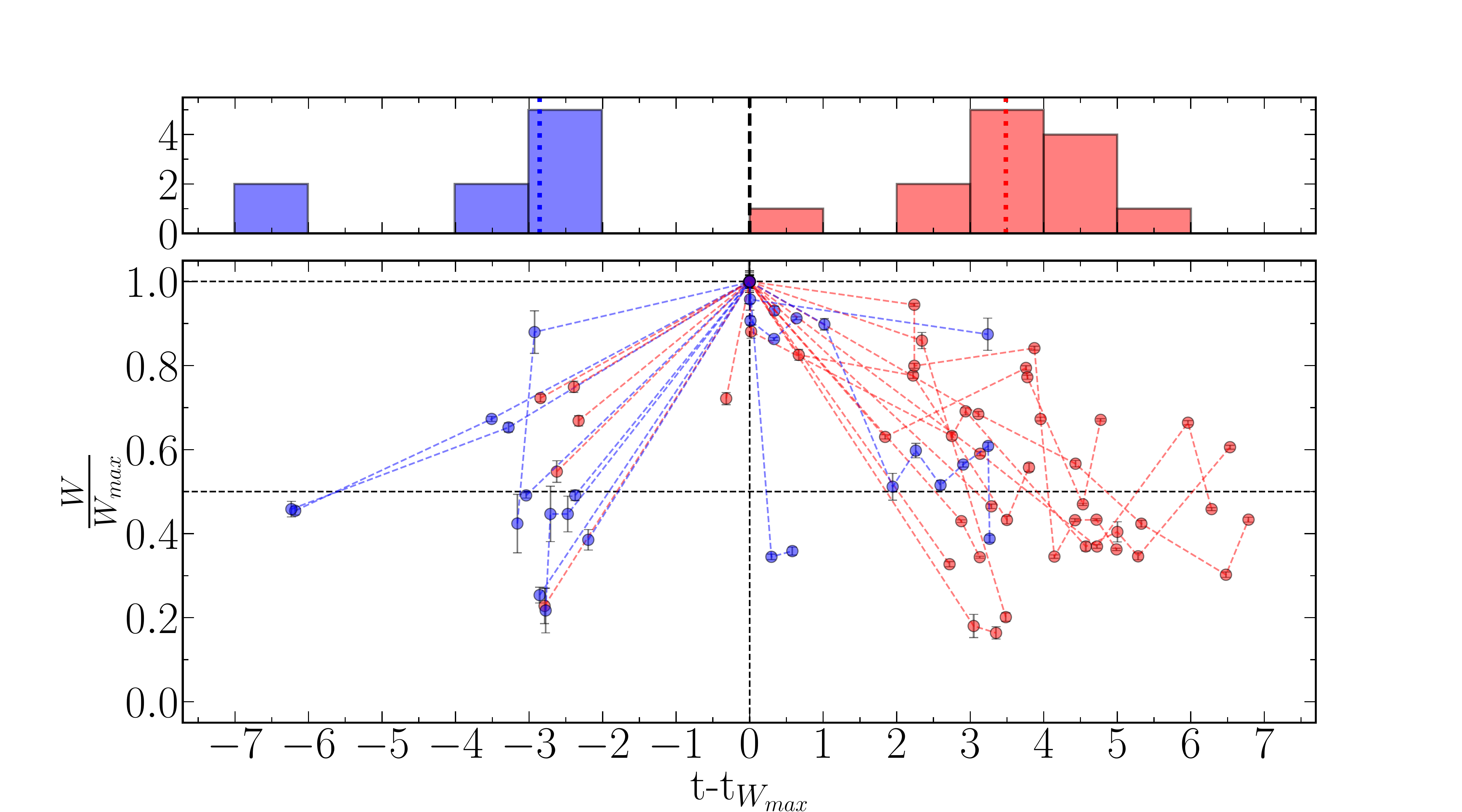}
    \caption{The \civ\ rest equivalent width (W) normalized by the maximum observed W (i.e., $\frac{W}{W_{max}}$) is plotted against the time difference (in year) from the epoch where it attained the maximum W value (i.e., t-t$_{W_{max}}$) for each BAL in the ``highly variable" sub-sample.  The blue points correspond to BALs which had the maximum change in $\frac{W}{W_{max}}$ in the increasing direction whereas the red correspond to the same in the decreasing direction. In the top panel, we plot the histogram of shortest t-t$_{W_{max}}$ at which $\frac{W}{W_{max}}$ crossed the 0.5 line with the median shown in dotted lines.
    }
    \label{fig:dw_doubling_dt}
\end{figure}

\subsection{Comparison with \citet{Filiz2013} results}
\label{subsec:filiz2013_comp}

\begin{figure}
    \centering
    \includegraphics[viewport=55 0 2400 680, width=\textwidth,clip=true]{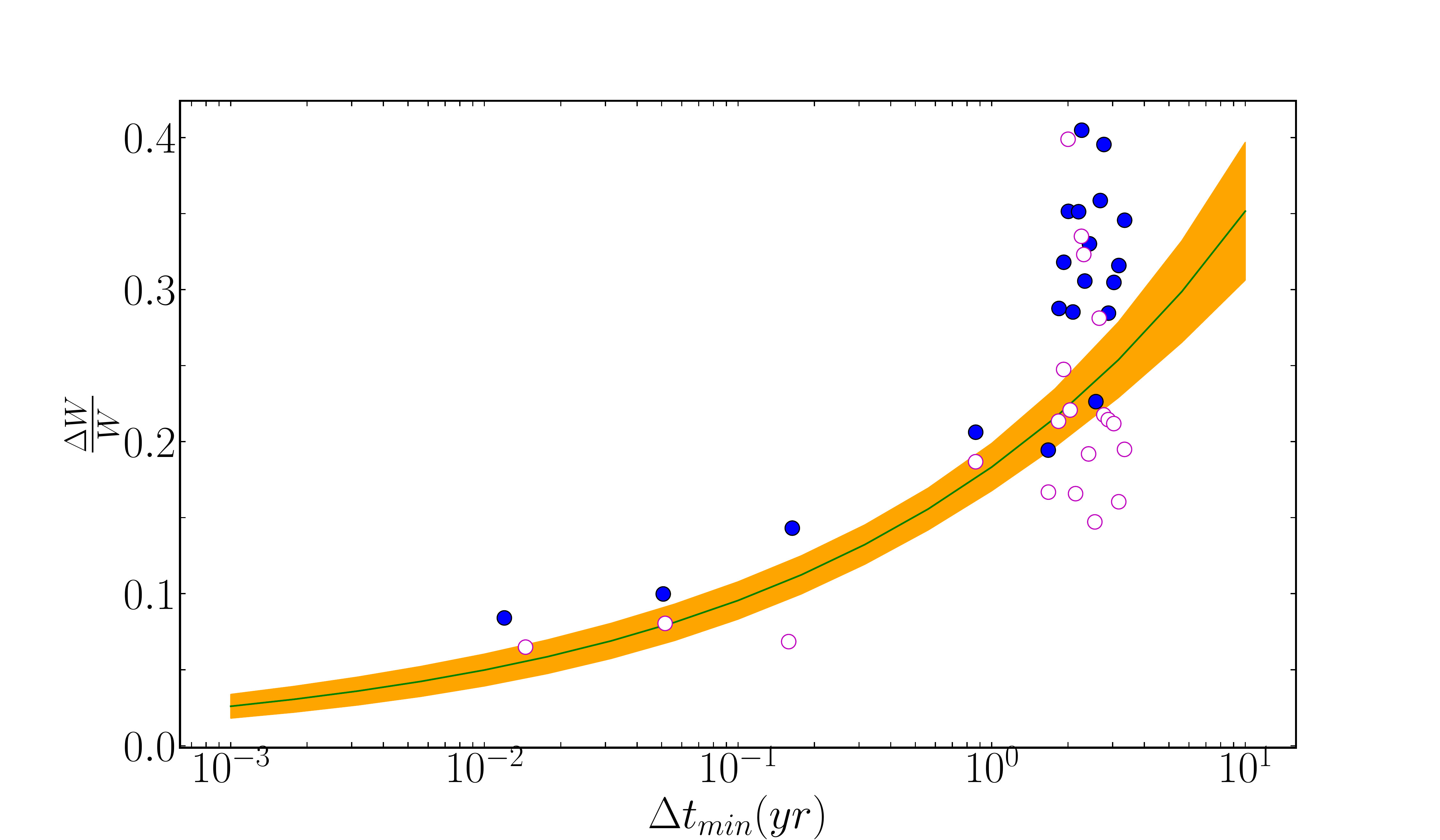}
    \caption{The fractional variation in W of \civ\ UFO BAL observed in our sample 
    (circles, see Section   \ref{subsec:filiz2013_comp}) 
    is plotted versus the minimum sampled rest-frame timescale
$\Delta t_{min}$. The best-fit relation obtained in \citet{Filiz2013} is overplotted as a blue curve and the shaded region shows its uncertainty.  
    The empty circles represent the same comparison after removing the ``highly variable" sources (as explained in Section~\ref{sec:asymmetry}) from the sample.
    }
    \label{fig:delw_vs_dmjd_filiz2013}
\end{figure}

\citet{Filiz2013} have obtained a relationship between 
the relative W variation \afdw\ and 
the minimum sampled rest-frame timescale
$\Delta t_{min}$ from the 428 BAL troughs identified in their sample (see equation 7 in their paper).  
As we noted in Fig.~\ref{fig:fdw_dw_distr}, our sample seems to show more scatter compared to this relationship. However, in order to perform a significant comparison we need to account for the differences in the distribution of minimum sampling time intervals ($\Delta t_{min}$) between their sample and our UFO sample (see Fig.~\ref{fig:del_t_min_distr}).
%
For this, we first calculate $\frac{\Delta W}{W}$ for all possible combinations of epochs for each source in the UFO sample and populate 
the \afdw\ vs. $\Delta t$ plane. Next, we resample 80 points from this plane
with the time sampling consistent with a normalized $\Delta t_{min}$ distribution as given in \citet{Filiz2013} for the 428 distinct BAL troughs in their sample. 
Following \citet{Filiz2013}, we also do an average over 4 time-ordered data points from the above sampled points. This results in a total of 21 points. We repeat the above steps 1000 times and take the mean of each point from the total number of iterations. Now we compare these points (filled blue circles) with the best-fit relation from \citet{Filiz2013} in Fig.~\ref{fig:delw_vs_dmjd_filiz2013}. We observe that most of the points are located at a few years time-scale and almost all of them show much higher \afdw\ values than what is expected from the fit. This confirms that the UFO BALs in our sample are highly variable compared to the fit from \citet{Filiz2013}.
This difference can not be attributed to the differences in the sampling of $\Delta t$.
 If we remove the 20 quasars with "highly variable" BALs
 and repeat the procedure, we derive measurements shown
 as open circles in Fig.~\ref{fig:delw_vs_dmjd_filiz2013}
 which are more consistent withe the \citet{Filiz2013} relation.
This confirms that the behavior of \afdw\ 
is significantly influenced by the ``highly variable" BALs.

\subsection{What drives the BAL variability ?}

In this section, we correlate \fdw\ with different BAL and quasar properties to investigate the physical conditions in UFO BALs and the possible origin of the equivalent width variations.

\subsubsection{Dependence on the BAL properties}
\label{sec:balprop}
\begin{figure}
    \centering
    \includegraphics[viewport=10 65 1600 1270, width=\textwidth,clip=true]{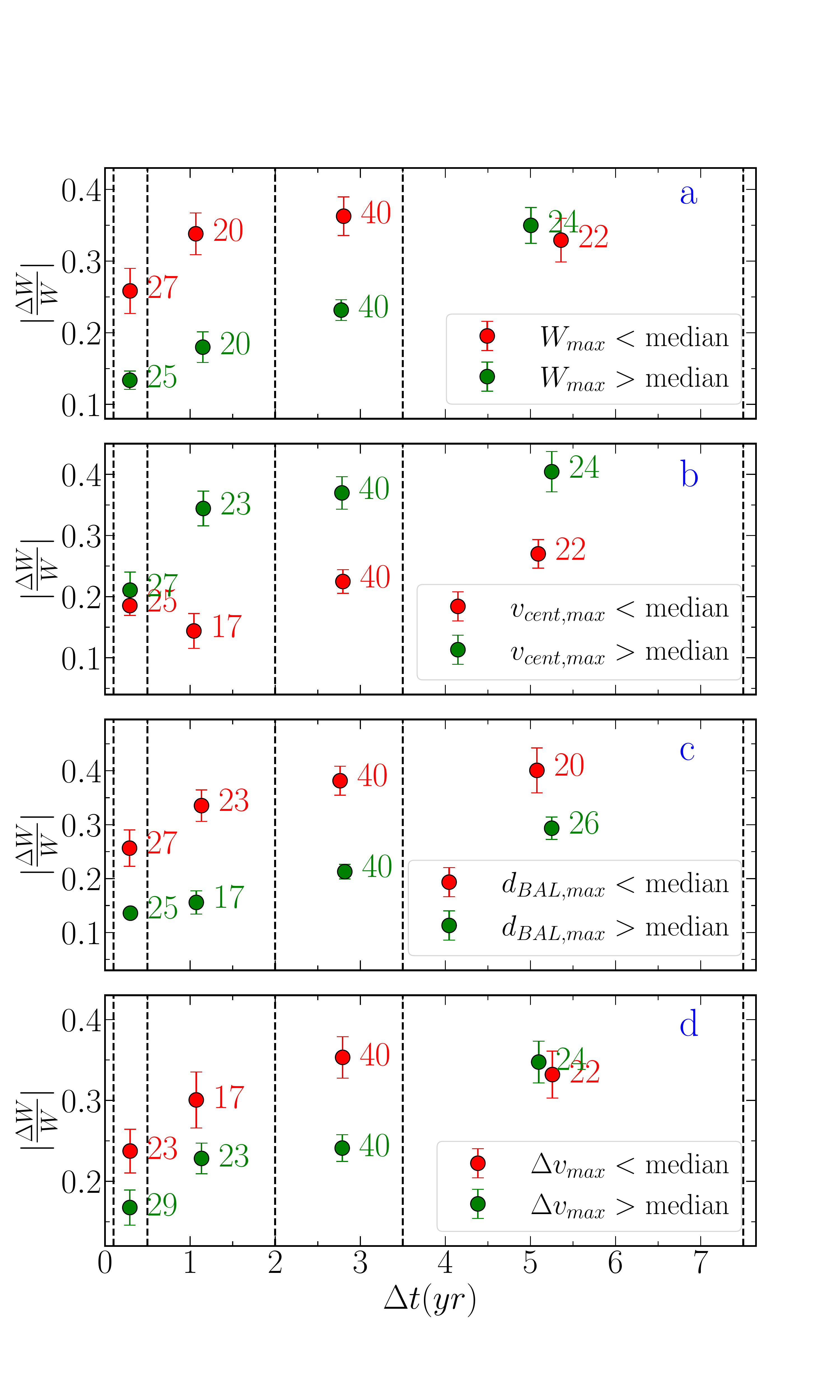}
    \caption{The average \afdw\ distribution of different sub-samples 
    ($W_{max}$ in panel a, optical depth weighted velocity centroid in panel b, maximum depth in panel c, and velocity width in panel d) defined below and above the median (see section~\ref{sec:balprop}) of different BAL properties for different time bins. Errors correspond to the standard deviation. The number of UFO BALs used for each measurement is also indicated in the figure.
    }
    \label{fig:delw_vs_dmjd_bal_prop}
\end{figure}

\begin{figure}
    \centering
    \includegraphics[viewport=30 65 1600 1150, width=\textwidth,clip=true]{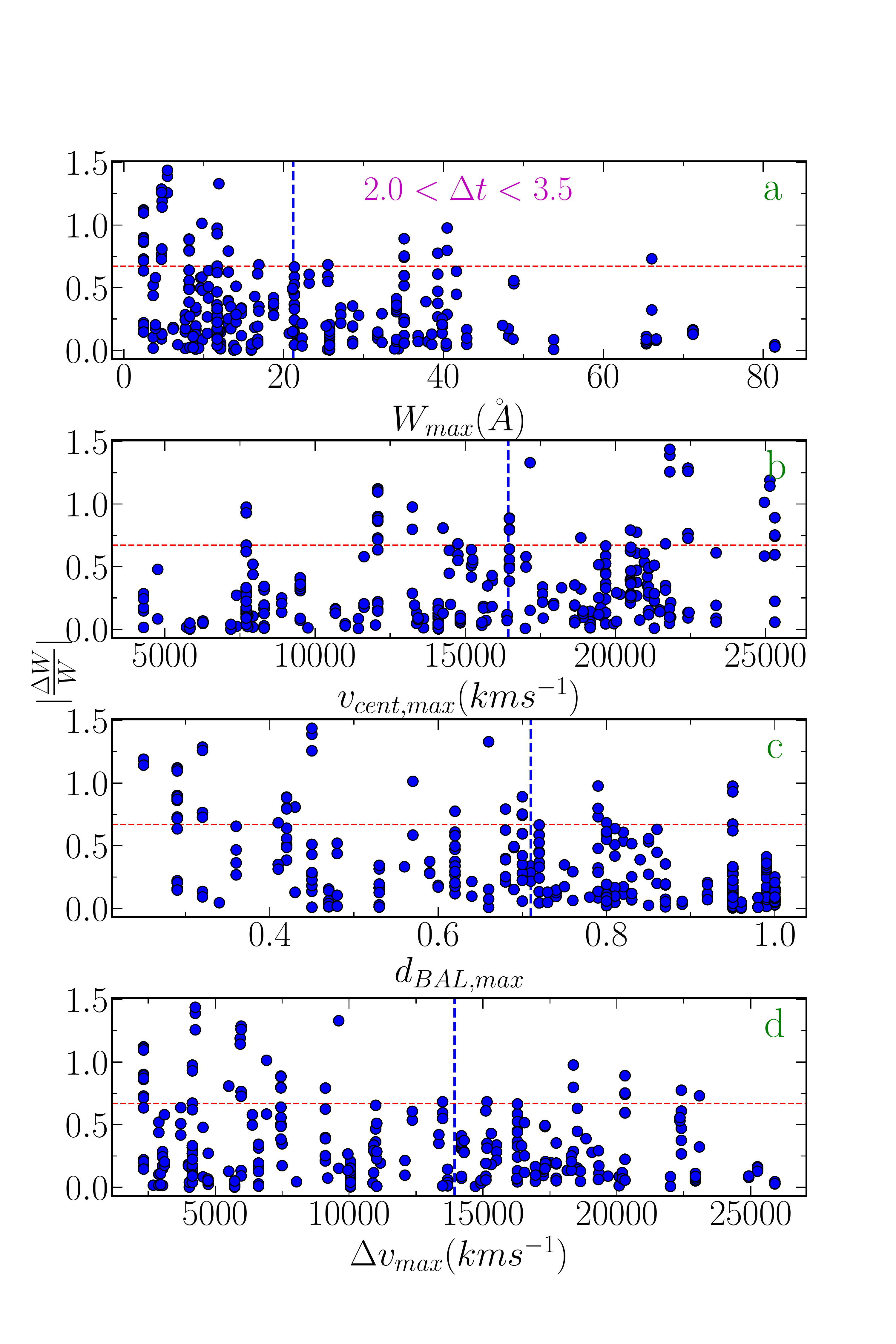}
    \caption{All the measured \afdw\ for each BAL component as a function of their properties for the time bin, 2.0 $\le \Delta t \le$ 3.5 yrs. The vertical dashed line in each panel indicates the median abscissa value. The horizontal line corresponds to \afdw=0.67. 
     }
    \label{fig:delw_vs_dmjd_bal_prop_full_pts}
\end{figure}

We consider four properties of the observed BAL troughs: the \civ\  equivalent width $W$, the optical depth weighted velocity centroid ($v_{cent}$), 
the maximum relative depth  $d_{BAL}$ and the velocity width ($\Delta V = v_{max}-v_{min}$) of the BAL trough.
For each of these parameters, we divide the BAL sample into two sub-samples of equal 
number including the BALs with parameter values larger and smaller than their median value.
The median values are 21.2 \AA, 16434 \kms, 0.71 and 13939 \kms  for  maxima of $W$, $v_{cent}$, $d_{BAL}$ and $\Delta v$ of the BAL respectively.


For each source in these sub-samples,  whenever available (as it is not necessary  that all objects contribute to all the time bins considered)
we randomly select a pair of epochs (and corresponding \afdw)  that falls 
in the four time bins considered here. 
We then measure the average \afdw\ for all sources in a sub-sample for each time bin.  We repeat the above process 100 times and obtain the mean and standard deviation of the average \afdw\ distribution of the sub-samples for each time bin.  Results are presented in Fig~\ref{fig:delw_vs_dmjd_bal_prop}.


In the first three time bins (i.e., $\Delta t\le$ 3.5 yrs) we note that the average \afdw\ is higher for the sub-sample with lower equivalent width (i.e., $W_{max} <21.2$\AA). The difference is larger than the 3.5 $\sigma$ level. This is not the 
case for larger time scales ($>$3.5 yrs).
This is consistent with what we find in the bottom panel of Fig.~\ref{fig:fdw_dw_distr}, where a large scatter in \fdw\ is seen for $W_{max}<5\AA$. 
In panel (a) of Fig.~\ref{fig:delw_vs_dmjd_bal_prop_full_pts}, we plot \afdw\ as a function of $W_{max}$ for all the observed pairs in the time bin 2.0$\le \Delta t \le$3.5 years (where all the 64 UFO BALs contribute). It is evident that the scatter in \afdw\ is larger when $W_{max}$ is below the median value indicated by the vertical dashed line. 
%
%
A similar trend 
is seen for $d_{BAL}$ (see panel (c) in Fig.~\ref{fig:delw_vs_dmjd_bal_prop} and Fig.~\ref{fig:delw_vs_dmjd_bal_prop_full_pts}). 
It is interesting to note that the trend is the same for the velocity widths ($\Delta v$).  Note that, in Fig.~\ref{fig:delw_vs_dmjd_bal_prop_full_pts}, the points look discretized due to the relatively small changes in these parameter values plotted in the x-axis for a given object for all the epoch pairs considered.
This means that absorption lines with low W having narrow and shallow absorption profiles tend to show large variability.

Some of the above findings are consistent with those of 
\citet{Filiz2013} (refer to subsection 4.5 therein). They find that shallower BALs show more variability and report  a highly significant correlation ($>$99 $\%$), between \dw, \fdw\ and average $W$ in their sample \citep[see also,][]{capellupo2011, Vivek2014, hemler2019}.
%
However, contrary to our results, they find that wider BALs vary more than narrower ones and state that this is expected since wider BAL troughs might have a better chance of containing variable regions. This difference could be related to the way our UFO BALs sample is constructed which avoids BALs that only have strong narrow absorption at low velocities (i.e.,  $v_{max} <15000$ \kms). We address this point in detail below.
Also, recall \citet{Filiz2013} have used one measurement per quasar obtained at the lowest time scale probed. Here, we use multiple epoch measurements for each quasar but sampled once for each time bin.

In low-resolution spectra, like the one we consider here, the equivalent width variations can be driven by (i) optical depth variations, (ii) variations in the covering factor \citep[$f_c$; see for example,][]{Muzahid2016} and/or (iii) changes in the line of sight density and velocity field due to transverse velocity component \citep[see for example,][]{aromal2022}. At this stage, there is no obvious way to
disentangle these possibilities which are all reasonably possible as
it is easier to change the property of weak components compared to strongly
saturated ones.


In panel (b) of Fig~\ref{fig:delw_vs_dmjd_bal_prop} we notice that the sub-sample with large $v_{cent}$
shows a larger variability (significant at  $>$3 $\sigma$ level) compared to the sub-sample with smaller $v_{cent}$ for all time-scales $\Delta t>0.5$ yrs.
In the lowest time bin, $0.1\le \Delta t (yrs) \le 0.5$, the difference between the two sub-samples is not statistically significant.
In panel (b) of Fig.~\ref{fig:delw_vs_dmjd_bal_prop_full_pts}, we plot \afdw\ as a function of maximum $v_{cent}$ for all the observed pairs in the time bin 2.0$\le \Delta t (yrs) \le $3.5 years. It is evident from this plot that the scatter in \afdw\ is larger for a higher value of maximum $v_{cent}$. Similarly highly variable UFO BALs are more frequent when  $v_{cent}$ is larger. 
%
This is consistent with the finding of \citet{Filiz2013} that the high-velocity BALs vary more than lower-velocity ones. They interpret this as a secondary effect based on the fact that higher-velocity absorption has lower $W$.  
However, in our sample, we do not find any significant anti-correlation between $W_{max}$ and $v_{cent}$ (with a Spearman's correlation coefficient of $-$0.02 and p-value of 0.85). Thus it is not obvious that the dependence of $v_{cent}$ is a secondary effect. 
%
To further investigate this, we note that ten out of 80 BAL components have $v_{cent}<8000$ \kms and tend to have low equivalent widths (i.e., $<$20~\AA). 
If we do not consider these 10 low-velocity components having low $W_{max}$ we do find indeed a moderate anti-correlation between  $W_{max}$ and $v_{cent}$ (with a Spearman correlation coefficient of $-$0.32 and p-value of 0.01). We come back to this in a bit more detail below.

Note that in the case of simple disk wind models (as well as when gas motions are assumed to be nearly Keplerian), large velocities are related to gas components ejected closer to the accretion disk \citep{arav1994b}.  In this kind of scenario $v_{cent}$ could well be the primary parameter of large \afdw. 
{ Further \citet{proga2012} demonstrated that \civ\ absorption produced by disk wind models from hydrodynamical simulations is quite complex and the  simulated \civ\ absorption profiles show large variability, especially at high velocities. They attribute this excess variability at large velocities to the emergence of very fast mass ejections from relatively large distances, where the gas is well shielded from X-ray radiation.  As we do not have independent distance measurements for different velocity components we will not  be able to test such a scenario using our data.}
%
%
To summarize the results presented here demonstrate that, in general, more fractional equivalent width variability is observed in weak, high-velocity, and shallow BALs. 

\subsubsection{Dependence on the quasar properties}

\begin{figure}
    \centering
    \includegraphics[viewport=20 55 1600 960, width=\textwidth,clip=true]{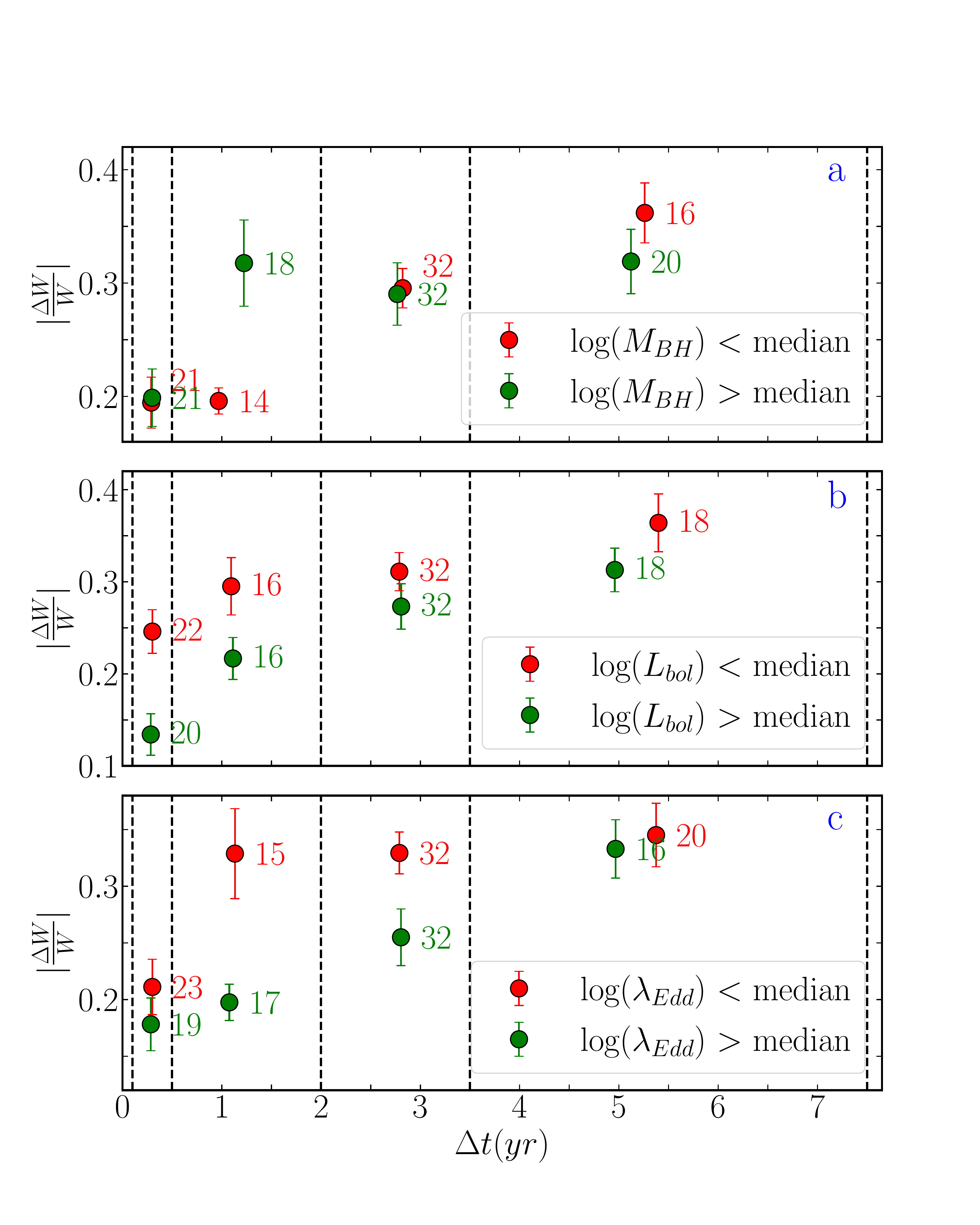}
    \caption{Same as Fig.~12 for \mbh (panel a), $L_{bol}$ (panel b), and \redd\ (panel c).
    }
    \label{fig:delw_vs_dmjd_quasar_prop}
\end{figure}

\begin{figure}
    \centering
    \includegraphics[viewport=17 0 1620 580, width=\textwidth,clip=true]{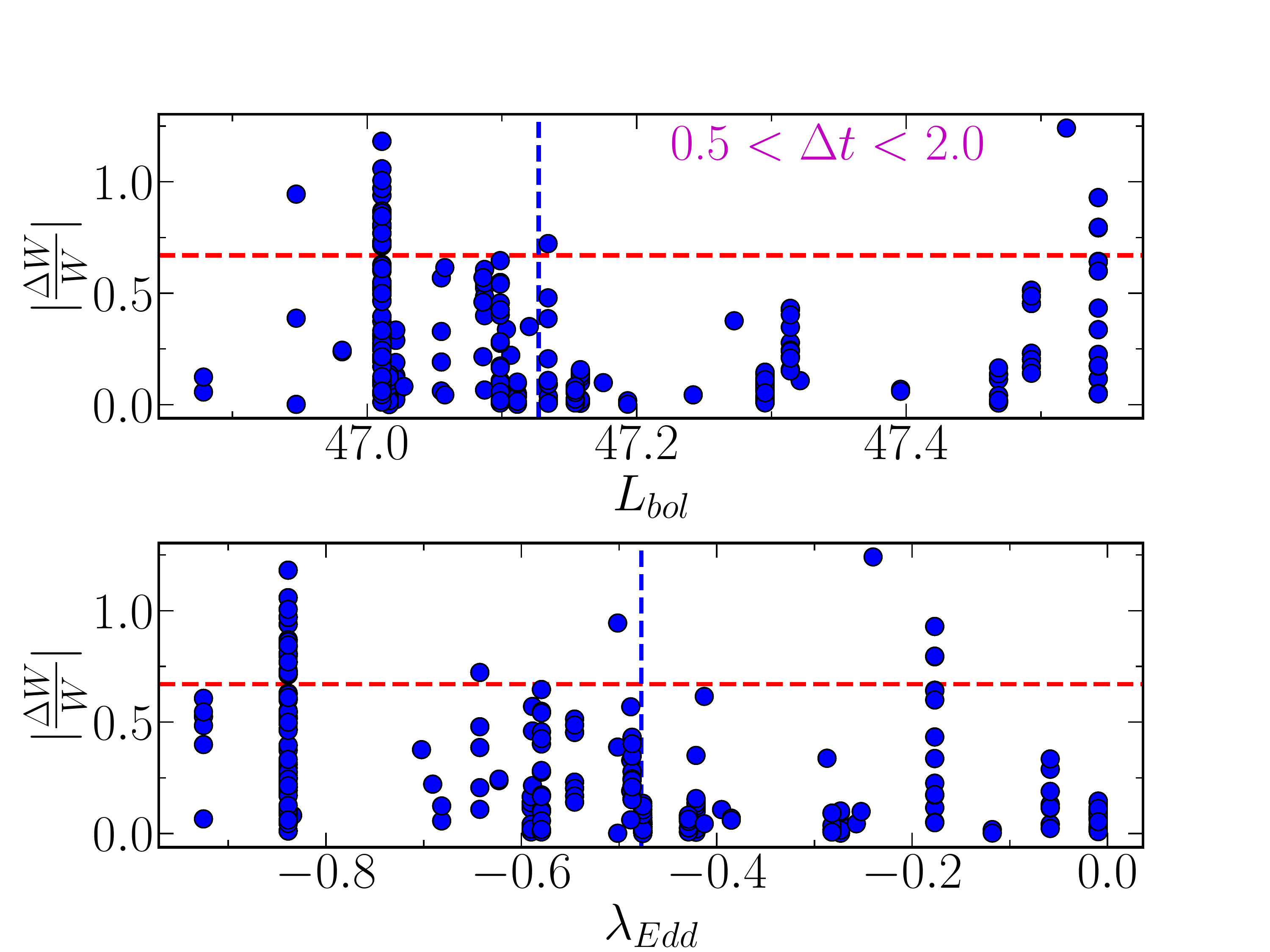}
    \caption{\fdw\ is shown for each UFO BAL as a function of $L_{bol}$ (top panel) and \redd\ (bottom panel) for 
    the time bin 0.5$< \Delta t <$2.0 yrs. The vertical dashed line indicates the median value of the quantity in the abscissa and the horizontal line indicates \afdw = 0.68. }
    \label{fig:delw_vs_lbol_redd}
\end{figure}

In this section, we explore whether the variations in \afdw\ are related to any of the 
following quasar properties: \mbh, \redd\ and bolometric luminosity $L_{bol}$.
For this, we again compare the \afdw\ distributions of two sub-samples 
defined as the objects having one of the above parameters 
smaller vs. larger than the median value of the whole UFO sample.
Following the same procedure as described above, 
we obtain the mean and standard deviation of \afdw\ in four time bins for each sub-sample. 
Results are summarized in Fig.~\ref{fig:delw_vs_dmjd_quasar_prop}. 

In the largest time bin (i.e., 3.5$\le t(yrs)\le$7.5) the variations of \afdw\ are  clearly independent of the quasar properties considered here. When we combine this with the discussions presented in the previous sub-section, it appears that for the largest time-scales 
only $v_{cent}$ seems to be related to the relative W variations indicating that
this may be an important parameter for the interpretation of these flows at large time scales (i.e., $t>3.5~yrs$). 

In the case of $L_{bol}$ and \redd, \afdw\ is found to be higher for low $L_{bol}$ and 
low \redd\ compare to high $L_{bol}$ and high \redd\ in all the three short time 
bins (i.e., for $\Delta t <3.5~ yrs$ in Fig.~\ref{fig:delw_vs_dmjd_quasar_prop}).
In particular for $0.5\le \Delta t <2.0~ yrs$, \afdw\ shows a large difference between sub-samples defined based on all three properties considered here (see Fig.~\ref{fig:delw_vs_dmjd_quasar_prop}). As mentioned before, this time bin gets contributions from only 32 quasars with 40 UFO BALs in our sample.
To explore this further, we plot \afdw\ vs. $L_{bol}$ (top panel) and 
\redd (bottom panel) in Fig.~\ref{fig:delw_vs_lbol_redd} for the time bin 0.5$\le \Delta t (yrs)\le$2.0. We do see the scatter to be larger towards lower $L_{bol}$ and \redd. 
However, we find (unlike in the case of $v_{cent}$ or $W_{max}$) no clear tendency for the ``highly variable" BAL components to have any preference for low $L_{bol}$ or \redd.
%
%
Also due to the limiting magnitude used for defining our sample  quasar properties of the objects in our sample may not 
span the full range spanned by the general population of quasars (see Fig.~\ref{fig:mbh_vs_redd}). 
%
Therefore, it would be good to confirm the differences seen in \afdw\ for different sub-sample using more measurements.

For the sake of comparison, we use an approach similar to that of
\citet{Filiz2013} and consider \afdw\ for the shortest time scale for each source. 
Except for $L_{bol}$ which shows a moderate anti-correlation (Spearman's coefficient = $-$0.31, p-value=0.01), both \mbh\ and \redd\ show no evidence of correlation. Similarly, 
when we consider the \afdw\ over the longest time scale for each source  we do not find any correlation with any of these quasar parameters. 
%
\citet{Filiz2013} found a large scatter in the \dw\ distribution for sources with 
low $L_{bol}$ on moderate (1-2.5 yr) and long ($>$2.5 yr) time scales compared 
to high $L_{bol}$ sources. However,  they could not find any significant correlation using rank-correlation analysis except for moderate time scales (1-2.5 yrs). 
They also did not find evidence for a correlation between \redd\ and 
scatter in either \dw\ or \fdw.

In the case of \mbh\ significant (i.e., at 3.6$\sigma$ level) difference is seen between 
our two sub-samples only in the $0.5\le\Delta t(yrs)\le 2.0$ time bin. 
\citet{Filiz2013} did not find any evidence for the presence of correlation between 
\mbh\ and \fdw\ in their sample. On the other hand, \citet{He2015} found that there is 
a medium strong negative correlation (r=$-$0.537, p=0.003) between \dw\ and \mgii\ based 
\mbh\ estimates in 28 BAL QSOs with variable BAL regions.

In summary,  in our sample, there is no strong evidence of any correlation between
the BAL variations and the characteristics of the quasars.

\subsubsection{Do highly variable UFO BAL quasars have different properties?}
\label{sub:qsoprop}

\begin{figure}
    \centering
    \includegraphics[viewport=50 50 2360 1150, width=\textwidth,clip=true]{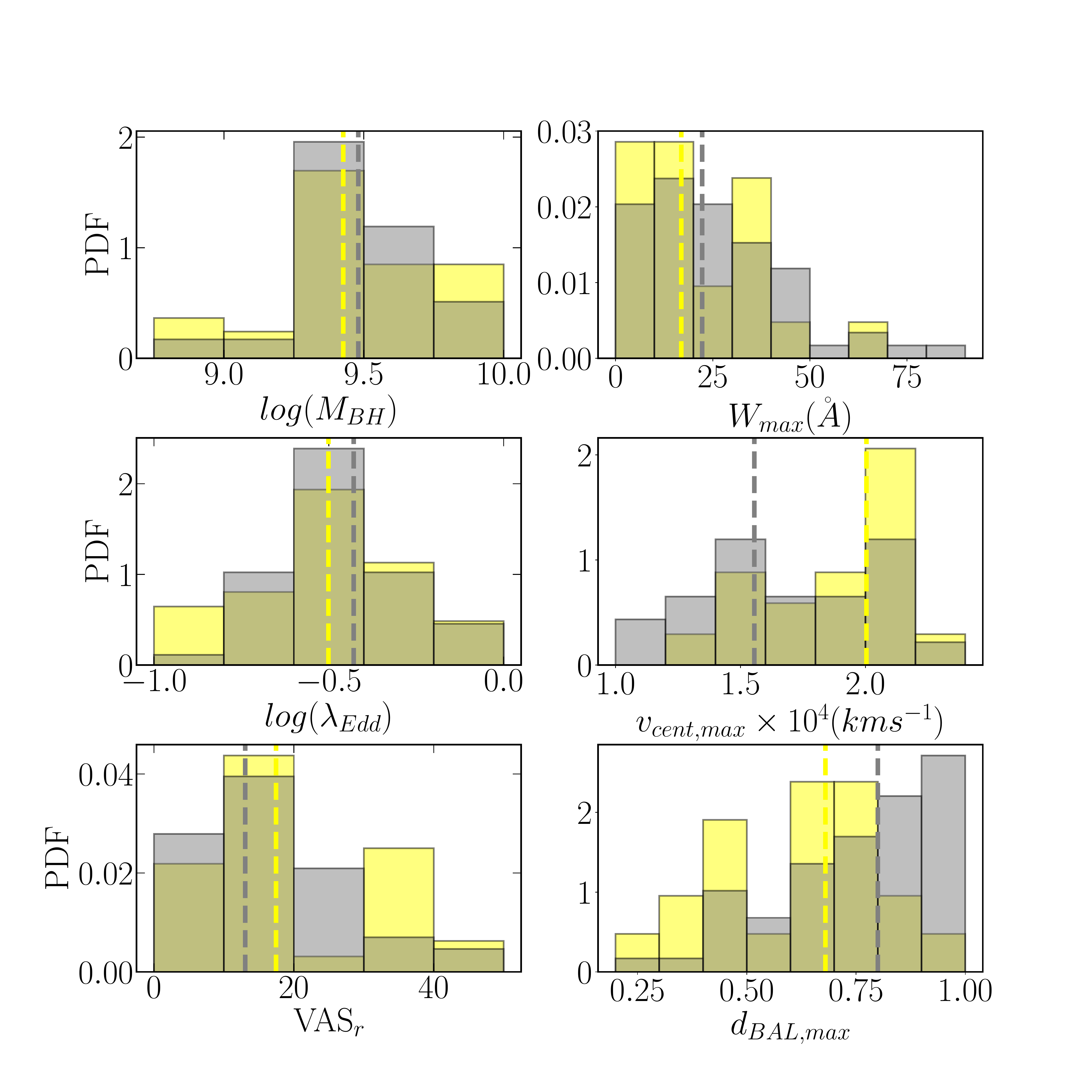}
    \caption{Distribution of quasar (left panels) and BAL (right panels) properties of ``highly variable" UFO BALs (shown in yellow) and the rest of the UFO sample (gray). 
    Vertical dashed lines indicate the median values for each histogram. 
    }
    \label{fig:high_var_sample_hist}
\end{figure}

As mentioned before, 24 BAL components in 21 UFO BAL quasars in our sample tend to be ``highly variable". 
In this section, we compare the properties of these ``highly variable" UFO BALs with 
those of the rest of the UFO BALs in our sample. In Fig.~\ref{fig:high_var_sample_hist} we compare 
the distributions of quasar properties (\mbh,  \redd\ and VAS$_{r}$ as introduced in  Section~\ref{sub:VAS})
and BAL properties ($W_{max}$, maximum of $v_{cent}$ and maximum of $d_{BAL}$)
for ``highly variable" BALs and the rest of our sample.  
The median value of each distribution is shown as a vertical (gray or yellow) 
dashed line in each panel. 


For quasar properties, we find that the \mbh\ and \redd\ distributions are similar for both sub-samples (p-value is $>$0.1 from the KS test) whereas for VAS$_{r}$ the p-value is 0.03 for the KS test. This indicates that occurrence of ``highly variable" BAL is not strongly coupled to \mbh\ or \redd.
However, there is an indication that the UFO BAL quasars showing ``highly variable" BALs tend to show slightly larger photometric variability compared to the rest of the UFO quasars.
This could either mean  the large equivalent width variabilities are driven by photoionization or disk instabilities that also results in photometric variability.
We come back to this discussion in more detail in section~\ref{Sec:photoionization}.
%
%

For BAL properties, the distributions of $W_{max}$ are identical as indicated by 
p-value $>$ 0.1 from the K-S test (see also Fig.~\ref{fig:high_var_sample_hist}).
However a larger fraction of ``highly variable" BALs have comparatively higher values of $v_{cent,max}$ and lower values of d$_{BAL,max}$.
K-S test p-values are 0.01 and 0.03 for  $v_{cent,max}$ and d$_{BAL,max}$ respectively.
In summary, we find that ``highly variable" BALs tend to have comparatively shallower absorption at higher velocities. 
A considerable fraction of UFO BALs with ``highly variable" components tend to show excess variability in their r-band light curves.

\subsection{Different classes of UFOs}
\label{subsec:class}

\begin{figure}
    \centering
    \includegraphics[viewport=90 10 2350 770, width=\textwidth,clip=true]{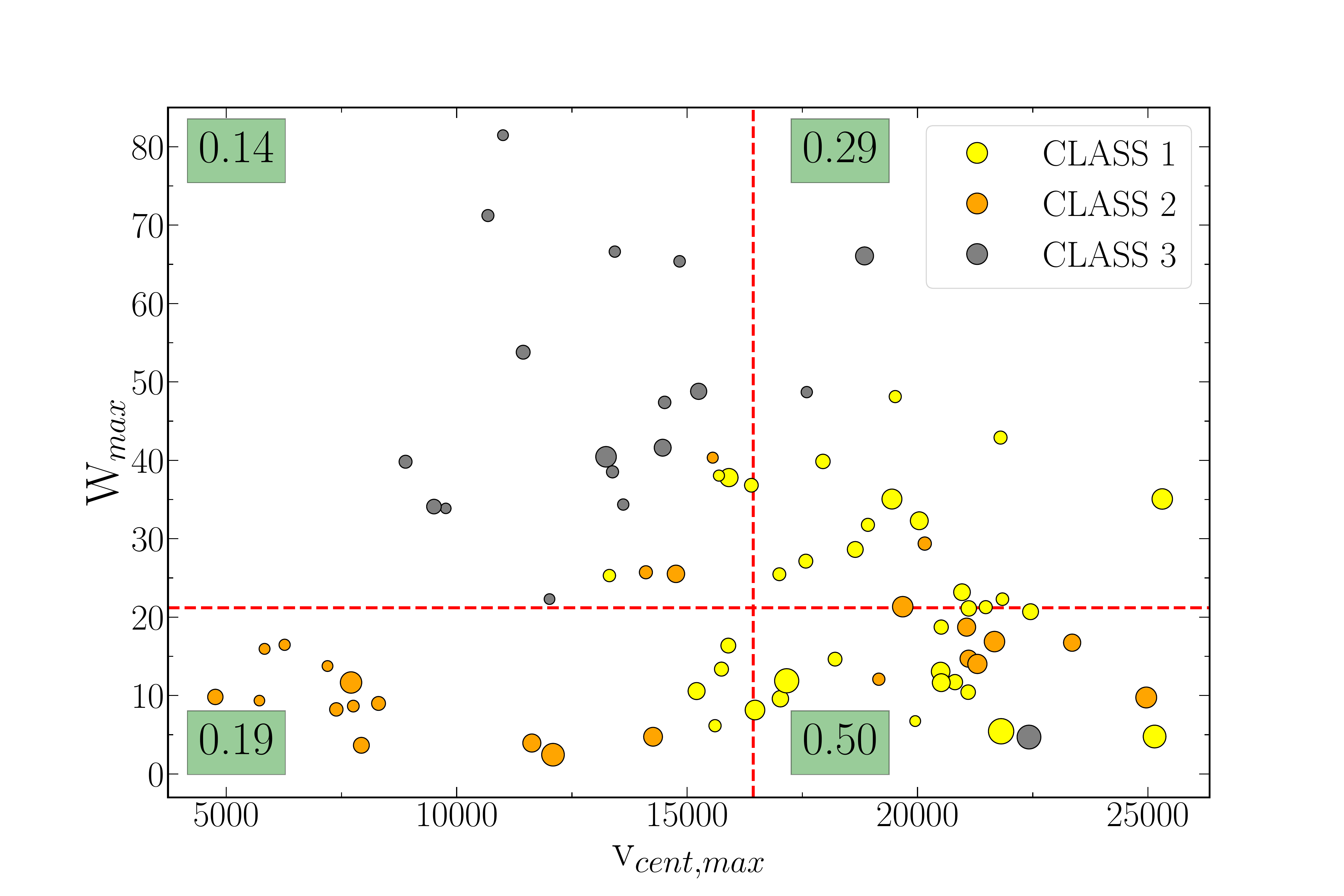}
    \caption{Maximum W ($W_{max}$) as a function of maximum $v_{cent}$ ($v_{cent,max}$) for all the UFO BALs in our sample. The class of each UFO BAL is indicated by the color as shown in the legend and the size of the circle scales with the maximum \afdw\ observed. The red dashed lines show the median along the respective axis and the fraction of sources with max(\afdw) $>$ 0.67 is indicated in green boxes for the corresponding regions of the plane.
    }
    \label{fig:wmax_vs_vmax}
\end{figure}


Here we investigate whether there is a relation between the  scatter in \afdw\ 
and the overall BAL profile. 
For this, we  classify quasars in our UFO sample into three classes based on the 
global absorption profile structure. Typical examples are shown 
in Fig.~\ref{fig:classification}. {\bf Class-1}: Includes sources with one or more 
\civ\ UFO BALs having $v_{\rm min}>8000$ \kms\ 
without any \civ\ broad absorption at lower velocities (top panel in Fig.~\ref{fig:classification}).
There are 33 sources in our sample that were classified as Class-1. In three of these cases (J0224-0528, J1054+0150 
and J1317+0100) there are two UFO BALcomponents with $v_{\rm min}>8000$ \kms.
In the remaining 30 sources there is only one \civ\ BAL complex. Basically, BALs in this 
class are well detached from the \civ\ emission line without any low-velocity absorption.
{\bf Class-2}: Includes only sources with multiple BAL troughs having at least one UFO BAL 
with $v_{\rm min}>8000$ \kms\ and one non-UFO BAL 
with $v_{\rm min}<8000$ \kms. In this case, the UFO BAL and non-UFO BAL components are distinguishable (middle panel of 
Fig.~\ref{fig:classification}). There are 13 objects in our sample that belongs to this class. In three cases (J0046+0104, 
J0242+0049 and J2352+0105) the low-velocity BAL complex has $v_{\rm max}<3000$ \kms and is not listed in Table~\ref{tab_sampledetails}. 
In two sources (J2310+0746 and J1322+0524) we have identified three distinct BAL complexes with one of them being a UFO BAL 
and the other two being present at lower velocities.
{\bf Class-3}: Contains sources with a single UFO BAL trough having $v_{\rm min}<8000$ \kms
(bottom panel of 
Fig.~\ref{fig:classification}). This class, by definition, has absorption spread over a wide range of velocities and absorption from distinct components are merged into a single BAL component when we follow the definition of \citet{Filiz2013}. There are 17 sources in our sample that are classified as Class-3.
Only one UFO BAL quasar, J0216+0115, could not be classified as the absorption profile does not fit with any of the definitions of the three classes mentioned above.
Column 8 in Table~\ref{tab_sampledetails} provides the classification for each quasar in our sample. 

Note, the classification scheme discussed here is motivated by the two-component (polar and equatorial) wind models discussed in \citet{Borguet2010}. In this model, absorption profiles consistent with that of Class-3 will have contributions from both fast-moving polar and slow-moving equatorial winds seen at low inclination angles. Detached profiles as those of Class-1 will mainly come from fast-moving polar components seen at high inclination angles to the disk plane. Thus the classification we adopt may reflect a classification based on the inclination angle with respect to the disk. While this gives a motivation for the above classification scheme, we are aware that a given absorption profile can be produced by models with a wide range of parameters.

The fraction of ``highly variable" sources 
are 0.30 (10/33), 0.54 (7/13), and 0.18 (3/17) for Class-1, Class-2 (considering only the UFO BAL components), and Class-3 respectively.  This indicates the possible connection between  profile shapes and variability of \civ\ absorption. 
To explore this further, we study how UFO BAL quasars belonging to different classes populate the $W_{max}$ vs. $v_{cent}$ plane (see Fig.~\ref{fig:wmax_vs_vmax}). 
The low-velocity absorption components of Class-2 populate mostly the bottom left part of the plane whereas the high-velocity components populate the bottom right part.
Class-3 populates mostly the upper left part of the plane as 
 expected because the BAL extends continuously to large velocities with $v_{cent}$ being typically in the lower side. Class-1 objects populate both the upper and lower quarters of the upper half of $v_{cent}$.
The size of each point in Fig.~\ref{fig:wmax_vs_vmax} is proportional to the amplitude of \afdw.
%
The fraction of UFO BALs showing ``highly variable" BALs are indicated in each quadrant in Fig.~\ref{fig:wmax_vs_vmax}. Nearly 50\% of the BALs in the high $v_{cent}$ and low $W_{max}$ show \afdw$>0.67$. 
This quadrant is mainly populated by Class-1 and high-velocity component of  Class-2 BALs in our sample. This implies that the relatively weaker and detached high-velocity BALs have a tendency to show more variability. On the other hand,
a lower fraction (i.e., 14\%) of UFO BALs seems to be ``highly variable" 
in the low $v_{cent}$ and high $W_{max}$ quadrant. 
This quadrant is predominantly occupied by Class-3 BALs. 
In the quadrant with low $W_{max}$ and low $v_{cent}$ we find only 19\% of the BALs are ``highly variable".  This quadrant is predominantly occupied by low-velocity components of Class-2 BALs in our sample. 
On the other hand, in the quadrant with high $v_{cent}$ and high $W_{max}$ nearly 29\% of BALs are ``highly variable".  {\it All this clearly imply that high \civ\ BAL variability is associated with both large $v_{cent}$ and  $W_{max}$ and
$W_{max}$ alone is not a primary driver of the large variability.}

\begin{figure}
    \centering
    \includegraphics[viewport=35 45 2400 1170, width=\textwidth,clip=true]{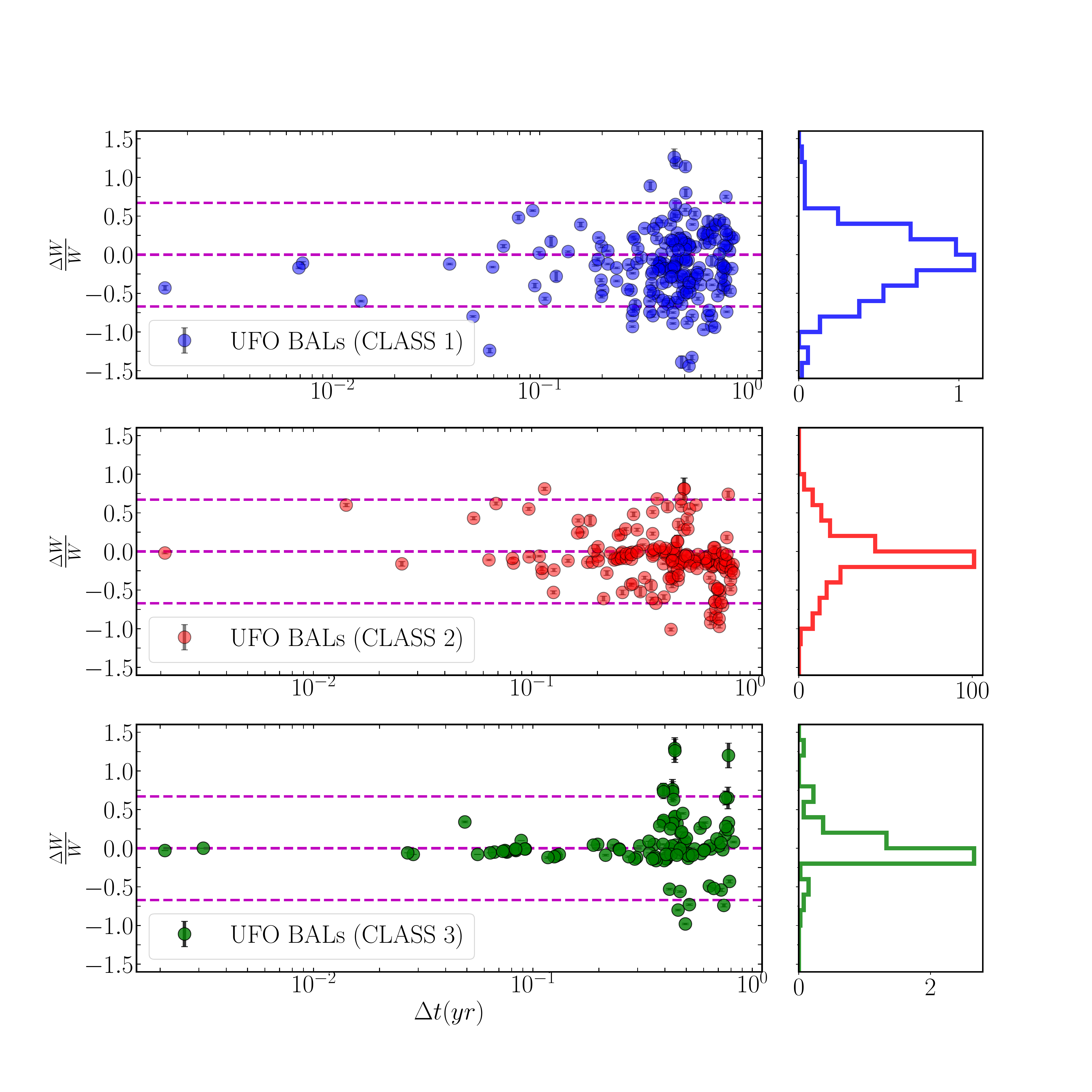}
    \caption{This figure shows the \fdw\ vs time for UFOs in Class 1 (top panel), Class 2 (middle panel), and Class 3 (bottom panel) sources. 
   The panels in the right side show the histogram distribution of \fdw. Horizontal dashed lines mark \fdw\ = 0 and $\pm$0.67. It is evident that Class-1 and Class-2 UFO BALs show larger scatter compared to that of Class-3.}
    \label{fig:w_class_I_and_II}
\end{figure}

Next, we study the \fdw\ distribution of UFO BALs in different classes as 
a function of time (see Fig.~\ref{fig:w_class_I_and_II}).
The histograms in the right panels of Fig.~\ref{fig:w_class_I_and_II} 
show the distributions of \fdw\ in each case.
It is apparent that the scatter in the distribution is decreasing from
Class-1 to Class-2 followed by Class-3. This result may be affected by the different number
of sources in each class.
To confirm the trend seen in Fig.~\ref{fig:w_class_I_and_II}, we measure IQR for different classes using method-I (see Section~\ref{sub:timedependence})
%
 for the four time bins (i.e., 0.1-0.5, 0.5-2.0, 2.0-3.5, and 3.5-7.5 yrs). IQR values obtained for Class-1 are 0.27$\pm$0.03, 0.45$\pm$0.10, 0.40$\pm$0.05, and 0.60$\pm$0.08 respectively. For Class-2, IQR values for the UFO BALs (i.e only for the high-velocity components) are 0.29$\pm$0.12, 0.46$\pm$0.15, 0.49$\pm$0.11, and 0.43$\pm$0.13 respectively. The IQR values for Class-1 and Class-2 objects are consistent within errors.
 However, for Class-3, IQR values are 0.06$\pm$0.01, 0.07$\pm$0.02, 0.22$\pm$0.06, and 
 0.61 $\pm$ 0.01 respectively. 
 Clearly, Class-3 objects show consistently lower IQR values (compared to the other two classes) except for the longest-time  bin of 3.5-7.5 years.

\begin{figure}
    \centering
    \includegraphics[viewport=45 25 2400 1150, width=\textwidth,clip=true]{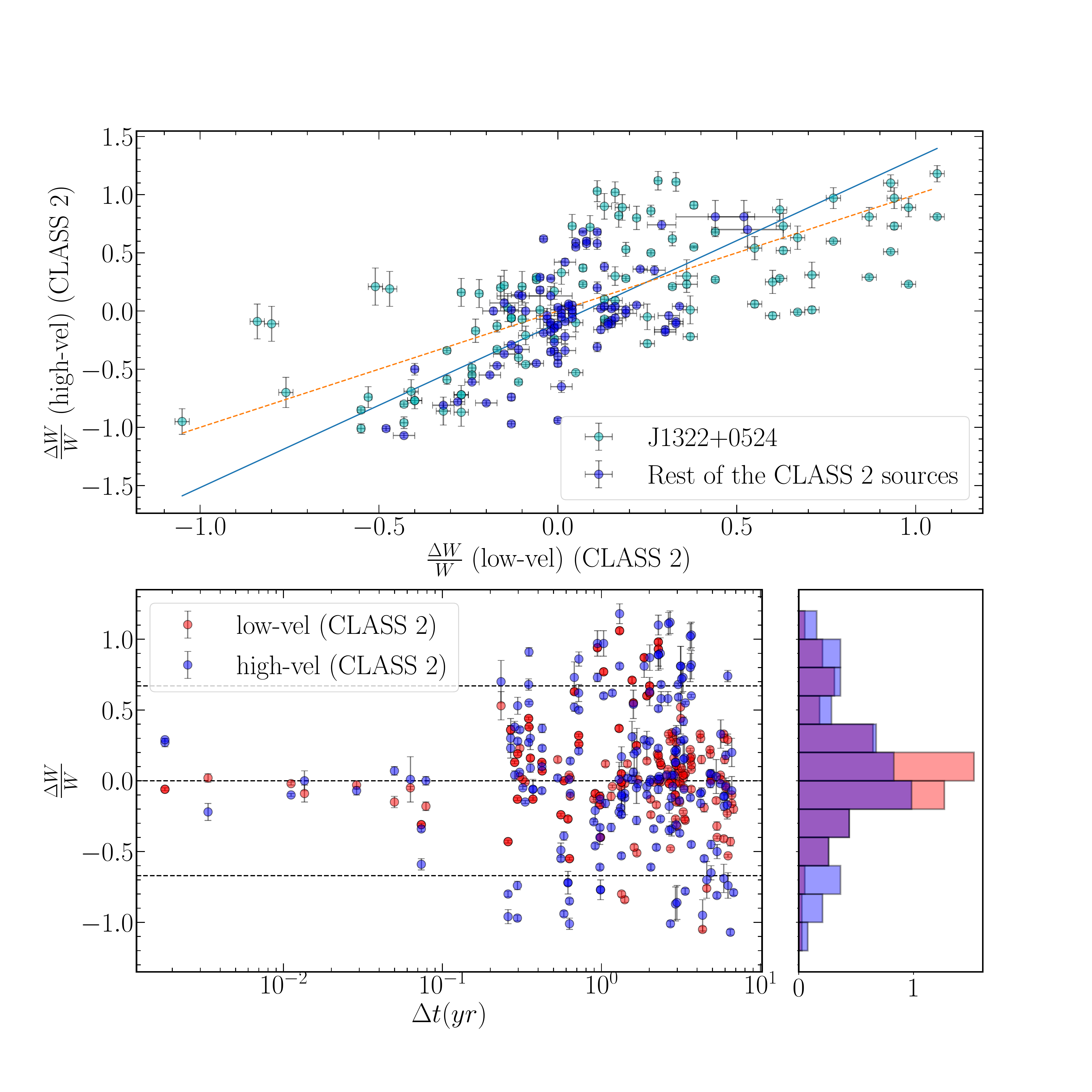}
    \caption{{\sl Top panel}: Comparison of \fdw\ of low-velocity non-UFO BAL troughs  to that of UFO BALs in Class-2 sources. The points contributed by J1322+0524 (cyan) are shown separately. The blue line is the best-fitted linear regression line. The equality line is shown by the orange dotted line. 
    {\sl Bottom panel}: The \fdw\ distribution of non-UFO (red) and UFO BALs (blue) in Class-2 sources as a function of time. Horizontal dashed lines mark \fdw\ = 0 and $\pm$0.67.
    }
    \label{fig:class2_low_high_vel_dw}
\end{figure}

\begin{figure}
    \centering
    \includegraphics[viewport=45 75 2400 1150, width=\textwidth,clip=true]{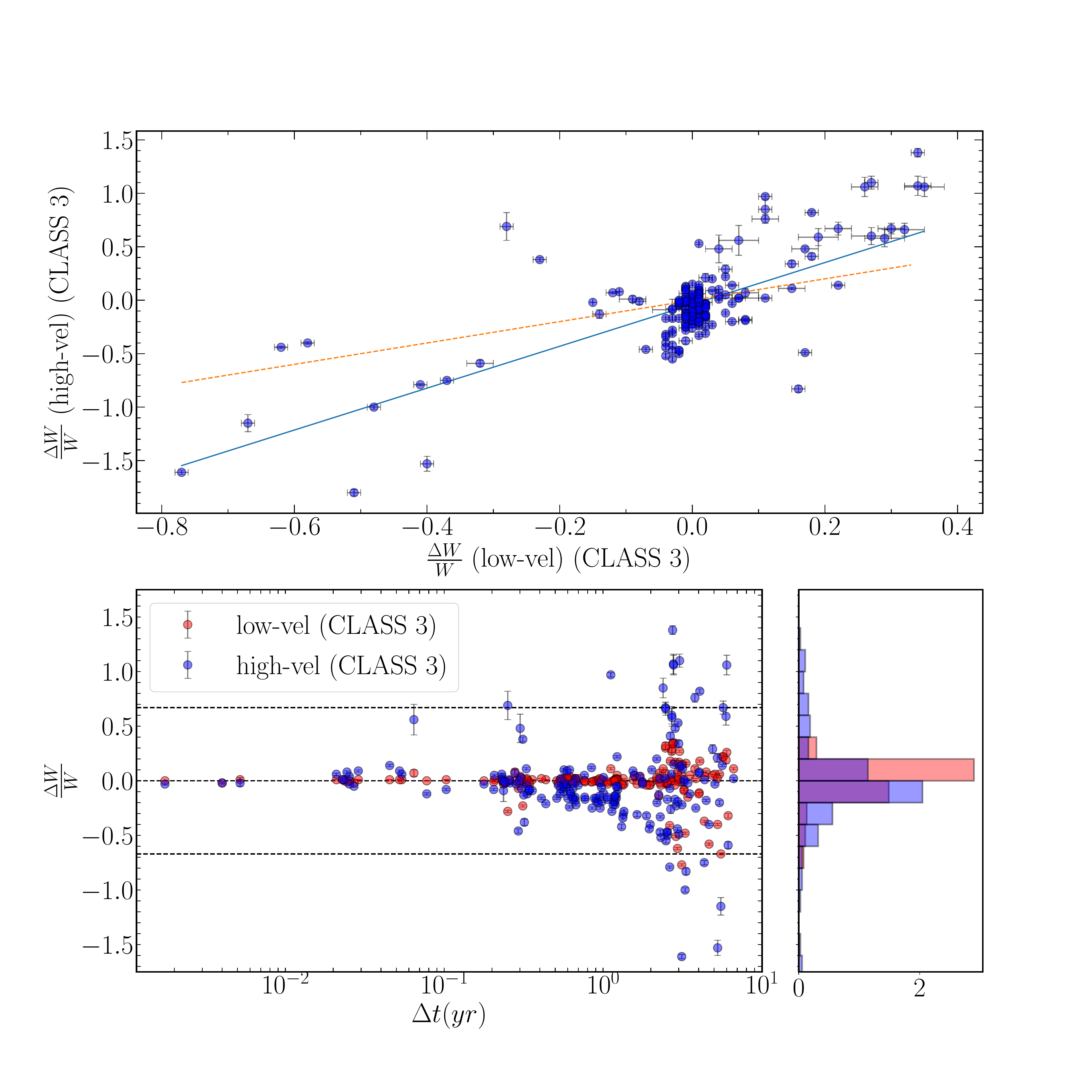}
    \caption{{\sl Top panel}: \fdw\ of low-velocity  vs. \fdw\ of high-velocity regions in quasars belonging to Class-3. The blue line shows the best-fitted linear regression line. The equality line is indicated by the orange dotted line. 
    {\sl Bottom panel}: \fdw\ of low- and high-velocity regions of BAL troughs belonging to Class-3 as a function of time. Horizontal dashed lines mark \fdw\ = 0 and $\pm$0.67.
    }
    \label{fig:class3_low_high_vel_dw}
\end{figure}


Next, we probe the possible correlation between the \fdw\ distributions of the 
distinct low- and high-velocity BAL troughs seen in Class-2 sources 
(i.e., UFO and non-UFO BALs).
%
{From the top panel of Fig~\ref{fig:class2_low_high_vel_dw}, it is clear that the variations in  high- and low-velocity BAL troughs are correlated. Also, using the linear regression, we find a slope of 0.95 considering all the class 2 sources whereas if we exclude BAL components towards J1322+0524 that contribute an appreciable number of points to this plot, the slope turns out to be 1.45. This indicates that the UFO BALs show more variability compared to non-UFO BALs in general, but in the case of J1322+0524, the UFO and non-UFO BALs vary with similar strength leading to a slope close to 1.}
Also, from the bottom panel of Fig~\ref{fig:class2_low_high_vel_dw}, we see that 
the \fdw\ distributions of high- and low-velocity BALs are considerably different (p-value = 0.04)  with high-velocity BALs showing more scatter than the low- velocity
ones. This is consistent with what was shown before: high-velocity BALs are more variable.

To extend this analysis to Class-3 objects, we divide the continuous absorption profile into two regions above and below the velocity mid point, $v_{mid} = \frac{v_{min} + v_{max}}{2}$. Even in this case (see Fig.~\ref{fig:class3_low_high_vel_dw}), we find a correlated variability between high- and low-velocity regions as in the case of Class-2 sources, but with a much steeper slope (1.95) indicating that high-velocity regions are much more variable compared to low-velocity ones. We should remember that this is due in part to the fact that low-velocity regions are more saturated.

\subsection{\civ\ emission lines and BAL connection }



As discussed in Section~\ref{sec:civprop}, we measured the equivalent width ($W_{BEL}$) and the blueshift (BS$_{BEL}$) of the \civ\ emission line in a 
sub-sample of 41 UFO BALs with $v_{min}$ larger than 6000 \kms\ so that the \civ\ emission profile is not affected by the absorption.
By definition, all 33 objects in Class-1 are part of this sample. In the remaining 8 sources, 2 and 6 sources belong to Class-2 and -3 respectively. 
As seen in Section~\ref{sec:civprop} and Fig.~\ref{fig:blueshift_comp}, BS$_{BEL}$ is much larger for UFO BAL quasars compared
to non-BAL quasars.
The measured large values of BS$_{BEL}$ in BAL quasars are assumed to be a
consequence of outflowing gas in the BLR.

We searched for correlations by calculating Spearman's correlation coefficient and corresponding p-values (see Table~\ref{tab_bal_bel_prop}) between $BS_{\rm BEL}$
or max($\Delta W_{\rm BEL}/W_{\rm BEL}$) with 
the BAL parameters $W_{max}$, $v_{cent,max}$, $v_{max}$, d$_{BAL,max}$ and $\Delta v_{max}$.
From Table~\ref{tab_bal_bel_prop}, we observe a 
strong correlation between BS$_{BEL}$ and  $W_{max}$, $v_{max}$ and d$_{BAL,max}$ whereas a moderate correlation is reported for $\Delta v_{max}$.
 These strong correlations  
are consistent with recent studies including \citet{rankine2020}
who found that BAL quasars with the highest $W_{BEL}$ and lowest BS$_{BEL}$ tend to show weaker, narrower, and comparatively low-velocity BAL troughs and vice-versa.
\citet{Hidalgo2022} also reported that the maximum velocity of  EHVO increases with blueshift as seen in our sample. This seems to suggest that the high velocity BAL phenomenon is
intimately related to the blueshift of the \civ\ emission line.

\begin{table*}
\begin{threeparttable}
    \centering
\caption{Dependence of BAL properties on properties of \civ\ broad emission line}
 \begin{tabular}{cccccc}
  \hline
   Sample & \multicolumn{5}{c}{Spearmann coefficient, p value}  \\ 
          & $W_{max}$ & $v_{cent,max}$ & $v_{max}$ & d$_{BAL,max}$ & $\Delta v_{max}$ \\
  \hline\hline
  $BS_\text{BEL}$ & 0.340, 0.021 & 0.093,0.538 & 0.351, 0.016 & 0.336, 0.022 & 0.272, 0.068 \\
  max($\frac{\Delta W_{BEL}}{W_{BEL}}$) & -0.447, 0.005 & 0.041, 0.805 & -0.151, 0.364 & -0.392, 0.015 & -0.417, 0.009\\
  \hline
 \end{tabular}

\label{tab_bal_bel_prop}
\end{threeparttable}
\end{table*}

For each pair of epochs, we measured the fractional equivalent width changes in \civ\ BAL and BEL.
The Spearman rank correlation test (with a coefficient  of $-$0.05 and p-value of 0.47) confirms the lack of any correlation between the two quantities. Note that for a couple of individual cases with enough observations, we do find a possible correlation between the two quantities \citep[see][]{Aromal2021, aromal2022}. The lack of a similar correlation for the full sample is probably related to the insufficient time sampling that fails to detect the delayed response from the BEL regions. Establishing such a correlation is important to confirm the  ionization-induced BAL variations.

Next, we divide our sample into two around the median BS$_{BEL}$. Then compared the mean value of \afdw\ in the four time-bins considered in this study. We find a significant difference only for the third time-bin, i.e., 2.0-3.5 yrs.  For this time bin, we looked at the \afdw\ distribution as a function of BS$_{BEL}$ as shown in Fig~\ref{fig:delw_vs_blueshift_timebin3}.
It is clear from the figure that the sources with less blueshift tend to show larger scatter in the \afdw\ distribution. The Spearman rank correlation test suggests a possible anti-correlation with a coefficient of $-0.25$ and a p-value of $<$0.001. This may be explained as the result of a moderate correlation of $W_{max}$ with blueshift from Table~\ref{tab_bal_bel_prop} combined with the fact that BALs with low $W$ show more variability. It is difficult to confirm these trends given the limited size of the sample, but future studies with larger sample sizes and better time sampling can help in reaching stronger conclusions.

\begin{figure}
    \centering
    \includegraphics[viewport=60 0 2400 650, width=\textwidth,clip=true]{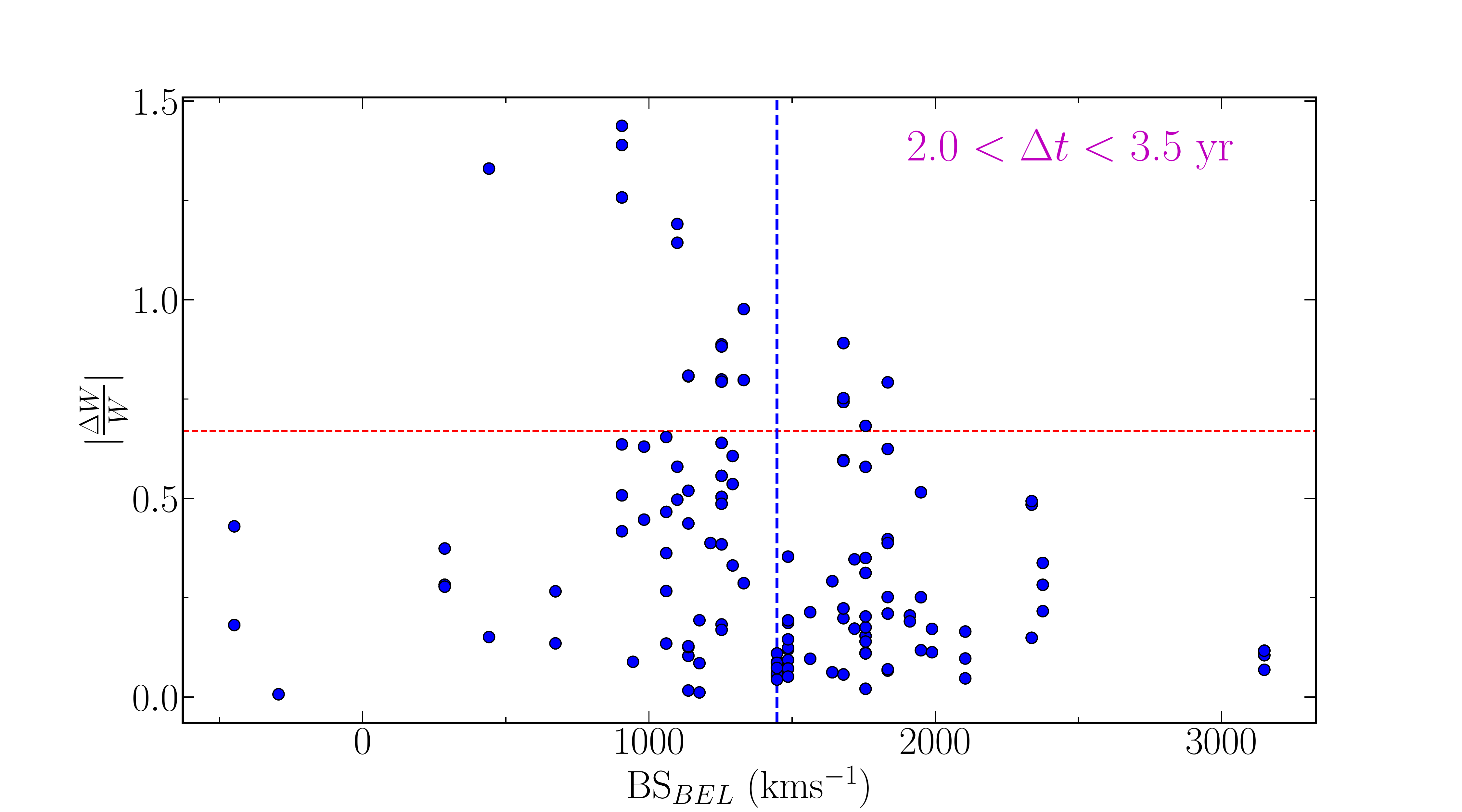}
    \caption{\fdw\ is shown for each UFO BAL source as a function of blueshift for 
    the time bin 2.0$< \Delta t <$3.5 yrs. The vertical dashed line indicates the median value of the quantity in the abscissa and the horizontal line indicates \afdw = 0.68. }
    \label{fig:delw_vs_blueshift_timebin3}
\end{figure}

\subsection{Are continuum variations related to BAL variability ?}
\label{Sec:photoionization}

\begin{figure}
    \centering
    \includegraphics[viewport=-5 10 2850 980, width=\textwidth,clip=true]{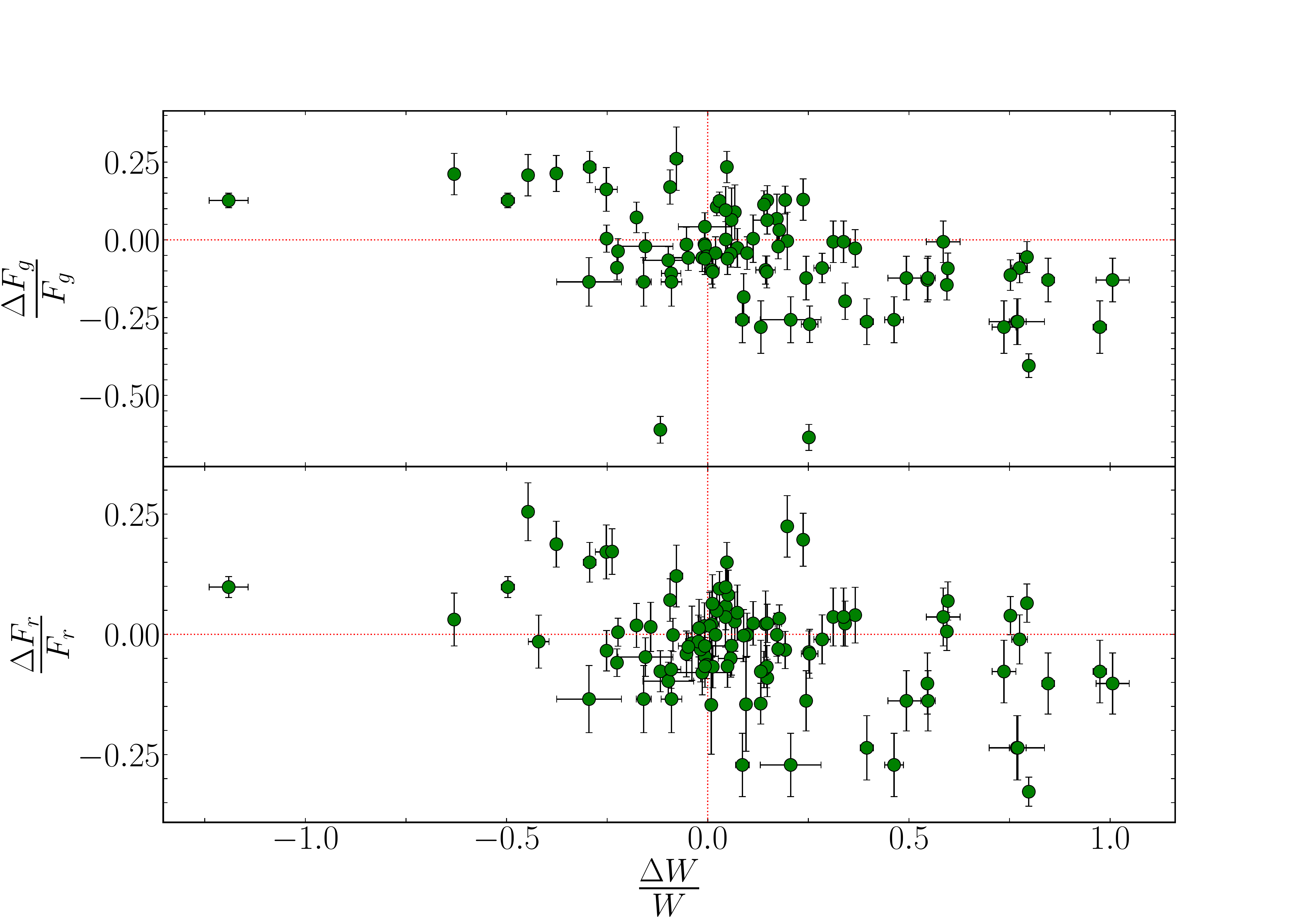}
    \caption{In this figure, we compare the fractional variations in total g- and r- band flux to that of the BAL EW in the top and bottom panels respectively.}
    \label{fig:delw_vs_dmjd_spec_bs}
\end{figure}

\begin{figure}
    \centering
    \includegraphics[viewport=60 0 2400 700, width=\textwidth,clip=true]{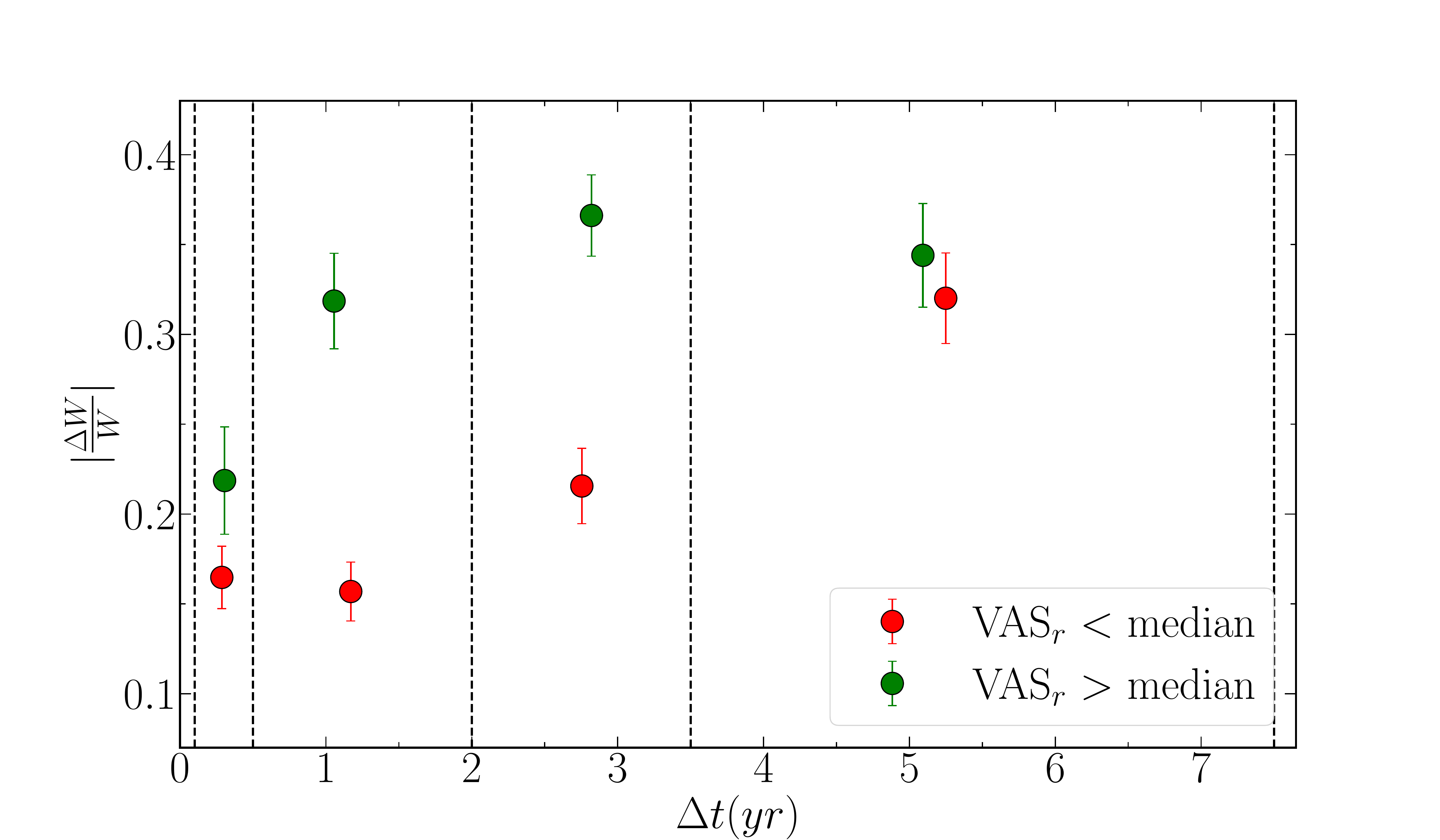}
    \caption{We plot the mean and standard deviation of the average \fdw\ distribution at the four different time-scales, i.e. short (0-0.5 yr), intermediate (0.5-2 yr) and two long ( 2-3.5 yr and >3.5 yr) time-scales,  for the two sub-samples having estimated VAS of the r-band light curve below and above its median.
    }
    \label{fig:delw_vs_dmjd_lc_vas}
\end{figure}

As mentioned before, we have obtained publicly available light curves to see the trends in continuum variations during our spectroscopic monitoring period. From Section~\ref{sub:VAS}, it is clear that the variability amplitude strength (VAS) follows the same distribution irrespective of the presence of BALs. This implies that the continuum variations of our UFO BAL quasars are not very different from the general quasar population. However, we also notice that the distribution of VAS in the r-band is different for ``highly variable" UFO BALs and the rest of the UFO BALs in our sample (see Fig.~\ref{fig:high_var_sample_hist}).
%

Here, we explore the possible connection between the continuum variability and the variability of \civ\ equivalent width in our UFO BAL sample.
We consider photometric epochs in g- and r-bands which are within 10 days in the observer's frame (roughly 3 days in the rest-frame) to our spectroscopic epochs whenever available and convert them to flux units. Using these, we compare the fractional variations in total g- and r- band flux (F$_{g}$ and F$_{r}$ respectively) to that of the rest equivalent width of the \civ\ BAL. As shown in Fig~\ref{fig:delw_vs_dmjd_spec_bs}, we see moderate anti-correlation with high significance between $\frac{\Delta F_{g }}{F_{g}}$ and \fdw\ with spearmann coefficient, r = $-$0.48, p-value = 3.50 $\times$ 10$^{-6}$ and weak anti-correlation with lesser significance between $\frac{\Delta F_{r}}{F_{r}}$ and \fdw\ with r = $-$0.25, p-value = 1.34 $\times$ 10$^{-2}$. 
In order to demonstrate that this effect is not dominated by the sources having a large number of epochs, we remove sources having more than 8 epochs of spectroscopic observations and carry out the same analysis to find even stronger anti-correlation signatures with r = $-$0.62, p-value = 1.27 $\times$ 10$^{-5}$ for g-band and r = $-$0.34, p-value = 1.92 $\times$ 10$^{-2}$ for r-band. While the anti-correlation is statistically significant, we do see the spread in the fractional variation in the continuum flux is smaller than that of \civ\ rest equivalent width.

The analysis indicates that the continuum flux variations may be responsible for the observed BAL variability where the W of the \civ\  BAL decreases as the continuum increases. This is in agreement with several other studies. \citet{Lu2018_cont_var} found the same trend with similar r-coefficient, r=-0.43 with p-value < 1e-44 between the fractional variations of the absorption \civ\ equivalent width and that of the monochromatic continuum luminosity at 1450 \AA\ calculated from the spectra. \citet{Misra2019} showed that the dimming in the continuum is associated with the appearance of new BALs.
\citet{Horiuchi2020} also looked at a few of the SDSS RM sources and found a possible presence of correlation between the BAL variability and photometric variability using iPTF and PanSTARRS surveys.

To ascertain this tendency, we also divided the sources into two sub-samples based on 
the VAS values in r-band (r-band is preferred over g-band since most BALs coincide with g-band region whereas r-band is little affected by the same) and looked at the average of \afdw\ in different time bins similarly to the analysis 
performed in Section~\ref{sub:qsoprop}. The results are shown in Fig.~\ref{fig:delw_vs_dmjd_lc_vas}. It is clear that except for the 4th time bin, sources with large photometric variability also show consistently higher BAL variability up to 3.5 years in rest-frame time scales. 

\citet{aromal2022} while analyzing the multiple epoch spectroscopic observations of three BAL components in J1322+0524 found that the amplitude of optical continuum flux variations is much smaller than what is needed to produce the observed equivalent width variations. Based on this and a large scatter in $W$ found for a given continuum flux they argued that \civ\ ionizing continuum and optical continuum need not vary in a correlated manner and the amplitude of ionizing continuum variation has to be much higher than what we see in the optical light curves. From Fig.~\ref{fig:delw_vs_dmjd_spec_bs},  we see scatter in the equivalent width fractional variations are much larger than that seen in the continuum. This implies the  conclusions drawn in the case of J1322+0524 may still be valid for most of the UFO BALs in our sample. 
Therefore, it will be interesting to probe the rest frame FUV variability of our UFO BAL quasars.

}

\section{Results and discussions}
\label{sec:discussions}

In this work we have presented an analysis of the \civ\ absorption variability of 80 distinct BAL components 
observed in the spectra of 64 UFO BAL quasars.
We used spectra from SDSS together with our own data from SALT. Our monitoring time-scale spans between a few months to $\sim$ 7 yrs in the quasar rest frame. Here, we mainly focus on the variability of the \civ\ rest equivalent width. We will present details of pixel-based optical depth analysis and a detailed study of interesting sources \citep[as in][]{Aromal2021,aromal2022} in future papers. The main results from this study are :

\vskip 0.1 in
\noindent {\bf 1) BAL variability fraction:} We find the \civ\ absorption in  $\sim95\%$ of  UFO BALs vary (by $\ge3\sigma$) at least once during our  monitoring. Roughly 33\% of quasars in our sample show ``highly variable" (i.e., \afdw$>$0.67) BAL components. Also, independent of the source
and time scale considered, 
$\sim80\%$ of epoch pairs show significant variations.
These percentages are higher than what is reported for the general BAL population in the literature.
For example, when we consider only the pair of observations separated by the shortest time-scale for each object, $\sim$70\% of UFO BAL QSOs and UFO BAL components show significant variability which is a higher occurrence than that found by \citet{Filiz2013}. The shortest time intervals probed here being shorter than that of \citet{Filiz2013} one may have expected less variability in our sample. All this indicates that UFO BALs are more variable compared to the normal BAL population. 

\vskip 0.1in
\noindent {\bf 2) Time dependence of BAL variability}: 
We study the variability of the \civ\ rest equivalent width in four different time-bins (i.e., 0.1-0.5, 0.5-2.0, 2.0-3.5 and 3.5-7.5 yrs) in the quasar rest frame. The inter quartile range (IQR) of \fdw\ increases with increasing time scale which means that the variations become larger with increasing time scale.
This result is not influenced by the different time-sampling achieved for different sources and is not related to any particular
behavior of systems with low ($W<5$\AA) or high ($W>55$\AA) \civ\ rest equivalent widths. 
We also note that the fraction of ``highly variable" BALs increases with increasing time scale. Such ``highly variable" \civ\ BALs dominate the large variability seen at long time scales. By comparing our data (with appropriate time sampling) with the \fdw vs. $t$ relationship of \citet{Filiz2013} we show our UFO BALs are more variable compared to the 
overall BAL population. 

A larger fraction of UFO BALs in our sample show a
negative \fdw in the largest time-bin (3.5-7.5 yrs). 
In addition, using the time evolution of \civ\ rest equivalent width in 21 BAL components that show large variability over a time-scale $>$ 2 yrs, we find that a decrease in rest equivalent width by a factor of two occurs over a longer time-scale (i.e., a median of $\sim$3.5 yrs) compared to an increase by the same amount (i.e., a median of $\sim$2.8 yrs). 
This is consistent with the finding that the average time scales are higher for BAL
disappearance events compared to the emergence events \citep[see,][]{mcgraw2017,Misra2019}.
Confirming our results for the full sample through our ongoing monitoring programme will provide a possible link between the large equivalent width variability and extreme events like emergence/disappearance of BALs.


\vskip 0.1in
\noindent {\bf 3) Dependence of BAL variability on quasar properties : } We find no significant dependence of BAL variability on quasar properties like \mbh, \lbol\ and \redd\ even though a moderate correlation is observed for \lbol\ on time scales less than 2 yrs. 
This is consistent with what has been found in the overall population of \civ\ BALs \citep[see for example,][]{Filiz2013}. 
We do notice that the properties of QSOs in our sample do not probe the same range as what we see in the general population of quasars. This may have limited our ability to detect any weak trend. While some quasars in our sample show large \civ\ BEL variations insufficient time-sampling related to the expected reverberation time-scales prevents us from correlating the \civ\ BEL and UFO BAL variabilities. In cases where such detailed investigation is possible we do see large BEL variations in quasars that show large UFO BAL variations \citep[see]
[]{Aromal2021,aromal2022}. Spectroscopic monitoring, at shorter time intervals, of a sub-sample of quasars in our sample that show significant \civ\ BEL variations would be very useful.

We find that on an average the optical photometric variability of quasars in our sample are not statistically distinguishable from that of our control sample. However, we do find an anti-correlation between the optical continuum flux variations  (in both g- and r-band) and variations in \civ\ BAL equivalent width. In addition, the scatter in \afdw\ is more for objects showing VAS value larger than the median of our sample compared to the rest of the objects. All these are consistent with the absorption variability being somehow linked to the continuum flux variations. However, the spread in the fractional flux variations is much smaller than that of $W$. In a simple photo-ionization scenario this would imply much larger amplitude for the variability of the \civ\ ionizing flux compared to that in the rest frame NUV range probed by the optical variability. It will be interesting to check the relationship between the variability NUV flux and \civ\ ionizing flux using direct measurements

\vskip 0.1in
\noindent {\bf 4) Dependence of BAL variability on BAL properties :} 
By studying the variability nature of different sub-samples based on the \civ\ absorption properties, 
we find that weak, high-velocity, shallow, and low-width BALs tend to show more variability than others. 
Similarly the ``highly variable" BAL components tend to be shallow having large velocities and narrow widths.
We find that detached \civ\ BAL components with large ejection velocities show larger equivalent width variations. 
In general \civ\ equivalent width variations are by and large correlated across the velocity profile. However, the amplitude of \fdw\ is larger in the case of high velocity components compared to their low-velocity counterparts. Based on this we suggest that both low-equivalent width and high-velocity are equally important to determine the strength of variability in a BAL.

\vskip 0.1in
\noindent {\bf 5) Reconciling with physical models? : } 
\citet{proga2000, Proga2004} use axisymmetric, time-dependent hydrodynamical simulations of radiation-driven disk winds in AGN to conclude that radiative line driving is efficient enough to accelerate disk winds to high velocities as seen in BALs. 
%
However the variable \civ\ absorption produced by such 
simulated disk winds depends on a complex velocity field 
and covering factor \citep{proga2012}.
%
We note that the  simulated \civ\ absorption profiles in their study show high variability, especially at high velocities. 
They attribute this excess of variability at large velocities to the emergence of very fast mass ejections from relatively large distances, where the gas is well shielded from X-ray radiation. 

The best way to constrain such models is to compare the gas density, covering factor and distance of the absorbing cloud from the continuum source in the model with the constrains coming from observation. For this we need high resolution spectroscopic observations of our objects covering a larger rest wavelength range \citep[][]{Srianand2000A}.
Alternatively, it will be interesting to ask whether observed properties like  (i) the \fdw\ vs. $t$ relationship, (ii) the time-scale of decreasing equivalent width being higher than that of increasing equivalent width, (iii) the lack of strong correlations with quasar properties,
(iv) the connection between BEL blueshift and 
\civ\ absorption line variability, (v) the observation of large equivalent variability when there is no large optical continuum variations  and (vi) the importance of low equivalent width and high velocity to produce high amplitude variation etc., can be reproduced by hydrodynamical simulations.
Additional constrains for models can be obtained from pixed optical depth analysis. We will be presenting results of such an analysis for our sample in our upcoming paper.

\section*{Acknowledgements}
We thank  Aseem Paranjape and K. Subramanian for useful discussions. PA thanks Labanya K Guha for helpful discussions on several python programming techniques used in this paper.
PPJ thanks Camille No\^us (Laboratoire Cogitamus) for 
inappreciable and often unnoticed discussions, advice and support. PPJ
is partly supported by the Agence Nationale de la Recherche under
contract ???.

Funding for SDSS-III has been provided by the Alfred P. Sloan Foundation, the Participating Institutions, the National Science Foundation, and the U.S. Department of Energy Office of Science. The SDSS-III web site is http://www.sdss3.org/.

SDSS-III is managed by the Astrophysical Research Consortium for the Participating Institutions of the SDSS-III Collaboration including the University of Arizona, the Brazilian Participation Group, Brookhaven National Laboratory, Carnegie Mellon University, University of Florida, the French Participation Group, the German Participation Group, Harvard University, the Instituto de Astrofisica de Canarias, the Michigan State/Notre Dame/JINA Participation Group, Johns Hopkins University, Lawrence Berkeley National Laboratory, Max Planck Institute for Astrophysics, Max Planck Institute for Extraterrestrial Physics, New Mexico State University, New York University, Ohio State University, Pennsylvania State University, University of Portsmouth, Princeton University, the Spanish Participation Group, University of Tokyo, University of Utah, Vanderbilt University, University of Virginia, University of Washington, and Yale University. 


\section*{Data Availability}
Data used in this work are obtained using SALT. Raw data will become available for public use 1.5 years after the observing date at https://ssda.saao.ac.za/.



\bibliographystyle{mnras}
\bibliography{mybib_bal} 



\appendix

\section{Estimation of quasar properties}

In this section, we explain how we estimate different quasar properties  and compare the same to \citet{shen2011}. We obtain $M_{\rm BH}$ using the FWHM 
of the \civ\ emission  measured from the spectrum using the empirical mass-scaling relationship given by \citet{vestergaard} and the bolometric luminosity using bolometric correction to the monochromatic luminosity at 1350 \AA\ derived from the composite SED from \citet{richards2006}. Note that for these measurements, we used the SDSS spectrum with the highest SNR available for each object.

Now, we compare our estimated values with that of \citet{shen2011} for both \mbh\ and \lbol\ for 38 sources in our UFO sample and 20 non-BAL quasars from our control sample as shown in Fig A1. It is clear that \citet{shen2011} tends to over-predict \mbh\ and under-predict \lbol\ for BAL quasars compared to their non-BAL counterparts as shown by the shift in respective distributions. We find that this may be due to
the continuum fits used in \citet{shen2011} which uses conventional line-free regions for the fitting that may still include BAL regions, leading to incorrect estimation of quasar properties. This mostly lead to overestimation \civ\ BEL FWHM and underestimation of L$_{1350 \AA}$ which leads to similar probelms in the calculation of \mbh\ and \lbol\ respectively. Hence, we did a detailed visual inspection to select line-free regions for each source in our sample and used PyQSOFit to calculate the quasar properties mentioned above.

\begin{figure}
    \centering
    \includegraphics[viewport=50 40 1850 900, width=\textwidth,clip=true]{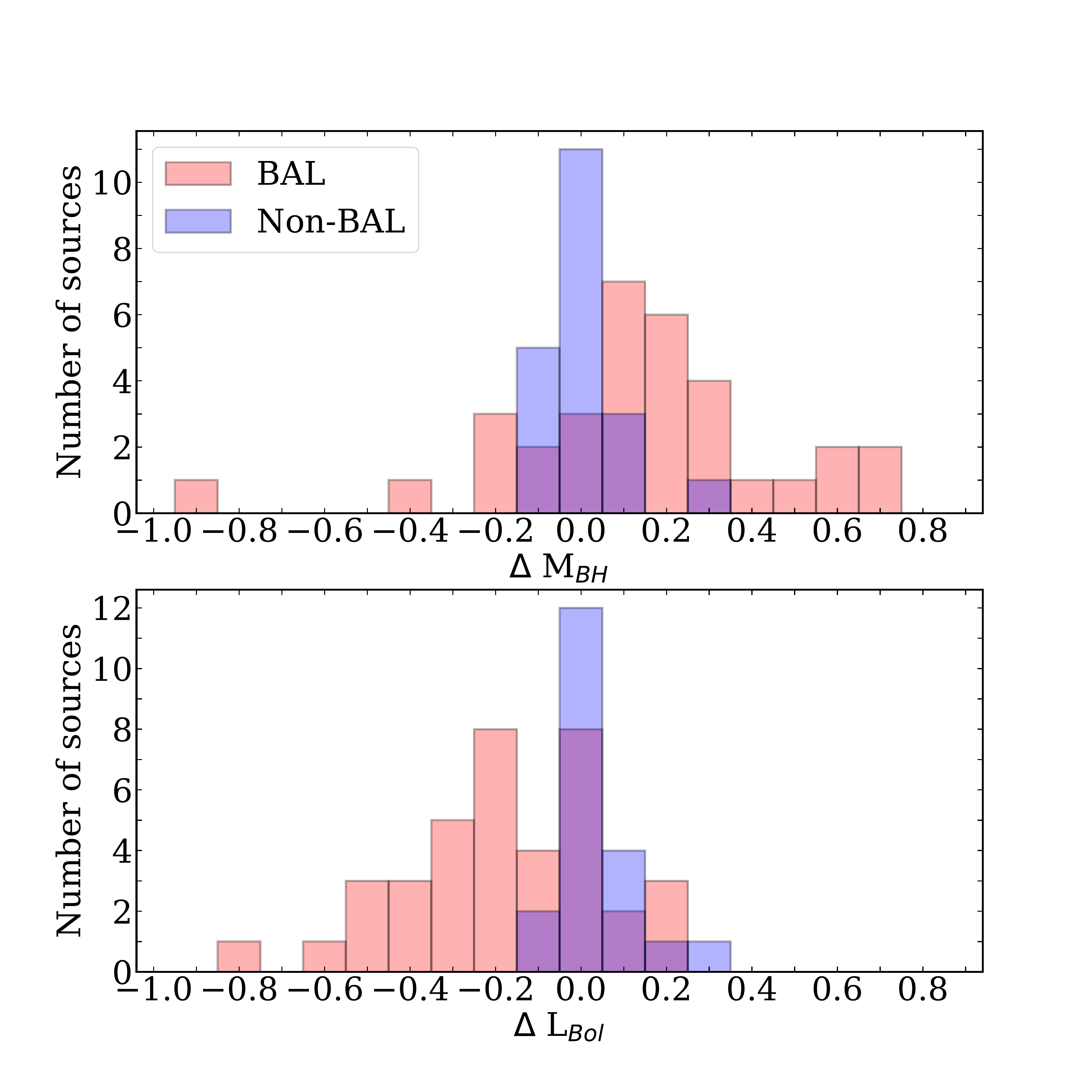}
    \caption{ In this figure, we show the distribution of differences in the estimated values of \mbh ($\Delta M_{BH}$) and \lbol\ ($\Delta L_{Bol}$) between \citet{shen2011} and this study for 38 UFO BAL sources and 20 non-BAL sources from the control sample.
    }
    \label{fig:quasar_prop_comp}
\end{figure}

\section{UFO BAL sample}
\label{sec:obs}
\begin{table*}

\begin{threeparttable}
    \centering
\caption{Log of sources, observations, details of spectra obtained at different epochs 
}
\begin{tabular}{cccccccc}
  \hline
   QSO & Telescope & Date & MJD  & Exposure & Wavelength & Spectral  & S/N \tnote{b} \\ 
       &  & (M/D/Y) & & time (s) & range (\AA)  & res. (\kms) &   \\
       \hline \hline
     J002809.45-053941.6 & SDSS & 10-06-2013 & 56571 & 4500 & 3562-10341 & 150 & 16.52 \\
     & SALT & 07-21-2020 & 59051 & 2400 & 4461-7524 & 304 & 33.54 \\
     & SALT & 06-20-2021 & 59385 & 2400 & 3321-6430 & 304 & 5.01 \\
     & SALT & 06-08-2022 & 59738 & 2400 & 3312-6426 & 304 & 22.42 \\
     &  &  &  &  &  \\
    J003419.19+030913.1 & SDSS & 12-01-2010 & 55531 & 3603 & 3589-10363 & 150 & 19.46 \\
\hline
 \end{tabular}
\label{tab_obs}
\end{threeparttable}
\end{table*}

\begin{table*}
\begin{threeparttable}
    \centering
 \begin{tabular}{cccccccc}
  \hline
   QSO & Telescope & Date & MJD  & Exposure & Wavelength & Spectral  & S/N \tnote{b} \\ 
       &  & (M/D/Y) & & time (s) & range (\AA)  & res. (\kms) &   \\
       \hline \hline
       
     & SALT & 09-12-2021 & 59469 & 2400 & 3613-6706 & 304 & 37.2 \\
     & SALT & 09-27-2022 & 59849 & 2400 & 3603-6703 & 304 & 48.38 \\
     &  &  &  &  &  \\
    J022431.59-052818.9 & SDSS & 10-17-2012 & 56217 & 4504 & 3561-10332 & 150 & 38.56 \\
     & SALT & 08-07-2021 & 59433 & 2200 & 3755-6843 & 304 & 73.21 \\
     & SALT & 08-21-2022 & 59812 & 2200 & 3754-6841 & 304 & 62.41 \\
\hline
 \end{tabular}

\end{threeparttable}
\end{table*}

\begin{table*}
\begin{threeparttable}
    \centering
 \begin{tabular}{cccccccc}
  \hline
   QSO & Telescope & Date & MJD  & Exposure & Wavelength & Spectral  & S/N \tnote{b} \\ 
       &  & (M/D/Y) & & time (s) & range (\AA)  & res. (\kms) &   \\
       \hline \hline
    J093251.98+023727.0 & SDSS & 02-25-2001 & 51965 & 4505 & 3815-9202 & 150 & 17.15 \\
     & SDSS & 04-09-2011 & 55660 & 7206 & 3593-10392 & 150 & 28.69 \\
     & SALT & 02-23-2021 & 59268 & 2400 & 3890-6684 & 304 & 34.29 \\
     & SALT & 01-01-2022 & 59580 & 2400 & 3896-6979 & 304 & 47.79 \\
     &  &  &  &  &  \\
    J095132.13-015706.9 & SDSS & 01-03-2011 & 55564 & 5405 & 3555-10315 & 150 & 11.62 \\
\hline
 \end{tabular}

\end{threeparttable}
\end{table*}

\begin{table*}
\begin{threeparttable}
    \centering
 \begin{tabular}{cccccccc}
  \hline
   QSO & Telescope & Date & MJD  & Exposure & Wavelength & Spectral  & S/N \tnote{b} \\ 
       &  & (M/D/Y) & & time (s) & range (\AA)  & res. (\kms) &   \\
       \hline \hline
       
    J133138.49+004221.1 & SDSS & 04-28-2000 & 51662 & 2700 & 3813-9234 & 150 & 9.03 \\
     & SDSS & 02-15-2001 & 51955 & 6300 & 3811-9215 & 150 & 12.53 \\
     & SDSS & 02-13-2011 & 55605 & 3603 & 3595-10363 & 150 & 12.1 \\
     & SDSS & 03-02-2011 & 55622 & 4504 & 3557-10320 & 150 & 10.73 \\
     &  &  &  &  &  \\
    J134112.37-011545.6 & SDSS & 06-02-2002 & 52427 & 3003 & 3816-9212 & 150 & 18.34 \\
\hline
 \end{tabular}

\end{threeparttable}
\end{table*}

\begin{table*}
\begin{threeparttable}
    \centering
 \begin{tabular}{cccccccc}
  \hline
   QSO & Telescope & Date & MJD  & Exposure & Wavelength & Spectral  & S/N \tnote{b} \\ 
       &  & (M/D/Y) & & time (s) & range (\AA)  & res. (\kms) &   \\
       \hline \hline

    J160650.08+074648.7 & SDSS & 04-01-2012 & 56018 & 2702 & 3564-10339 & 150 & 20.98 \\
     & SALT & 06-26-2020 & 59026 & 2300 & 4321-7389 & 304 & 36.35 \\
     & SALT & 06-20-2021 & 59385 & 2300 & 3198-6294 & 304 & 20.44 \\
     &  &  &  &  &  \\
    J160956.17+052651.2 & SDSS & 08-15-2007 & 54327 & 11111 & 3817-9200 & 150 & 22.47 \\
     & SDSS & 05-27-2011 & 55708 & 4504 & 3560-10322 & 150 & 17.41 \\
\hline
 \end{tabular}
 \begin{tablenotes}\footnotesize
    \item[a] 
    \item[b] 

 \end{tablenotes}
\end{threeparttable}
\end{table*}

\begin{table*}
\begin{threeparttable}
    \centering
 \begin{tabular}{cccccccc}
  \hline
   QSO & Telescope & Date & MJD  & Exposure & Wavelength & Spectral  & S/N \tnote{b} \\ 
       &  & (M/D/Y) & & time (s) & range (\AA)  & res. (\kms) &   \\
       \hline \hline

    J231027.23+074658.2 & SDSS & 09-17-2012 & 56187 & 4504 & 3611-10392 & 150 & 23.26 \\
     & SALT & 07-25-2020 & 59055 & 2300 & 4178-7252 & 304 & 28.78 \\
     & SALT & 07-18-2021 & 59413 & 2300 & 3175-6290 & 304 & 37.9 \\
     & SALT & 06-27-2022 & 59757 & 2300 & 3160-6283 & 304 & 29.01 \\
     &  &  &  &  &  \\
    J235238.08+010552.3 & SDSS & 09-01-2000 & 51788 & 1800 & 3794-9196 & 150 & 19.94 \\
\hline
 \end{tabular}
 \begin{tablenotes}\footnotesize
    \item[a] 
    \item[b] 

 \end{tablenotes}

\end{threeparttable}
\end{table*}

\begin{table*}
\begin{threeparttable}
    
    \centering
\caption{BAL properties at different epochs 
}
 \begin{tabular}{ccccccccccc}
  \hline
   QSO & \zem\ & BAL & \zabs\  & $v_{min}$ & $v_{max}$ & MJD  & $v_{cent}$ & W & $\sigma_{W}$ & $d_{BAL}$ \\ 
       &  & & & \kms\ & \kms\ & & \kms & \AA\ & \AA\ &   \\
       \hline \hline

    J0028-0539 	&	 2.5584 	&	 1 	&	 2.36 	&	 10378 	&	 25096 	&	 56571 	&	 17001 	&	 25.29 	&	 0.25 	&	 0.66 \\ 
    &	  	&	  	&	 	&	 	&	 	&	 59051 	&	 16302 	&	 19.93 	&	 0.13 	&	 0.57 \\ 
    &	  	&	  	&	 	&	 	&	 	&	 59738 	&	 16333 	&	 25.47 	&	 0.19 	&	 0.66 \\ 
    J0034+0309 	&	 2.3329 	&	 1 	&	 2.14 	&	 10799 	&	 26307 	&	 55531 	&	 17575 	&	 19.30 	&	 0.23 	&	 0.59 \\ 
    &	  	&	  	&	 	&	 	&	 	&	 58077 	&	 16669 	&	 27.14 	&	 0.22 	&	 0.71 \\ 
    &	  	&	  	&	 	&	 	&	 	&	 59046 	&	 16659 	&	 25.65 	&	 0.14 	&	 0.65 \\
\hline
 \end{tabular}
\label{tab_appendix2}
\end{threeparttable}
\end{table*}

\begin{table*}
\begin{threeparttable}
    \centering

 \begin{tabular}{ccccccccccc}
  \hline
   QSO & \zem\ & BAL & \zabs\  & $v_{min}$ & $v_{max}$ & MJD  & $v_{cent}$ & W & $\sigma_{W}$ & $d_{BAL}$ \\ 
       &  & & & \kms\ & \kms\ & & \kms & \AA\ & \AA\ &   \\
       \hline \hline

J0220-0812 	&	 2.0095 	&	 1 	&	 1.86 	&	 9509 	&	 29709 	&	 52162 	&	 15612 	&	 29.76 	&	 0.39 	&	 0.88 \\ 
 	&	  	&	  	&	 	&	 	&	 	&	 54821 	&	 14903 	&	 26.78 	&	 0.20 	&	 0.90 \\ 
 	&	  	&	  	&	 	&	 	&	 	&	 55833 	&	 16113 	&	 33.45 	&	 0.27 	&	 0.88 \\ 
 	&	  	&	  	&	 	&	 	&	 	&	 59469 	&	 15649 	&	 31.23 	&	 0.15 	&	 0.90 \\ 
 	&	  	&	  	&	 	&	 	&	 	&	 59849 	&	 16394 	&	 36.82 	&	 0.10 	&	 0.92 \\ 
J0224-0528 	&	 2.0845 	&	 1 	&	 1.91 	&	 14234 	&	 20623 	&	 56217 	&	 16954 	&	 5.28 	&	 0.09 	&	 0.32 \\
\hline
 \end{tabular}
\label{tab_appendix2}
\end{threeparttable}
\end{table*}

\begin{table*}
\begin{threeparttable}
    \centering

 \begin{tabular}{ccccccccccc}
  \hline
   QSO & \zem\ & BAL & \zabs\  & $v_{min}$ & $v_{max}$ & MJD  & $v_{cent}$ & W & $\sigma_{W}$ & $d_{BAL}$ \\ 
       &  & & & \kms\ & \kms\ & & \kms & \AA\ & \AA\ &   \\
       \hline \hline

J0932+0237 	&	 2.1679 	&	 1 	&	 1.96 	&	 13042 	&	 30000 	&	 51965 	&	 20678 	&	 33.00 	&	 0.34 	&	 0.81 \\ 
 	&	  	&	  	&	 	&	 	&	 	&	 55660 	&	 21275 	&	 38.93 	&	 0.20 	&	 0.80 \\ 
 	&	  	&	  	&	 	&	 	&	 	&	 59268 	&	 21801 	&	 37.14 	&	 0.18 	&	 0.78 \\ 
 	&	  	&	  	&	 	&	 	&	 	&	 59580 	&	 21676 	&	 42.90 	&	 0.10 	&	 0.79 \\ 
J0951-0157 	&	 3.2553 	&	 1 	&	 2.97 	&	 14909 	&	 27271 	&	 55564 	&	 20964 	&	 23.18 	&	 0.38 	&	 0.82 \\ 
 	&	  	&	  	&	 	&	 	&	 	&	 59229 	&	 20496 	&	 13.38 	&	 0.07 	&	 0.70 \\
\hline
 \end{tabular}
\label{tab_appendix2}
\end{threeparttable}
\end{table*}

\begin{table*}
\begin{threeparttable}
    \centering

 \begin{tabular}{ccccccccccc}
  \hline
   QSO & \zem\ & BAL & \zabs\  & $v_{min}$ & $v_{max}$ & MJD  & $v_{cent}$ & W & $\sigma_{W}$ & $d_{BAL}$ \\ 
       &  & & & \kms\ & \kms\ & & \kms & \AA\ & \AA\ &   \\
       \hline \hline

J1331+0042 	&	 2.4341 	&	 1 	&	 2.35 	&	 6555 	&	 9453 	&	 51662 	&	 7810 	&	 2.11 	&	 0.21 	&	 0.22 \\ 
 	&	  	&	  	&	 	&	 	&	 	&	 51955 	&	 7931 	&	 3.65 	&	 0.16 	&	 0.39 \\ 
 	&	  	&	  	&	 	&	 	&	 	&	 55605 	&	 7727 	&	 3.59 	&	 0.15 	&	 0.48 \\ 
 	&	  	&	  	&	 	&	 	&	 	&	 55622 	&	 7745 	&	 3.29 	&	 0.16 	&	 0.37 \\ 
 	&	  	&	 2 	&	 2.27 	&	 11383 	&	 16893 	&	 51662 	&	 14261 	&	 2.01 	&	 0.33 	&	 0.24 \\ 
 	&	  	&	  	&	 	&	 	&	 	&	 51955 	&	 13900 	&	 4.17 	&	 0.24 	&	 0.32 \\
\hline
 \end{tabular}
\label{tab_appendix2}
\end{threeparttable}
\end{table*}

\begin{table*}
\begin{threeparttable}
    \centering

 \begin{tabular}{ccccccccccc}
  \hline
   QSO & \zem\ & BAL & \zabs\  & $v_{min}$ & $v_{max}$ & MJD  & $v_{cent}$ & W & $\sigma_{W}$ & $d_{BAL}$ \\ 
       &  & & & \kms\ & \kms\ & & \kms & \AA\ & \AA\ &   \\
       \hline \hline

J1621+0758 	&	 2.1394 	&	 1 	&	 1.98 	&	 12355 	&	 19811 	&	 53501 	&	 15544 	&	 3.14 	&	 0.20 	&	 0.19 \\ 
 	&	  	&	  	&	 	&	 	&	 	&	 56016 	&	 15876 	&	 8.15 	&	 0.12 	&	 0.33 \\ 
 	&	  	&	  	&	 	&	 	&	 	&	 57181 	&	 15684 	&	 7.32 	&	 0.11 	&	 0.42 \\ 
 	&	  	&	  	&	 	&	 	&	 	&	 58245 	&	 15757 	&	 4.17 	&	 0.26 	&	 0.26 \\ 
 	&	  	&	  	&	 	&	 	&	 	&	 58606 	&	 16325 	&	 4.87 	&	 0.14 	&	 0.24 \\ 
 	&	  	&	  	&	 	&	 	&	 	&	 58991 	&	 16164 	&	 4.20 	&	 0.09 	&	 0.25 \\
\hline
 \end{tabular}
\label{tab_appendix2}
\end{threeparttable}
\end{table*}

\begin{table*}
\begin{threeparttable}
    \centering

 \begin{tabular}{ccccccccccc}
  \hline
   QSO & \zem\ & BAL & \zabs\  & $v_{min}$ & $v_{max}$ & MJD  & $v_{cent}$ & W & $\sigma_{W}$ & $d_{BAL}$ \\ 
       &  & & & \kms\ & \kms\ & & \kms & \AA\ & \AA\ &   \\
       \hline \hline

    &	  	&	 3 	&	 2.08 	&	 14232 	&	 29382 	&	 56187 	&	 21494 	&	 8.29 	&	 0.21 	&	 0.30 \\ 
 	&	  	&	  	&	 	&	 	&	 	&	 59055 	&	 21669 	&	 16.88 	&	 0.19 	&	 0.41 \\ 
 	&	  	&	  	&	 	&	 	&	 	&	 59413 	&	 21310 	&	 5.82 	&	 0.11 	&	 0.20 \\ 
 	&	  	&	  	&	 	&	 	&	 	&	 59757 	&	 20810 	&	 6.05 	&	 0.16 	&	 0.29 \\ 
J2352+0105 	&	 2.1513 	&	 1 	&	 1.92 	&	 14600 	&	 29706 	&	 51788 	&	 23040 	&	 16.74 	&	 0.26 	&	 0.80 \\ 
 	&	  	&	  	&	 	&	 	&	 	&	 55471 	&	 23278 	&	 15.78 	&	 0.15 	&	 0.76 \\
\hline
 \end{tabular}
\label{tab_appendix2}
\end{threeparttable}
\end{table*}

\bsp	
\label{lastpage}
\end{document}